\documentclass[prb,aps,twocolumn,10pt,floatfix,nofootinbib,superscriptaddress]{revtex4-2}

\usepackage{amsmath}
\usepackage{amssymb}
\usepackage{mathtools}

\usepackage{amsthm}
\usepackage{stackrel}
\usepackage[normalem]{ulem}
\usepackage[flushleft]{threeparttable}

\usepackage{amsfonts}
\usepackage{txfonts}
\usepackage{dsfont}   
\usepackage{pifont}                % double stroke font
\usepackage{bbold}                  % \mathbb{0}

\usepackage{array}
\usepackage{dcolumn}                % cell alignment at decimal point
\usepackage{makecell}               % linebreak in table cell
\usepackage{multirow}

\usepackage{enumerate}
\usepackage[shortlabels]{enumitem}

\usepackage[usenames,dvipsnames]{xcolor}
\usepackage{graphicx}
\graphicspath{{figs/}} % Setting the graphics path

\usepackage[version=4]{mhchem}      % chemical formulae
\usepackage{physics}                % physics notation
\usepackage{csquotes}               % quotations

\usepackage[colorlinks=true,citecolor=cyan,linkcolor=magenta,filecolor=magenta]{hyperref}
\usepackage[capitalize]{cleveref}
\usepackage{orcidlink}
\usepackage{mathrsfs}
\usepackage{placeins} 

\newcommand{\e}{\mathrm{e}}         % Euler number / exponential function
\DeclareMathAlphabet{\mathbbold}{U}{bbold}{m}{n}
\def\abs#1{\left|{#1}\right|}      	% absolute value, \abs{x} gives |x|
\def\bs#1{\boldsymbol{#1}}			% shorted version of "boldsymbol"
\def\imi{\mathrm{i}}				% imaginary i
\def\mcH{\mathcal{H}}					% Hamiltonian
\def\mcT{\mathcal{T}}					% Time-reversal symmetry operator
\def\ztwo{\mathbbold{Z}_2}					% parity group
\newcommand{\cmark}{\ding{51}}%
\newcommand{\xmark}{\ding{55}}%
	
\definecolor{TB}{rgb}{1,0.5,0}

\definecolor{PL}{rgb}{0,0.8,0.4}

\begin{document}

\title{Symmetry and topology of hyperbolic Haldane models}

    \author{Anffany Chen\,\orcidlink{0000-0002-0926-5801}
    }
    \email{anffany@ualberta.ca}
    \affiliation{Department of Physics, University of Alberta, Edmonton, Alberta T6G 2E1, Canada}
    \affiliation{Theoretical Physics Institute, University of Alberta, Edmonton, Alberta T6G 2E1, Canada}
    
    \author{Yifei Guan\,\orcidlink{0000-0001-7778-4083}
    }
    \affiliation{Institute of Physics, École Polytechnique Fédérale de Lausanne (EPFL), CH-1015 Lausanne, Switzerland}

    \author{Patrick M. Lenggenhager\,\orcidlink{0000-0001-6746-1387}
    }
    \affiliation{Department of Physics, University of Zurich, Winterthurerstrasse 190, 8057 Zurich, Switzerland}
    \affiliation{Condensed Matter Theory Group, Paul Scherrer Institute, 5232 Villigen PSI, Switzerland}
    \affiliation{Institute for Theoretical Physics, ETH Zurich, 8093 Zurich, Switzerland}
    
    \author{Joseph Maciejko\,\orcidlink{0000-0002-6946-1492}}
     \affiliation{Department of Physics, University of Alberta, Edmonton, Alberta T6G 2E1, Canada}
    \affiliation{Theoretical Physics Institute, University of Alberta, Edmonton, Alberta T6G 2E1, Canada}

    \author{Igor Boettcher\,\orcidlink{0000-0002-1634-4022}
    }
     \affiliation{Department of Physics, University of Alberta, Edmonton, Alberta T6G 2E1, Canada}
    \affiliation{Theoretical Physics Institute, University of Alberta, Edmonton, Alberta T6G 2E1, Canada}
    
    \author{Tom\'{a}\v{s} Bzdu\v{s}ek\,\orcidlink{0000-0001-6904-5264}
    }
    \email{tomas.bzdusek@uzh.ch}
    \affiliation{Condensed Matter Theory Group, Paul Scherrer Institute, 5232 Villigen PSI, Switzerland}
    \affiliation{Department of Physics, University of Zurich, Winterthurerstrasse 190, 8057 Zurich, Switzerland}

\date{\today}

\begin{abstract}
Particles hopping on a two-dimensional hyperbolic lattice feature unconventional energy spectra and wave functions that provide a largely uncharted platform for topological phases of matter beyond the Euclidean paradigm. Using real-space topological markers as well as Chern numbers defined in the higher-dimensional momentum space of hyperbolic band theory, we construct and investigate hyperbolic Haldane models, which are generalizations of Haldane's honeycomb-lattice model to various hyperbolic lattices. 
We present a general framework to characterize point-group symmetries in hyperbolic tight-binding models, and use this framework to constrain the multiple first and second Chern numbers in momentum space. 
We observe several topological gaps characterized by first Chern numbers of value $1$ and $2$. The momentum-space Chern numbers respect the predicted symmetry constraints and agree with real-space topological markers, indicating a direct connection to observables such as the number of chiral edge modes. 
With our large repertoire of models, we further demonstrate that the topology of hyperbolic Haldane models is trivialized for lattices with strong negative curvature. 
\end{abstract}

\maketitle
%%%%%%%%%%%%%%%%%%%%%%%%
%%%    MAIN TEXT     %%%
%%%%%%%%%%%%%%%%%%%%%%%%

\section{Introduction}
In a series of recent experiments, two-dimensional (2D) hyperbolic lattices emulating negatively curved space have been realized in circuit quantum electrodynamics and with topolectrical circuits~\cite{Kollar:2019,Lenggenhager:2021,Zhang:2022,Zhang:2023,Chen2023}.
These experimental breakthroughs have opened up a new research direction, namely, the study of \emph{hyperbolic matter}. 
The pivotal question permeating this arena is whether inclusion of negative curvature as a further tunable ingredient in condensed-matter setups facilitates unique physical phenomena or advantages over flat (Euclidean) systems.
%In particular, 
On the one hand, the great versatility of the electrical circuit platform enabled the realization of topological models~\cite{Zhang:2022,Zhang:2023}, in which a macroscopic fraction of all eigenstates are found to participate in the topological edge mode~\cite{Urwyler:2022}.
%opening the frontier of table-top implementation and design of novel topological phases of matter in curved space~\cite{Zhang:2022,Zhang:2023}. 
The above experimental achievements stimulated ample theoretical research into the design and characterization of Hamiltonians in curved spaces.
First theoretical investigations of topological models on hyperbolic lattices appeared in Refs.~\onlinecite{Urwyler:2022,Liu:2022,Chen2023,Zhang:2023,Yu:2020}. 
Other recently investigated aspects of hyperbolic lattices include hyperbolic band theory~\cite{Maciejko:2021,Maciejko:2022} and crystallography~\cite{Boettcher:2022}, non-Abelian Bloch states~\cite{Maciejko:2022,Cheng:2022,Kienzle:2022,nagy2022,Lenggenhager:2023}, periodic boundary conditions~\cite{sausset2007,Maciejko:2022,Zhu:2021,Lux:2022,Lux:2023} and the thermodynamic limit~\cite{Lux:2022,Lux:2023,Mosseri2023}, Hofstadter spectra~\cite{Yu:2020,Stegmaier:2021,Ikeda:2021,Ikeda:2021b}, effects of strong correlations~\cite{Daniska:2016,Zhu:2021,Bienias:2022,Gluscevich:2023}, continuum approximation~\cite{Boettcher:2020}, exact trace formulas~\cite{Attar:2022}, elastic vibrations~\cite{Ruzzene:2021}, 
flat bands~\cite{Kollar:2019,Saa:2021,Urwyler:2021,Bzdusek:2022,Mosseri:2022}, connections to holography \cite{PhysRevD.102.034511,PhysRevD.103.094507,PhysRevLett.130.091604,Basteiro:2022a,basteiro2022aperiodic}, and higher-order topological phenomena~\cite{Tao:2022,Liu:2022b}.
It is by now established that hyperbolic lattices exhibit single-particle dynamics and many-body phenomena richly distinct from those observed in Euclidean lattices.

In this work, we investigate the interplay of symmetry and topology in tight-binding models on hyperbolic lattices.
Particular attention is given to generalizations of the seminal Haldane model, originally defined on the $\{6,3\}$ honeycomb lattice~\cite{Haldane:1988}, to a variety of regular \{$p,q$\} hyperbolic lattices. 
Here, the Schl\"{a}fli symbol $\{p,q\}$ stands for a tessellation of the 2D plane by regular $p$-sided polygons (or ``tiles'') with $q$ of them meeting at each vertex.
For $(p-2)(q-2)=4$, this comprises a tessellation of the Euclidean plane, whereas in the infinite number of cases with $(p-2)(q-2)>4$, we have a tessellation of the hyperbolic plane with constant negative curvature. 
The original Haldane model transforms the ground state of the non-interacting tight-binding model for graphene from a semimetal to a Chern insulator by introducing complex-valued second-neighbor hopping amplitudes $t_2\, e^{\pm \imi \phi}$, where $\pm\phi$ correspond to alternating magnetic fluxes through parts of the tiles, in addition to a sublattice mass $m$ which trivializes the Chern insulator for large $m$~\cite{Semenoff1984, Haldane:1988}.
Analogously, on a general \{$p,q$\} lattice, complex-valued Haldane hoppings can be consistently assigned to all second-neighbor pairs, and the sublattice mass can be realized by staggered on-site potentials $\pm m$ if the lattice is bipartite (i.e., if it has an even value of $p$).

The key aspect that makes the characterization of band topology in hyperbolic lattices challenging is the non-commutative nature of translations in curved space. 
Specifically, the translation symmetry groups of 2D hyperbolic lattices are non-Abelian groups known as Fuchsian groups, which contrast with the commutative translation group $\mathbb{Z}^2$ in the 2D Euclidean case. 
Hyperbolic band theory (HBT) and crystallography~\cite{Maciejko:2021,Boettcher:2022,Maciejko:2022} imply that the non-commutativity of the translation generators necessitates a $2\mathfrak{g}$-dimensional Brillouin zone (BZ) torus spanned by momenta $\{k^i\}_{i=1}^{2\mathfrak{g}}$, each component valued in the range $[-\pi,+\pi]$, where $\mathfrak{g}$ is the genus of the compactified hyperbolic unit cell and is strictly greater than $1$.
In addition, higher-dimensional irreducible representations (IRs) also exist in non-toroidal representation spaces~\cite{Maciejko:2022} and provide an ansatz for further eigenstates~\cite{Cheng:2022}.
The characterization of hyperbolic band structures by topological invariants therefore presents new opportunities and challenges. 
For one, higher-dimensional topological invariants such as the second Chern number, conventionally incompatible with 2D systems, become accessible within the higher $2\mathfrak{g}$-dimensional momentum space~\cite{Zhang:2023}.
On the other hand, there exist $\binom{2\mathfrak{g}}{2} = \mathfrak{g}(2\mathfrak{g}{-}1)>1$ %independent 
first Chern numbers $C_{ij}$, one for each subtorus
$(k^i,k^j)$ of the hyperbolic BZ, despite the system being 2D in real space. 
This calls for a new interpretation to relate the multiple first Chern numbers to %a single 
observables such as the Hall conductance or the number of chiral edge modes.

The principal results of our work include the systematic mathematical characterization of Chern invariants in hyperbolic $\{p,q\}$ lattices, as well as numerically determined phase diagrams of hyperbolic Haldane models. Our analysis begins in Sec.~\ref{sec:real-space} with the construction of Haldane models on finite portions of $\{p,q\}$ hyperbolic lattices in the real coordinate space, henceforth called \emph{hyperbolic flakes}, for a large selection of integers $p$ and $q$. Using exact diagonalization, we calculate the bulk density of states (bulk DOS) as a function of the hopping phase $\phi$ for generic parameter choices. 
The bulk DOS of many \{$p, q$\} Haldane models exhibits gaps in which real-space Chern numbers~\cite{Kitaev:2006} are quantized to~$1$. 
In addition, we observe an intriguing universality in a qualitative feature of the DOS which is characterized solely by $p$, independent of $q$. 
Namely, the peaks and valleys in the DOS are centered around the same energies for models with the same $p$. Lastly, we uncover a correlation between the topology of hyperbolic Haldane models and the geometry of the underlying lattices: Haldane models on weakly curved hyperbolic lattices predominantly exhibit nontrivial topological gaps, while those with strong curvature typically remain gapless.

In Sec.~\ref{sec:k-space} we complement our real-space approach based on hyperbolic flakes by studying hyperbolic Haldane models on $\{p,q\}$ lattices with periodic boundary conditions using HBT. According to the automorphic Bloch theorem,
the eigenstates of such hyperbolic lattice models obey a Bloch
ansatz that is either Abelian~\cite{Maciejko:2021} or non-Abelian~\cite{Maciejko:2022}.
In this work, when applying HBT, we mean considering exclusively %only
the Abelian Bloch states $\psi_{\bs{k}}(z)$, which transform according to 1D IRs of the Fuchsian translation group:
\begin{equation}
    \psi_{\bs{k}}(\gamma^{-1}_j (z))=e^{{\rm i} k^j}\psi_{\bs{k}}(z).\label{eqn:Abelian-Bloch-ansatz}
\end{equation} 
Here, $\gamma_j$ is a generator of the non-commutative translation group and the hyperbolic crystal momentum $\bs{k}=(k^1,\dots,k^{2\mathfrak{g}})^\top$ is an element from a higher-dimensional $\textrm{BZ}\cong [-\pi,+\pi]^{2\mathfrak{g}}$. 
Abelian Bloch states and the related notion of hyperbolic crystallography are directly relevant to realistic experimental realizations of finite hyperbolic lattices~\cite{Kollar:2019,Lenggenhager:2021,Zhang:2022,Chen2023,Zhang:2023}. In Ref.~\onlinecite{Chen2023}, the Abelian Bloch spectrum was simulated using tunable-phase circuit elements, and in Ref.~\onlinecite{Zhang:2023}, hyperbolic circuits were engineered with periodic boundary conditions designed specifically to realize exclusively Abelian Bloch states.

In the present theoretical analysis, 
we focus on several hyperbolic lattices associated 
with small $\mathfrak{g}=2$ and $\mathfrak{g}=3$, which, respectively, possess 4D and 6D BZs. 
We tabulate the key information needed for implementing HBT in these lattices in Table~\ref{table:unit-cells} and Fig.~\ref{fig:unit-cells}, both shown on page~\pageref{fig:unit-cells}.
Abelian Bloch energy bands computed from HBT were previously found to very accurately capture energy spectra of non-topological nearest-neighbor tight-binding models on several hyperbolic lattices~\cite{Chen2023,Bzdusek:2022}. 
For the topological Haldane models, we find that the Abelian DOS from HBT agrees equally well with the real-space analysis, especially for lattices with weaker curvature. 
We expect the remaining deviations to be mainly attributable to non-Abelian Bloch states belonging to higher-dimensional IRs of the Fuchsian symmetry.
Such states are present in the real-space lattices but are not captured in the momentum space associated with HBT. 
Let us point out that the characterization of non-Abelian Bloch states on hyperbolic lattices is a topic of active ongoing research~\cite{Lux:2022,Lux:2023,Mosseri2023,Lenggenhager:2023}.

To characterize the topology of the hyperbolic Bloch bands from the vantage point of HBT, we derive constraints on the first and second Chern numbers on various subtori of the high-dimensional BZ, which are imposed by crystalline symmetry. %the symmetries of the hyperbolic Haldane models. 
The general form of these constraints appears in  Eqs.~(\ref{eqn:C1-constraint}--\ref{eqn:C2-constraint-6D}), and their specific application for Haldane models on $\{p,q\}$ lattices with $\mathfrak{g}=2$ and $\mathfrak{g}=3$ is summarized in Table~\ref{table:Cherns}. 
Next, we verify that the numerically computed first Chern numbers at all observed gaps obey these symmetry constraints, and that they match the real-space Chern numbers in cases where there is a unique symmetry-inequivalent first Chern number. 
All second Chern numbers are found to vanish for the hyperbolic Haldane models studied here, thus trivially satisfying the derived constraints. 
We further compute phase diagrams in terms of parameters $t_2$ and $m$ at half-filling and at the maximally time-reversal-breaking magnetic flux $\phi=\pi/2$, revealing additional insulating phases with Chern number $2$.
We note in passing that the derived relations among the Chern numbers potentially also apply to other platforms characterized with a higher-dimensional BZ, such as superlattices and quasicrystals~\cite{Lohse:2018,Petrides:2018,Su:2020,Koshino:2022}, and systems with synthetic dimensions~\cite{Zilberberg:2018,Weisbrich:2021,Chen:2021}.

Finally, before concluding the paper, we indicate in Sec.~\ref{sec:flux-patterns} how the symmetry analysis of momentum-space Chern numbers can be extended to Haldane-like models with a modified pattern of alternating magnetic fluxes. 
In particular, we reveal certain flux patterns that possess a reflection (i.e., mirror) symmetry, for which the symmetries allow for a non-vanishing momentum-space Chern number while simultaneously imposing a strictly zero real-space Chern number.
We summarize our results with a discussion of open problems in Sec.~\ref{sec:conclusion}, followed by Appendices~\ref{app:flake_Hh}--\ref{app:complete-Chern-matrices} which provide detailed mathematical derivations and a description of the numerical algorithms supporting the discussions in the main text. 
Supplementary code and data are available online as Ref.~\onlinecite{Chen:2023:SDC}.

\section{Hyperbolic Haldane models in real space}\label{sec:real-space}

In this section, we analyze a large selection of \{$p,q$\} hyperbolic Haldane models in real space. In Sec.~\ref{sec:flake-model-construction}, we briefly describe the numerical procedure for constructing the hyperbolic flakes and the corresponding tight-binding Haldane models, relegating details to Appendix~\ref{app:flake_Hh}. In Sec.~\ref{sec:bulk-dos}, we define and compute the bulk DOS to identify bulk gaps induced by the complex hopping terms. We observe bulk gaps in some of the models, while others remain gapless. In Sec.~\ref{sec:Cr}, we compute the real-space Chern numbers \cite{Kitaev:2006} at $\phi$=$\pi/2$ and find that the bulk gaps have Chern number 1. In Sec.~\ref{sec:curvature}, we infer an inverse 
correlation between the topology of hyperbolic Haldane model and the Gaussian curvature of the underlying lattice.

\subsection{Flake model construction}\label{sec:flake-model-construction}
We generalize the Haldane model to various  \{$p,q$\} hyperbolic lattices with the integer $p$ ranging from 5 to 12 and the integer $q<p$ ranging from 3 to 6. 
Hyperbolic lattices with $p < 5$ are not suitable for realizing the Haldane model, as the second-neighbor hoppings do not form closed paths inside each tile. 
We further focus on lattices with $q<7$, as we observe that
larger values of $q$ suppress gap opening from the complex Haldane terms. 

\begin{figure}[t]
\includegraphics[width=0.7\linewidth]{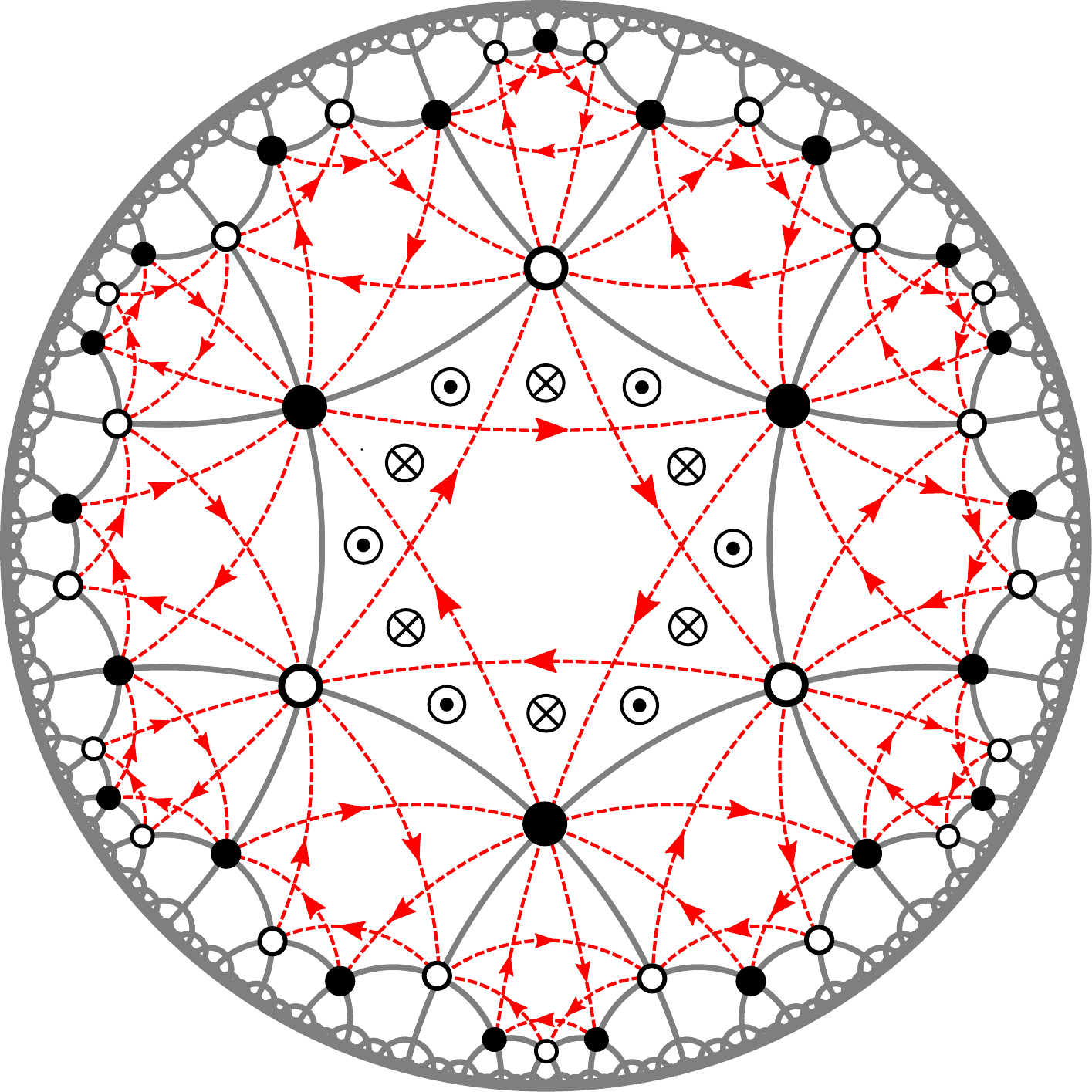}
\caption{\textbf{Hyperbolic Haldane models.} 
Here we portray the tight-binding Haldane model on the $\{6, 4\}$ lattice, with Hamiltonian given by Eq.~\eqref{eq:Hh}. 
The nearest-neighbor hopping is along the edges of the lattice graph (in gray). 
The sublattices A and B are marked by black and white circles. 
The second-neighbor hopping gains a phase $+\phi$ along the red arrows, which are clockwise within the polygons.
Physically, the hopping phases $\pm \phi$ correspond to perpendicular magnetic fluxes which alternate in sign (symbols $\otimes$ and $\odot$) across parts of the polygon, and which add up to zero within each polygon.}
\label{fig:hoppingdiagram}
\end{figure}

We numerically generate finite-sized flakes with open boundary conditions on the Poincar\'{e} disk  using the so-called vertex-inflation tiling procedure~\cite{Jahn2020}. 
For this, we start with a central $p$-polygon, or tile, and then iteratively attach additional ``rings'' of tiles.
At each iteration, we ensure that all open vertices are each 
equipped with $q$ adjacent tiles. 
We end the iterative process when the number of sites reaches a value between $\sim 10^3$ to $10^4$. 
For flakes constructed via vertex inflation, it is natural to define the outermost ring as the boundary and all the remaining inner rings as the bulk. The former is known to realize a 1D quasicrystal with discrete conformal symmetry~\cite{Boyle2020}.

Once the flakes are constructed, we use their adjacency matrices and site coordinates to define the \textit{hyperbolic Haldane models}, which are tight-binding models with Hamiltonian given by 
\begin{equation}
\begin{split}
     \mathcal{H}_{\rm H}(m,t_{1},t_{2},\phi) & = 
   t_{1}\underset{\langle i,j\rangle}{\sum}(c_{i}^{\dagger}c_{j}+c_{j}^{\dagger}c_{i})\\
   &+t_{2}\underset{\overrightarrow{ij}}{\sum}(e^{{\rm -i}\phi}c_{i}^{\dagger}c_{j}+e^{{\rm i}\phi}c_{j}^{\dagger}c_{i}) \\
   & +m\underset{i\in A}{\sum}c_{i}^{\dagger}c_{i}-m\underset{i\in B}{\sum}c_{i}^{\dagger}c_{i}.
\label{eq:Hh}
\end{split}
\end{equation}
Here, $t_1=-1$ is the nearest-neighbor hopping amplitude, $t_2$ ($\pm \phi$) is the magnitude (phase) of the second-neighbor hopping amplitude,  $\overrightarrow{ij}$ denotes pairs of second neighbors $i$ and $j$ such that the arrow from $i$ to $j$ is clockwise inside a tile (marked by the arrows in Fig.~\ref{fig:hoppingdiagram}), and $m$ is the strength of the staggered on-site potential, with opposite signs on sublattices~A and~B. 
In our convention, the sign in front of $\phi$ is positive for clockwise second-neighbor hopping. In addition, lattices with odd values of $p$ are not bipartite, in which case the staggered on-site potential $\pm m$ is omitted from the Hamiltonian~(\ref{eq:Hh}).
All numerical algorithms used for constructing the real-space hyperbolic Haldane models are detailed in Appendix~\ref{app:flake_Hh}.

\subsection{Bulk DOS \texorpdfstring{$\rho_\textrm{bulk}(E,\phi)$}{rho(E,phi)}} \label{sec:bulk-dos}
 
With only nearest-neighbor hoppings, $t_2=m=0$, the bulk DOS (defined below) is gapless. 
To identify bulk gaps induced by a nonzero $t_{2}$, we compute the DOS as a function of energy $E$ and phase $\phi$ with tight-binding parameters initially set to $m=0$ 
and $t_{2}=0.5$. 
In order to minimize the boundary contribution to the DOS, we define the bulk DOS as 
\begin{equation} \rho_{\text{bulk}}(E,\phi)\equiv\stackrel[j=1]{N}{\sum}\underset{z\in S_{\!\textrm{bulk}}}{\sum}|\psi_{j}(z,\phi)|^{2}f_{\eta}(E-\varepsilon_{j}), 
\label{eq:bdos}
\end{equation}
where $S_{\!\textrm{bulk}}$ is the set of lattice sites in the bulk (consisting of all the inner rings of the lattice obtained via vertex inflation), $N$ is the number of sites in the flake, and $\psi_{j}(z,\phi)$ is a single-particle eigenstate of Eq.~\eqref{eq:Hh} with eigenvalue $\varepsilon_{j}$. 
Further, we utilize a Gaussian smearing function $f_{\eta}(\varepsilon)=\frac{1}{\eta\sqrt{2\pi}}\exp(-\frac{\varepsilon^{2}}{2\eta^{2}})$ with parameter $\eta=0.063$ to obtain a smooth spectrum and reduce finite-size effects~\cite{Urwyler:2022}. 
A selection of numerically computed functions $\rho_{\text{bulk}}(E,\phi)$ is shown in Fig.~\ref{fig:dos}, plotted in red tones (see Fig.~\ref{supp_fig:flake-dos} in Appendix~\ref{app:flake_Hh} for the remaining figures).

In all $\{p,q\}$ models considered, we observe an intriguing $\phi$-dependence of $\rho_{\text{bulk}}(E,\phi)$ reminiscent of Hofstadter's butterfly. Interestingly, the functions $\rho_{\text{bulk}}(E,\phi)$ for Haldane models with the same value of $p$ bear similar qualitative features as regards the location of regions of low or vanishing DOS, irrespective of the value of $q$. 
Such universality in the bulk DOS characterized by $p$ alone is also observed in hyperbolic Hofstadter butterflies~\cite{Stegmaier:2021}, and even in the DOS of hyperbolic tight-binding models with only nearest-neighbor hopping (see Fig.~\ref{supp_fig:Hnn-dos} in Appendix~\ref{app:flake_Hh}).

\makeatletter\onecolumngrid@push\makeatother
\begin{figure*}[htbp]
\includegraphics[width=0.84\linewidth]{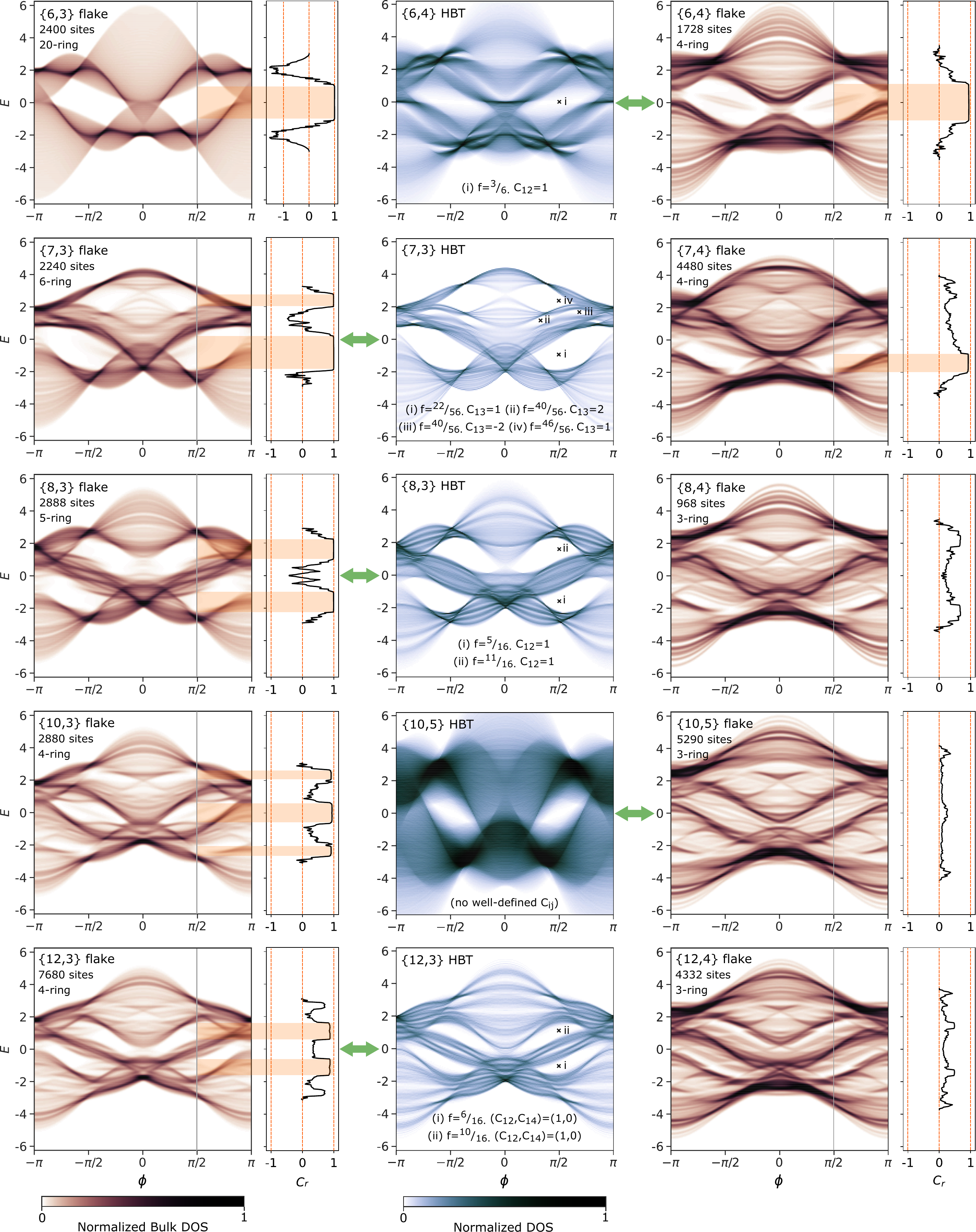}
\caption{\textbf{DOS of hyperbolic Haldane models.} {\it Columns 1 and 3, plotted in red tones}: 
The normalized bulk DOS $\rho_{\text{bulk}}(E,\phi)$ of various $\{p,q\}$ hyperbolic Haldane models on flakes are shown here with $(t_2,m)$=$(0.5,0)$. 
White pockets indicate potentially gapped regions. 
The real-space Chern number $C_{r}(\mu)$ is computed at $\phi=\pi/2$, with nonzero quantized $C_{r}$ plateaus (indicating Chern-insulator phases) observed in some of the models. 
Orange highlights are added to emphasize that  
plateaus coincide with the white pockets in $\rho_{\text{bulk}}(E,\phi)$. 
Models with the same $p$ display similar qualitative features in the DOS and $C_{r}$, pointing to a universality characterized by $p$ alone.
{\it Column 2, plotted in blue tones:}
The DOS obtained from the HBT characterization of the Haldane models are shown alongside the bulk DOS of the corresponding flake models. The HBT DOS displays a good agreement with the flake bulk DOS, especially for lattices with weaker curvatures.
In each identified gapped region (marked by ``$\times$''), the first Chern numbers $C_{ij}$ are computed on all 
subtori $(k^i,k^j)$ of the 4D or 6D hyperbolic BZ, and reduced by point-group symmetry to one or two independent first Chern numbers (Sec.~\ref{eqn:Chern-theory}).  In cases with only one independent first Chern number, the momentum-space Chern numbers agree with the real-space Chern numbers. The HBT DOS for $\{8,4\}$ and $\{12,4\}$ are shown in Fig.~\ref{supp_fig:hbt-dos}, and do not exhibit energy gaps.
\label{fig:dos}}
\end{figure*}
\clearpage
\makeatletter\onecolumngrid@pop\makeatother

Low-DOS regions, visible as white pockets in Fig.~\ref{fig:dos} and Fig.~\ref{supp_fig:flake-dos}, are observed in some of the hyperbolic Haldane models. 
In particular, $\{p,3\}$ models have the most distinguishable low-DOS regions, which, however, shrink or vanish as $q$ increases. 
Note that $\rho_{\text{bulk}}(E,\phi)$ includes residual contributions from the boundary modes leaking into the bulk of the flakes. Therefore, we consider all white pockets with small but non-vanishing bulk DOS as potentially fully gapped regions.

\subsection{Real-space Chern numbers} \label{sec:Cr}

We use Kitaev's real-space Chern number~\cite{Kitaev:2006} as a real-space topological marker to determine the topology of the observed low-DOS regions~\cite{Urwyler:2022}. 
To that end, we first divide the bulk region of a flake into three wedge-shaped regions $A$, $B$, and $C$, arranged in counterclockwise order, and define the projector onto the Hilbert space spanned by the occupied states~by
\begin{equation} 
\mathbf{P}^{\mu}=\underset{E_{n}<\mu}{\sum}|\psi_{n}\rangle\langle\psi_{n}|,
\end{equation} 
where $\mu$ is a chemical potential set to lie within the low-DOS region.
The real-space Chern number of the corresponding, potentially insulating region is then given by
\begin{equation} 
C_{r}(\mu)=12\pi{\rm i}\underset{a\in A}{\sum}\underset{b\in B}{\sum}\underset{c\in C}{\sum}(\mathbf{P}_{ab}^{\mu}\mathbf{P}_{bc}^{\mu}\mathbf{P}_{ca}^{\mu}-\mathbf{P}_{ac}^{\mu}\mathbf{P}_{cb}^{\mu}\mathbf{P}_{ba}^{\mu}).
\end{equation} 
In systems with Euclidean translation symmetry, $C_r(\mu)$ becomes the momentum-space Chern number or TKNN invariant~\cite{Thouless:1982} when expressed in the momentum representation. One can further show that $C_r(\mu)$ is a topological invariant even in the absence of Euclidean translation symmetry, such that it is quantized in the infinite-system limit and independent of the detailed shape of the regions $A$, $B$, and $C$~\cite{Kitaev:2006}.

If the system is not gapped at the chemical potential $\mu$, $C_r(\mu)$ can still be defined and computed, but is not quantized. In particular, in a finite system, boundary states with long penetration depth can trivialize $C_r(\mu)$. To minimize this effect, we define the bulk and boundary regions so that the bulk region is maximized without overlapping significantly with the supports of boundary states. Our definition of the boundary as the last ring during vertex inflation leads to optimal real-space Chern numbers in this sense. 
We compute $C_r(\mu)$ at $\phi = \pi/2$ for $(t_2,m)$=$(0.5,0)$ in the hyperbolic Haldane models and show the main results in Fig.~\ref{fig:dos}. (The rest is shown in Fig.~\ref{supp_fig:flake-dos}). We observe quantized plateaus with $C_r=1.0\pm0.1$ in $\{p,3\}$ models for $6\le p\le 12$ and in $\{p,4\}$ models for $5\le p\le 7$. This is a strong indication that the low-DOS regions in these models represent Chern-insulating phases with a single chiral edge mode. The remaining models considered do not exhibit quantized $C_r$ plateaus, so we conclude that they are gapless. We finally investigate the gap opened by nonzero $m$ at $\mu=0$ and $\phi=\pi/2$ for the bipartite models. This gap is topologically trivial when $t_2=0$. We find that those Haldane models without topological gaps for $(t_2,m)$=$(0.5,0)$ generally do not exhibit nontrivial $C_r$ for $0\le t_2\le 2$ and $m > 0$. 

\subsection{Correlation with curvature}\label{sec:curvature}

By using the same hopping amplitudes $t_{1}$ and $t_{2}$ for all $\{p,q\}$ models in the analysis above, we implicitly assume that the lattice constants $a$ are the same in all models. The Gaussian curvature $\kappa$ of a $\{p,q\}$ lattice is thus most appropriately measured in units of the inverse squared lattice constant:
\begin{equation} \kappa a^{2}=-\frac{a^2}{\mathcal{R}^2}=-\left(\text{arcosh}\left(1+\frac{2|z_{1}-z_{2}|^{2}}{(1-|z_{1}|^{2})(1-|z_{2}|^{2})}\right)\right)^2, \end{equation}
where $\mathcal{R}$ is the curvature radius and $z_{1}$ and $z_{2}$ are the Poincar\'{e} disk coordinates of any two nearest neighbors, see Eqs.~\eqref{eq:app_zn} and \eqref{eq:app_r0}.

\begin{figure}
\includegraphics[width=\linewidth]{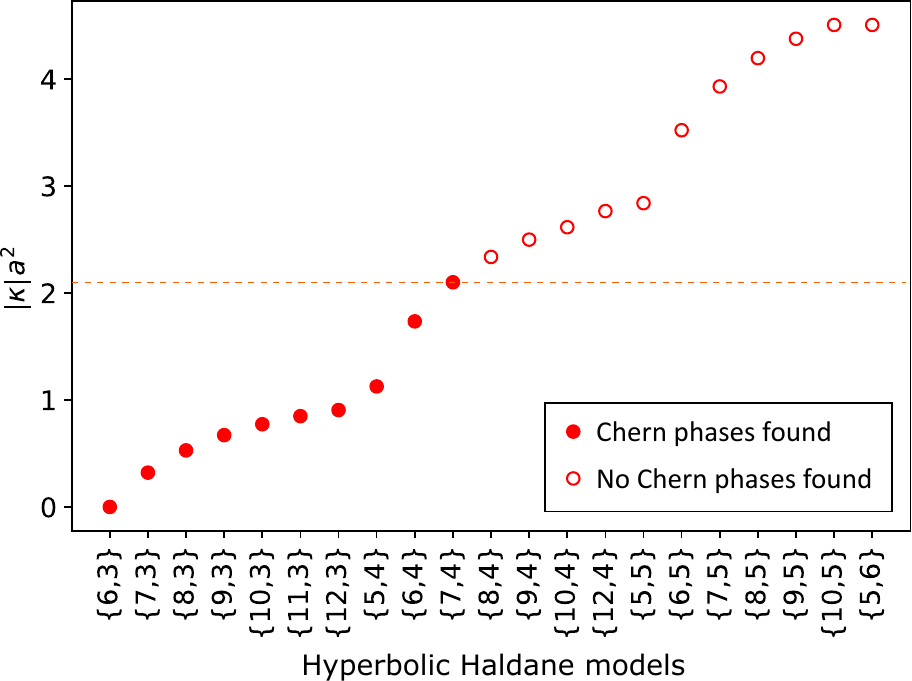}
\caption{Topology of hyperbolic Haldane models on the  $\{p,q\}$ lattice, plotted against $|\kappa|a^2$, the magnitude of the Gaussian curvature $\kappa~<~0$ in units of the inverse squared lattice constant $1/a^2$. Only those models with $|\kappa|a^2 \lesssim 2$ exhibit bulk gaps with nonzero quantized $C_r(\mu)$ plateaus at  $(t_2,m,\phi)=(0.5,0,\pi/2)$.}
\label{fig:Cr-trend}
\end{figure}

In Fig.~\ref{fig:Cr-trend}, we sort our hyperbolic Haldane models by increasing curvature magnitude $|\kappa|$. Incidentally, the models in which we observe Chern phases (marked by solid dots) are the ones with weaker curvatures. Further study is required to explain this correlation, conceivably in terms of the effect of spatial curvature on the formation of anomalous cyclotron orbits \cite{Comtet:1987}. 
Note that $\{p,3\}$ lattices up to $p=23$ have smaller curvatures than the $\{5,4\}$ lattice, which is the least curved $\{p,4\}$ lattice. 
Therefore the $\{p,3\}$ lattices are generally flatter and more suitable candidates for realizing nontrivial Chern-insulating phases in hyperbolic space.

\section{Hyperbolic Haldane models \texorpdfstring{\\}{} in momentum space}\label{sec:k-space}

In this section, we study selected hyperbolic Haldane models from Sec.~\ref{sec:real-space} using the perspective of HBT.
In this approach, one first identifies the Bravais lattice or, equivalently, the hyperbolic translation group of the corresponding $\{p,q\}$ lattice~\cite{Maciejko:2021,Maciejko:2022,Boettcher:2022}, together with its 1D IRs which are labelled by a momentum $\bs{k}$ residing in a higher-dimensional BZ. 
We provide a summary of selected aspects of hyperbolic crystallography in Sec.~\ref{sec:main-crystallography}, with the key information summarized in Fig.~\ref{fig:unit-cells} and Table~\ref{table:unit-cells}, while relegating further details to Appendix~\ref{sec:64-12-HBT}.
In a second step, one uses these 1D IRs as an ansatz to construct a family of eigenstates $\psi_{\bs{k}}$, hyperbolic Bloch 
states, to construct the Bloch
Hamiltonian $H(\bs{k})$ of the system. 
The eigenvalues of $H(\bs{k})$ are the Bloch energy bands of the underlying model.
The details of this construction, together with a discussion of the computed DOS functions and the identification of bulk energy gaps, are the subject of Sec.~\ref{sec:HBT-DOS}.

Having constructed the Bloch
Hamiltonians, we subsequently compute the momentum-space topology of the associated Bloch energy bands, namely their first and second Chern numbers.
We first analyze how the Chern numbers on the various subtori of the high-dimensional BZ are constrained by the symmetry of the hyperbolic lattices. 
While first considerations along those directions for the $\{8,3\}$ lattice appeared in Ref.~\onlinecite{Urwyler:2022}, our present analysis, detailed in Appendices~\ref{app:derive-M-matrices} and~\ref{app:Chern-constraints}, provides a major generalization that encompasses all symmetry generators of the seven $\{p,q\}$ lattices considered within the framework of HBT in this work.
A summary of the analysis of these symmetry constraints appears in Sec.~\ref{eqn:Chern-theory} and in Table~\ref{table:Cherns}, with further details provided in Appendix~\ref{app:complete-Chern-matrices}.
Finally, in Sec.~\ref{sec:Chern-phases} we numerically compute the momentum-space Chern numbers of the hyperbolic Bloch eigenstates of the Haldane models on the seven considered lattices. We verify that the explicitly computed Chern numbers respect the theoretically predicted symmetry constraints, and we present phase diagrams illustrating the dependence of the topological invariants on the choice of parameters $t_2$ and $m$ at half-filling.

\subsection{Hyperbolic crystallography and band theory}\label{sec:main-crystallography}

Among the lattices considered in Sec.~\ref{sec:real-space}, an HBT description with 4D or 6D BZs applies
for $\{8,3\}$, $\{6,4\}$, $\{8,4\}$, $\{10,5\}$, $\{7,3\}$, $\{12,3\}$, and $\{12,4\}$~\cite{Maciejko:2021,Boettcher:2022,Cheng:2022,Bzdusek:2022,Chen2023}.
These seven lattices are thus amenable for a comparison of the real-space and momentum-space topology, and thus will be our principal focus within the entire Sec.~\ref{sec:k-space}.

The construction of the Bloch Hamiltonian for a given Haldane model within HBT can be broken down into three principal steps.
First, given a $\{p,q\}$ lattice, we specify its lattice symmetry, which in mathematical terms is the so-called triangle group $\Delta(2,q,p)$~\cite{Boettcher:2022}. 
In the present context, this group adopts the role of a \emph{hyperbolic space group}. 
The second step is to identify its \emph{hyperbolic translation group} $\mathsf{T}(2,q,p)$ defined via the following attributes~\cite{Boettcher:2022}: it is the (\emph{i}) largest (\emph{ii}) normal subgroup of (\emph{iii}) orientation-preserving transformations in $\Delta(2,q,p)$ that (\emph{iv}) contains no rotation elements. 
While this identification has previously been carried out for the $\{8,3\}$, $\{8,4\}$, $\{7,3\}$, $\{12,4\}$, and $\{10,5\}$ lattices~\cite{Boettcher:2022}, our work extend this list to include also the $\{6,4\}$ and $\{12,3\}$ lattices. In the last step, given the translation group, we construct all its 1D IRs, which are labelled by a crystal momentum $\bs{k}$ in a 4D or 6D BZ for the $\{p,q\}$ lattices considered here. The unit cells and generators for their translation groups are summarized in Fig.~\ref{fig:unit-cells}, with certain supplementary information on the group structure listed in Table~\ref{table:unit-cells}. 
In addition, for the symmetry analysis in Sec.~\ref{eqn:Chern-theory}, it is convenient to also perform a fourth step: construct the \emph{hyperbolic point group} $\mathsf{P}(2,q,p)$ as the quotient of $\Delta(2,q,p)$ by $\mathsf{T}(2,q,p)$. 

In Secs.~\ref{sec:Delta-groups}--\ref{sec:BZ-1DIRs} below, we provide a basic description of these four steps, which is necessary to understand how Chern numbers in hyperbolic Haldane models are constrained by lattice symmetry. 
We further provide a detailed summary of the key notions of hyperbolic crystallography in Appendix~\ref{app:crystallography}, apply those ideas to the $\{6,4\}$ and $\{12,3\}$ lattices in Appendices~\ref{app:6-4-symmetry} and~\ref{app:12-3-symmetry}, and showcase the hyperbolic point group of all seven lattices in Appendix~\ref{sec:hyper-PGs}. 
Although we focus on seven specific lattices in this paper, we stress that all $\{p,q\}$ lattices are amenable to an HBT description, albeit with potentially higher-dimensional BZs (see Table~\ref{table:p-g-genie} in Appendix~\ref{app:crystallography}).

\begin{figure}[t]
\includegraphics[width=0.9\linewidth]{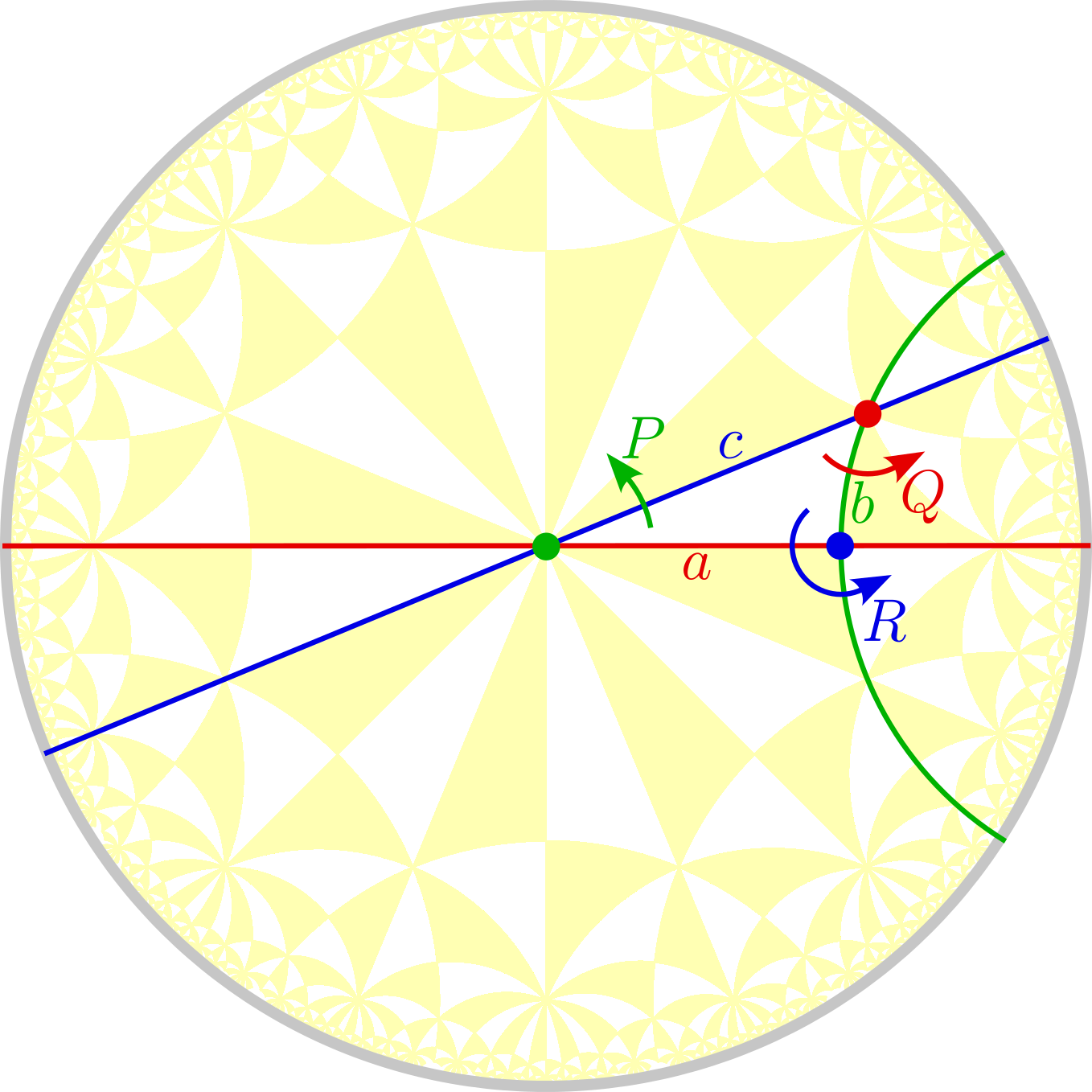}
\caption{
Tessellation of the hyperbolic plane by Schwarz triangles of $\Delta(2,q,p)$, illustrated for $p=8$, $q=4$. 
Schwarz triangles come in two orientations: positive (white) and negative (yellow). 
When plotting the various $\{p,q\}$ lattices (e.g.,~Fig.~\ref{fig:unit-cells}) we always position a white Schwarz triangle such that (\emph{i}) the $\pi/p$ angle touches the center of the Poincar\'{e} disk, and (\emph{ii}) the $\pi/2$ angle is to the right horizontally. 
We define reflections $a,b,c$ and rotations $P,Q,R$ with respect to this particular Schwarz triangle.
Specifically, $c$ is the reflection across its hypothenuse, and $a$ ($b$) is the reflection across its horizontal (vertical) edge.
By composing pairs of reflections we obtain $P,Q,R$, which are respectively the counter-clockwise rotations by $2\pi/p$, $2\pi/q$, $\pi$ about the triangle corners.
}
\label{fig:Schwarz-notation}
\end{figure}

\subsubsection{Hyperbolic space groups}\label{sec:Delta-groups}

The $\{p,q\}$ lattice is invariant under the action of the so-called triangle group $\Delta(2,q,p)$. 
This group is obtained by subdividing each $p$-gon into $2p$ right triangles known as \emph{Schwarz triangles}, and considering the reflections (and their compositions) across the edges of these triangles. 
A tessellation of the hyperbolic plane with Schwarz triangles related by elements of $\Delta(2,q,p)$ for $\{p,q\}=\{8,4\}$ is shown in Fig.~\ref{fig:Schwarz-notation}.
The triangles appear in two orientations. They have interior angles $\pi/2,\pi/q,\pi/p$ if read counter-clockwise for positively oriented triangles, shown in white (resp.~if read clockwise for negatively oriented triangles, shown in yellow).

The generators of $\Delta(2,q,p)$ are reflections across the edges of a given positively oriented Schwarz triangle. In our notation, we always take $a$ ($b$) to be the reflection across the edge of the Schwarz triangle whose adjacent angles are $\pi/2$ and $\pi/p$ ($\pi/q$), and $c$ to be the reflection across its hypotenuse. Rotations around the corners of the Schwarz triangle are obtained by composing the above reflections.\footnote{\label{eqn:foot-group-action}In this work, we always assume the \emph{left action} of symmetry groups. In particular, the product $gh$ acts first with $h$, then with $g$.
}
Namely $R = ab, Q = bc, P = ca$ are respectively counterclockwise rotations by $\pi$, $2\pi/q$, and $2\pi/p$ around the three corners of the Schwarz triangle. 
We indicate the underlying Schwarz triangles for all $\{p,q\}$ lattices considered in Fig.~\ref{fig:unit-cells}.

The Haldane models (even for $m=0$) explicitly break certain lattice symmetries.
Namely, the reflections $a,b,c$ (and all other orientation-reversing elements of the group) flip the sign of the fluxes $\pm \phi$.
This is corrected if each reflection is composed with the antiunitary operation of time reversal ($\mcT$). Therefore, the symmetry $\mathsf{M}(2,q,p)$ of Haldane models with $m=0$ is a \emph{magnetic hyperbolic space group} generated by $\mcT a$, $\mcT b$ and $\mcT c$. 
When one further assumes a sublattice mass $m\neq 0$, which is possible in models with $p$ even, then the reflection $\mcT a$ ceases to be a symmetry.
In its place, one supplements the antiunitary generators $\mcT b$ and $\mcT c$ with $(ca)^2=:P^2$, which corresponds to a counter-clockwise $4\pi/p$-rotation around a $p$-sided polygon. 
We denote the resulting magnetic hyperbolic space group as~$\widetilde{\mathsf{M}}(2,q,p)$. 

\subsubsection{Hyperbolic translation and point groups}\label{sec:T-groups}

In line with earlier works~\cite{Maciejko:2021,Maciejko:2022,Bzdusek:2022}, we define the hyperbolic translation group $\mathsf{T}(2,q,p)$ as the largest torsion-free normal subgroup of orientation-preserving elements in $\Delta(2,q,p)$.
The unit cell is subsequently defined as the collection of Schwarz triangles which tessellate the hyperbolic plane without overlaps or gaps if translated by all elements $\mathsf{T}(2,q,p)$.
In the above specification of $\mathsf{T}(2,q,p)$, one needs to clarify the meaning of the attributes (\emph{i}) ``largest'', (\emph{ii}) ``normal'', (\emph{iii}) ``orientation-preserving'', and (\emph{iv}) ``torsion-free''. 

Starting with (\emph{iv}), note that a characteristic feature of any translation group is the absence of elements of finite order, which are elements $g\neq 1$ such that $g^n = 1$ for some $n>1$, such as rotations and reflections. 
Elements of finite order are also called torsion elements, and a group without torsion elements is labelled torsion-free.
For (\emph{ii}), the requirement to be normal allows us to construct the hyperbolic point group as the quotient $\mathsf{P}(2,q,p)=\Delta(2,q,p)/\mathsf{T}(2,q,p)$. 
The order, i.e., number of elements, of the quotient is equal to the \emph{index of the subgroup} $\mathsf{T}$ in $\Delta$. For (\emph{iii}) note that the triangle group splits into transformations that either preserve or reverse orientation, analogous to proper and improper rotations.
Finally, concerning (\emph{i}), it turns out that $\Delta(2,q,p)$ has an infinite number of torsion-free normal subgroups of finite index, a result known as Fenchel's conjecture~\cite{bundgaard1951,fox1952,mennicke1967,chau1983,Siran:2001}. 
The physical significance of this mathematical result is that all $\{p,q\}$ lattices admit a Bravais lattice and a finite-order point group.
We select $\mathsf{T}(2,q,p)$ to be the largest such subgroup, i.e., of the smallest index, such that the order of $\mathsf{P}(2,q,p)$ is smallest. 
We show in Appendix~\ref{app:crystallography} that $\mathsf{T}(2,q,p)$ as defined above is generally also the largest torsion-free normal subgroup of the magnetic group $\mathsf{M}(2,q,p)$ and, for the lattices considered here, also of $\widetilde{\mathsf{M}}(2,q,p)$.

To identify the hyperbolic translation group $\mathsf{T}(2,q,p)$ in practice, we utilize a tabulated list of quotients of hyperbolic triangle groups that act on Riemann surfaces with genus $2\leq \mathfrak{g} \leq 101$~\cite{Conder:2007}. 
These quotients are exactly the (orientation-preserving) hyperbolic point groups, and from the knowledge of both $\mathsf{P}(2,q,p)$ and $\Delta(2,q,p)$ it is possible to reconstruct the translation subgroup $\mathsf{T}(2,q,p)$.
Furthermore, note that the translation group partitions all sites of the hyperbolic lattice into orbits. 
We define \emph{sites per unit cell} by selecting for each orbit under $\mathsf{T}(2,q,p)$ exactly one site which is located inside the unit cell or on its boundary.

\subsubsection{Hyperbolic Brillouin zones}\label{sec:BZ-1DIRs}

The point group $\mathsf{P}(2,q,p)$ acts on the (cosets of) Schwarz triangles that constitute the unit cell, which is a (not necessarily regular) polygon with $2\mathfrak{e}$ edges. 
The translation group $\mathsf{T}(2,q,p)$ is generated by elements $\{\gamma_j\}_{j=1}^{\mathfrak{e}}$ which relate pairs of edges of the unit cell.
Identifying these pairs yields a Riemann surface which is characterized by its genus $\mathfrak{g} \leq \mathfrak{e}/2$.\footnote{\label{foot:corner-Euler}This follows from computing the Euler characteristic $\mathfrak{\chi}$ of the compactified unit cell, which has $\mathfrak{f}=1$ face, $\mathfrak{e}$ distinct edges, and $\mathfrak{v}=1+\delta\mathfrak{v}$ distinct vertices (where ``$1$'' is the smallest possible value, relevant if all unit cell vertices are identified as the same vertex). From the definition $\chi = \mathfrak{f}-\mathfrak{e}+\mathfrak{v}$ combined with the Gauss-Bonnet theorem $\chi=2(1-\mathfrak{g})$ one derives $\mathfrak{g}=\mathfrak{e}/2 -\delta\mathfrak{v}/2$.}
Hyperbolic momenta correspond to fluxes that can be threaded through the 2$\mathfrak{g}$ holes of the genus-$\mathfrak{g}$ Riemann surface, implying the appearance of a $2\mathfrak{g}$-dimensional BZ of 1D IRs of the translation group~\cite{Maciejko:2021}.
More precisely, given a model invariant under $\mathsf{T}(2,q,p)$, every choice of a $2\mathfrak{g}$-component momentum $\bs{k}\in\textrm{BZ}$ is an ansatz to construct eigenstates obeying certain twisted boundary conditions across the $\mathfrak{e}$ unit cell edge pairs.

\makeatletter\onecolumngrid@push\makeatother

\begin{table*}[p]
\caption{Translation group for the $\{p,q\}$ lattices studied with HBT in Sec.~\ref{sec:k-space}, with number of sites ($\{p,q\}$ vertices) per unit cell.
The geometric meaning of the generators $\{\gamma_j\}_{j=1}^{\mathfrak{e}}$ is shown in Fig.~\ref{fig:unit-cells}.
The relators, products of generators set to unity in the respective translation group, imply that momenta appearing in the 1D IRs $\gamma_j \mapsto \e^{\imi k^j}$ are constrained as listed. 
The linearly independent momenta determine the dimension of the higher-dimensional BZ, which can be interpreted as the number $2\mathfrak{g}$ of fluxes threaded through the non-contractible cycles of a compactified unit cell. 
The compactification is achieved through identification of unit cell edges related by translation generators as shown in Fig.~\ref{fig:unit-cells}.
The last column identifies the resulting genus-$\mathfrak{g}$ Riemann surface by either its name or the equation of its complex algebraic curve ($w,z\in\mathbb{C}$)~\cite{Wolfart:2005}.
}
\begin{ruledtabular}
\begin{tabular}{ccccccc}
lattice &
sites per cell &
generators &
relators &
constraints on momentum coordinates &
BZ &
unit cell
\tabularnewline
\hline 
$\{8,3\}$ &
$16$ &
$\gamma_{1,\ldots,4}$ &
$\gamma_4\gamma_3^{-1}\gamma_2\gamma_1^{-1}\gamma_4^{-1}\gamma_3\gamma_2^{-1}\gamma_1$ &
(none)  &
4D   &
Bolza
\tabularnewline
$\{6,4\}$ &
$6$  &
$\gamma_{1,\ldots,6}$ &
$\gamma_1\gamma_3\gamma_5\,,\;\gamma_2\gamma_4\gamma_6\,,\; \gamma_1\gamma_2\gamma_3\gamma_4\gamma_5\gamma_6$ &
$k^1+k^3+k^5 = k^2+k^4+k^6 = 0$  &
4D &
$w^2 = z^6 - 1$
\tabularnewline
$\{8,4\}$ &
$4$ &
$\gamma_{1,\ldots,4}$ &
$\gamma_4\gamma_3^{-1}\gamma_2\gamma_1^{-1}\gamma_4^{-1}\gamma_3\gamma_2^{-1}\gamma_1$ &
(none)  &
4D &
Bolza
\tabularnewline
$\{10,5\}$ &
$2$  &
$\gamma_{1,\ldots,5}$ &
$\gamma_5\gamma_4^{-1}\gamma_3\gamma_2^{-1}\gamma_1\,,\; \gamma_1\gamma_2^{-1}\gamma_3\gamma_4^{-1}\gamma_5$ &
$k^1 - k^2 + k^3 - k^4 + k^5 =0$  &
4D &
$w^2 = z^5 - 1$
\tabularnewline
$\{7,3\}$ &
$56$  &
$\gamma_{1,\ldots,7}$ &
$\gamma_1 \gamma_3 \gamma_5 \gamma_7 \gamma_2 \gamma_4 \gamma_6\, , \; \gamma_4 \gamma_7 \gamma_3 \gamma_6 \gamma_2 \gamma_5\gamma_1$ &
$k^1 + k^2 + k^3 + k^4 + k^5 + k^6 + k^7 =0$  &
6D &
Klein
\tabularnewline
\multirow{ 2}{*}{$\{12,3\}$} &
\multirow{ 2}{*}{$16$} &
\multirow{ 2}{*}{$\gamma_{1,\ldots,12}$} &
$\gamma_{10} \gamma_7 \gamma_4 \gamma_1\, , \; \gamma_{11} \gamma_8 \gamma_5 \gamma_2\, , \; \gamma_{12} \gamma_9 \gamma_6 \gamma_3$ &
$k^1 {+} k^5 {+} k^9  = k^2 {+} k^6 {+} k^{10} = k^3 {+} k^7 {+} k^{11} = k^4 {+} k^8 {+} k^{12} = 0$  &
\multirow{ 2}{*}{6D}   &
\multirow{ 2}{*}{{$\textrm{M}(3)$}}
\tabularnewline
& & & 
$\gamma_9\gamma_5\gamma_1\, , \; \gamma_{10}\gamma_6\gamma_2\, , \; \gamma_{11}\gamma_7\gamma_3\, , \; \gamma_{12}\gamma_8\gamma_4$ &
$k^1 {+} k^4 {+} k^7 {+} k^{10} = k^2 {+} k^5 {+} k^8 {+} k^{11} = k^3 {+} k^6 {+} k^9 {+} k^{12} =0$ & &
\tabularnewline
$\{12,4\}$ &
$6$  &
$\gamma_{1,\ldots,6}$ &
$\gamma_6\gamma_5^{-1}\gamma_4\gamma_3^{-1}\gamma_2\gamma_1^{-1}\gamma_6^{-1}\gamma_5\gamma_4^{-1}\gamma_3\gamma_2^{-1}\gamma_1$ &
(none)  &
6D &
$w^2 = z^7 - z$
\end{tabular}\end{ruledtabular} 
\label{table:unit-cells} 
\end{table*}

\begin{figure*}[p]
\includegraphics[width=0.85\linewidth]{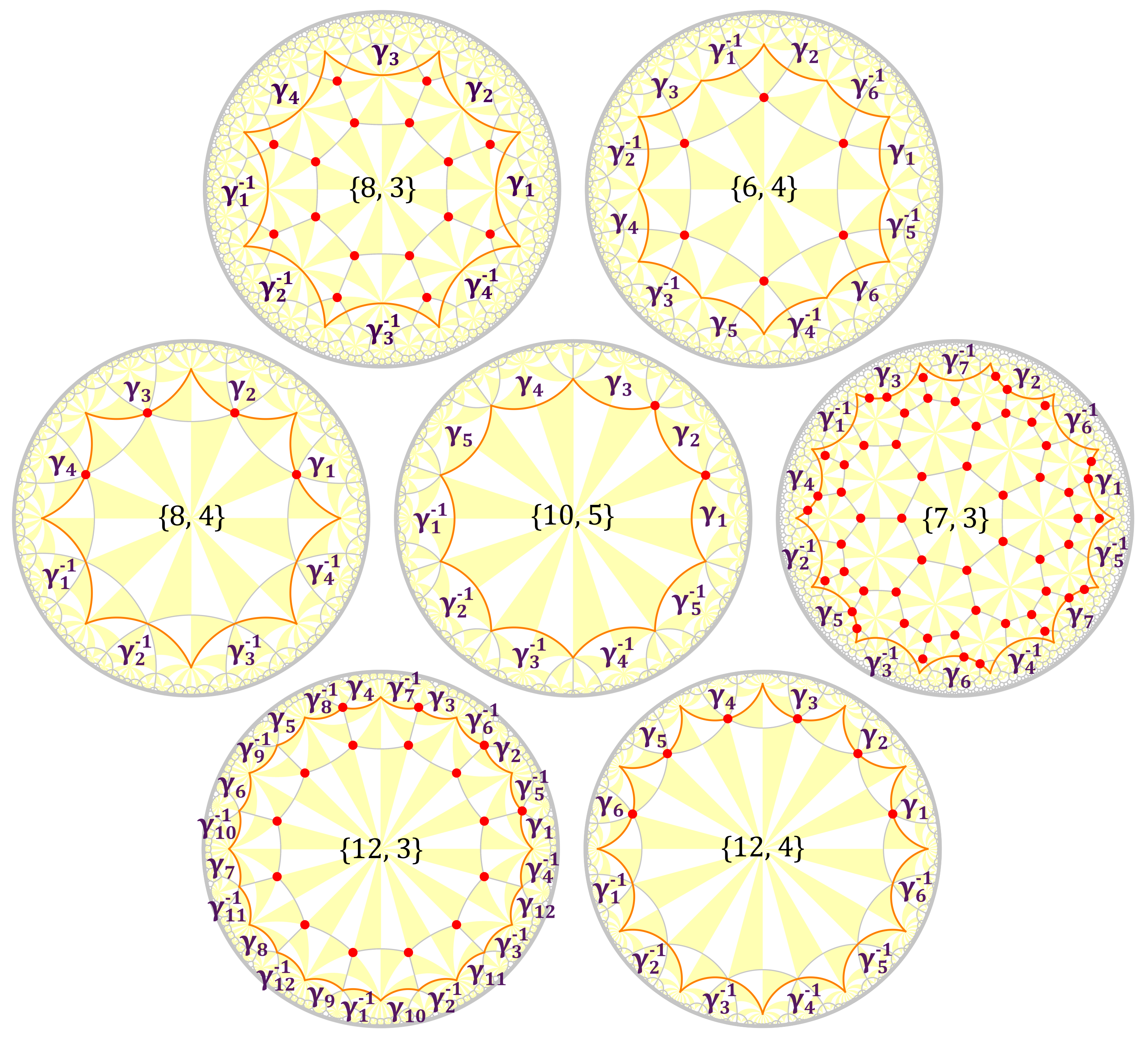}
\caption{Unit cell (red contour) and sites per unit cell (red dots) for each of the hyperbolic $\{p,q\}$ lattices (grey lines) investigated via HBT in Sec.~\ref{sec:k-space}. 
The white/yellow triangles are the Schwarz triangles of opposite orientations. 
The labels $\gamma_j^{(-1)}$, $1 \leq j \leq \mathfrak{e}$, shown along the unit cell boundary, indicate how the adjacent unit cells are reached from the outlined unit cell by acting with the generators of the translation groups. 
The relators among the translation generators are listed in Table~\ref{table:unit-cells}.
Identification of edges labelled by $\gamma_j$ and $\gamma_j^{-1}$ results in a compactified unit cell, which is a Riemann surface with genus $\mathfrak{g}=2$ (leading to a 4D BZ) or $\mathfrak{g}=3$ (leading to a~6D~BZ).}
\label{fig:unit-cells} 
\end{figure*}

\clearpage
\makeatletter\onecolumngrid@pop\makeatother

It is important to note that simply representing the $\mathfrak{e}$ generators by $\gamma_j \mapsto \e^{\imi k^j}$ may result in more momentum components than can be fitted into a $2\mathfrak{g}$-dimensional BZ.
The reason is that the translation generators $\{\gamma_j\}_{j=1}^{\mathfrak{e}}$ are not independent, but constrained via \emph{relators}, i.e., particular nontrivial compositions of the generators that are  equivalent to identity.
By inspecting the relators, it is possible to identify the $2\mathfrak{g}$ independent momenta $\{k^j\}_{j=1}^{2\mathfrak{g}}$ and to express the remaining momenta $\{k^j\}_{j=2\mathfrak{g}+1}^{\mathfrak{e}}$ as unique linear combinations of the previous ones~\cite{Boettcher:2022}.

Let us now summarize the main findings relevant for the subsequent analysis of hyperbolic Haldane models. 
By searching in Ref.~\onlinecite{Conder:2007}, we find that the only $\{p,q\}$ lattices
which admit a characterization via HBT with 4D BZ are $\{8,3\}$, $\{6,4\}$, $\{8,4\}$, $\{10,5\}$, $\{6,6\}$, $\{8,8\}$, plus their dual lattices with exchanged $p{\,\leftrightarrow\,} q $. 
Those with 6D BZ are $\{7,3\}, \{12,3\}, \{12,4\}, \{14,7\}, \{12,12\}$, plus their duals. 
Motivated by Fig.~\ref{fig:Cr-trend}, we limit our attention to seven of the listed lattices which have a comparatively small curvature (for fixed lattice constant), such as to be likely to support a gapped topological phase. 
Their hyperbolic translation groups $\mathsf{T}(2,q,p)$ are specified in Table~\ref{table:unit-cells} and Fig.~\ref{fig:unit-cells}. 
More precisely, Fig.~\ref{fig:unit-cells} illustrates the hyperbolic unit cell and the generators $\{\gamma_j\}_{j=1}^\mathfrak{e}$ of each translation group.
To complement this information, Table~\ref{table:unit-cells} lists the relators of $\mathsf{T}(2,q,p)$, and indicates how the substitution $\gamma_j \mapsto \e^{\imi k^j}$ results in concrete expressions for the momenta with $j > 2\mathfrak{g}$.

The translation group and the Brillouin zone for $\{8,3\}$, $\{8,4\}$, $\{10,5\}$, $\{7,3\}$, and $\{12,4\}$ have been introduced in Ref.~\onlinecite{Boettcher:2022}, whereas the HBT for $\{6,4\}$ and $\{12,3\}$ is developed in Appendices~\ref{app:6-4-symmetry} and~\ref{app:12-3-symmetry}. 
We note that Ref.~\onlinecite{Boettcher:2022} assumed a different compactification of the unit cell edges for $\{7,3\}$ than that shown in our Fig.~\ref{fig:unit-cells}, resulting in a different definition of momenta. 
However, this choice does not correspond to choosing a \emph{normal} subgroup $\mathsf{T}(2,3,7)$ of $\Delta(2,3,7)$, which makes the construction of a point group impossible. 
In contrast, the definition of $\mathsf{T}(2,3,7)$ in the more recent Refs.~\onlinecite{Cheng:2022,Bzdusek:2022} is compatible with ours.
In addition to the main-text discussion, Table~\ref{table:p-g-genie} in Appendix~\ref{app:crystallography} lists all hyperbolic $\{p,q\}$ lattices which per our construction result in a $2\mathfrak{g}$-dimensional BZ with $2 \leq \mathfrak{g}\leq 10$. 
Additional discussion of the hyperbolic point groups for the seven lattices we study appears in Appendix~\ref{sec:hyper-PGs}, with key findings summarized in Table~\ref{table:PG-summary}.

\subsection{HBT model construction and DOS}\label{sec:HBT-DOS}

The crystallography of a $\{p,q\}$ hyperbolic lattice specifies a Bravais unit cell with $N$ sites and a Fuchsian translation group $\mathsf{T}(2,q,p)$ with generators $\{\gamma_j\}_{j=1}^{\mathfrak{e}}$. The Fuchsian-group symmetry allows us to describe various tight-binding models on a periodic $\{p,q\}$ hyperbolic lattice by momentum-space Hamiltonians confined to the unit cell (see Ref.~\onlinecite{Chen2023} and Supplementary Material therein for proof in the case of nearest-neighbor-hopping models). With the Bloch 
ansatz, the momentum-space Hamiltonians are $N\times N$ matrices in the position-state basis formed by the unit-cell sites. Hoppings within the unit-cell are prescribed by the tight-binding model. Hopping from site $z_a$ in the unit cell to site $z_b$ outside of the unit cell is replaced by hopping from $z_a$ to $z_c$ in the unit cell multiplied by some phase factor $e^{{\rm i}f(\boldsymbol{k})}$, where $z_b$ and $z_c$ belong to the same orbit under $\mathsf{T}(2,q,p)$ (i.e., the same equivalence class modulo all translations)
~\cite{Urwyler:2022,Chen2023}. Specifically, to determine the function $f(\boldsymbol{k})$, one looks for the unique combination of generators $\gamma_{j_{1}}\cdots\gamma_{j_{n}}$ (with possible repetitions) such that $z_b=\gamma_{j_{1}}\cdots\gamma_{j_{n}}z_c$.\footnote{Searching for the unique combination of generators which translates site $i$ to site $j$ can be done either by hand with Fig.~\ref{fig:unit-cells} or by numerical ground search. The latter involves representing the generators by  $\mathsf{PSU}(1,1)$ matrices and applying arbitrary combinations of them to the Poincar\'{e} disk coordinate of site $i$, until a specific combination is found to bring it to site $j$ (see Ref.~\onlinecite{Chen2023} and Supplementary Material therein for examples).} Representing each $\gamma_{j}$ by its 1D IR $e^{{\rm i}k^{j}}$, one obtains
\begin{equation}
f(\boldsymbol{k})=\stackrel[\alpha=1]{n}{\sum}k^{j_{\alpha}},
\end{equation}
which can be expressed in terms of the linearly independent momenta through the relators in Table~\ref{table:unit-cells}. Using this approach, we derive the momentum-space version of Eq.~\eqref{eq:Hh} for our seven $\{p,q\}$ lattices.

The DOS $\rho(E,\phi)$ of each HBT Haldane model at $(t_2,m)$=$(0.5,0)$ is computed by diagonalizing the Hamiltonian at each fixed $\phi$ over a fine grid of momentum points ($\,{\sim}\,10^6$ points) in the 4D or 6D hyperbolic BZ, then sorting all the eigenvalues into histogram bins at different energies. 
Figure~\ref{fig:dos} shows $\rho(E,\phi)$ for models \{$6,4$\}, \{$7,3$\}, \{$8,3$\}, \{$10,5$\}, and \{$12,3$\}, while \{$8,4$\} and \{$12,4$\} are shown in Fig.~\ref{supp_fig:hbt-dos} in the Appendix. 
Models \{$7,3$\}, \{$8,3$\}, and \{$12,3$\} exhibit distinctive gapped regions at various filling fractions $f$ and ranges of $\phi$, while \{$6,4$\}, \{$8,4$\}, \{$12,4$\}, and \{$10,5$\}  appear gapless, only showing regions of low DOS.

Comparing with the real-space analysis, the DOS of HBT Haldane models \{$6,4$\}, \{$7,3$\}, \{$8,3$\}, and \{$12,3$\} demonstrate excellent agreement with the bulk DOS of the corresponding \{$p,q$\} models in the flake geometry, whereas there is a clear disagreement in the cases of \{$8,4$\} and \{$10,5$\}.
We expect the discrepancy to be mainly attributable to the Abelian Bloch ansatz [Eq.~(\ref{eqn:Abelian-Bloch-ansatz})] in our momentum-space analysis, whereas the real-space models in the flake geometry also support non-Abelian Bloch states transforming in higher-dimensional representations~\cite{Maciejko:2022,Cheng:2022}. 
%It is worth noting that the finite-size effects in the real-space models do not contribute significantly to the discrepancy, as it has been shown that the DOS of a hyperbolic flake is qualitatively independent of the system size \cite{Chen2023}.
Incidentally, lattices \{$8,4$\} and \{$10,5$\} have stronger Gaussian curvature $\kappa$ than the other lattices. 
The inverse correlation between curvature and agreement in the DOS comparison (also observed in hyperbolic tight-binding models with only nearest-neighbor hoppings \cite{Chen2023}) calls for future works to determine whether non-Abelian Bloch states play an increasingly important role in the spectrum of periodic lattices with stronger curvature.
Crucial steps in this direction have been achieved in very recent works~\cite{Lux:2022,Lux:2023,Mosseri2023,Lenggenhager:2023}.

\subsection{Symmetry constraints on Chern numbers}\label{eqn:Chern-theory}

Having discussed hyperbolic band structures and the DOS, we now turn our attention to momentum-space topological invariants.
In order to characterize the topology of gapped regions, we need a way to identify the independent Chern numbers arising in the 4D and 6D hyperbolic BZs. 
We first discuss the case of first Chern numbers defined on 2D subtori of the BZ, before later focusing on second Chern numbers defined on 4D subtori of the BZ.
The main text only presents a brief summary of a mathematically rather involved discussion whose details extend over Appendices~\ref{app:derive-M-matrices}, \ref{app:Chern-constraints} and~\ref{app:complete-Chern-matrices}. 
These detailed derivations and results constitute one of the major achievements of our present work.

Given a gapped band structure in $\textrm{n}$-dimensional hypercubic BZ, with $\textrm{n}=2\mathfrak{g}$, it is possible to define a first Chern number $C_{ij}$ on each 2D subtorus of the BZ spanned by two distinct momentum coordinates $(k^i,k^j), i{\neq}j$. 
We refer to them as the \emph{coordinate subtori} and we call $C_{ij}$ the \emph{coordinate Chern numbers}.
Note that exchanging $i{\leftrightarrow}j$ results in a change of orientation on the coordinate subtorus, which (as discussed in Appendix~\ref{app:Chern-defs}) implies a sign flip, $C_{ji} = -C_{ij}$. 
It is therefore natural to arrange the $\binom{\textrm{n}}{2} = \textrm{n}(\textrm{n}{-}1)/2$ potentially distinct coordinate Chern numbers into a skew-symmetric $\textrm{n} \times \textrm{n}$ matrix, which we label~$C_{(1)}$. (Note that there are additional 2D subspaces of an $\textrm{n}$-dimensional torus that are not (in the homotopical sense) deformable to any of the coordinate subtori; however, they can be decomposed (in a homological sense) into a linear combination of the coordinate subtori. Therefore, the Chern number on any such subspace is equal to a certain specific linear combination of the coordinate Chern numbers~$C_{ij}$.) 

If translations $\gamma_i$  are the \emph{only} symmetries of a model on a hyperbolic lattice, then all the coordinate Chern numbers can be tuned independently. 
However, the presence of additional point-group symmetries is expected to significantly reduce the number of independent coordinate Chern numbers.
For example, although the 4D BZ of the $\{8,3\}$-Haldane model contains six coordinate subtori, it was shown in Ref.~\onlinecite{Urwyler:2022} that there are, in fact, no more than three independent first Chern numbers. 
Here, we complete the symmetry characterization of the $\{8,3\}$ Haldane model, and we generalize the arguments to Haldane models on all hyperbolic $\{p,q\}$ lattices listed in Fig.~\ref{fig:unit-cells}.

To proceed with the analysis, we utilize the fact, derived in Appendix~\ref{app:Chern-first}, that in the presence of a symmetry $g$,
\begin{equation}
C_{(1)} = \varsigma_g M_g^\top C_{(1)} M_g^{\phantom{\top}}\label{eqn:C1-constraint}
\end{equation}
where the point-group matrix $M_g$ describes the transformation of momentum $\bs{k}=(k^1,k^2,\ldots,k^{\rm n})^\top$ under the action of $g$, i.e. $\bs{k}\mapsto M_g\bs{k}$, and the sign $\varsigma_g\in\{+1,-1\}$ is positive (negative) if the symmetry is unitary (antiunitary).
It is worth emphasizing that, somewhat non-intuitively, the hyperbolic point-group matrices are not necessarily orthogonal (i.e., they may not preserve angles) but rather are elements of the general linear group $\mathsf{GL}(\textrm{n},\mathbb{Z})$.
To obtain all the constraints on the first Chern numbers, it is sufficient to study Eq.~(\ref{eqn:C1-constraint}) for the generators of the symmetry group of the corresponding model; these are $a\mcT,b\mcT,c\mcT$ (resp.~$P^2,b\mcT,c\mcT$) in the absence (presence) of sublattice mass~$m$. 

\begin{table}[t]
\caption{
Number of linearly independent first and second momentum-space Chern numbers for lattice models having the symmetry of hyperbolic Haldane models on the selected $\{p,q\}$ lattices.
We present the results separately for models without (\xmark) and with (\cmark) sublattice mass $m$. 
}
\begin{ruledtabular}
\begin{tabular}{ccccccccc}
$\{p,q\}$ & $m$ & BZ & & $\# C_{(1),ij}$ & Eqs.& & $\# \mathscr{C}_{(2)}^{ijk\ell}$ & Eqs.
\tabularnewline
\hline 
$\{8,3\}, \{6,4\}, \{8,4\}$ & \xmark & 4D & & $1$ & (\ref{eqn:C83,M=0},\ref{eqn:C64,M=0}) & & $1$ & $-$
\tabularnewline
$\{8,3\}, \{6,4\}, \{8,4\}$ & \cmark & 4D & & $2$ & $\!\!\!$(\ref{eqn:C83,M!=0},\ref{eqn:C64,M!=0},\ref{eqn:C84,M!=0})$\!\!\!$ & & $1$ & $-$
\tabularnewline
$\{10,5\}$ & \xmark/\cmark & 4D & & $2$ & (\ref{eqn:C105,both-M}) & & $1$ & $-$
\tabularnewline
$\{7,3\}$ & \xmark & 6D & & $1$ & (\ref{eqn:C73}) & & $1$ & (\ref{eqn:secondC73})
\tabularnewline
$\{12,3\}$, $\{12,4\}$ & \xmark & 6D & & $2$ & (\ref{eqn:C1mat_m=0},\ref{eqn:C1mat_m=0_124}) & &  $2$ & $\!\!\!$(\ref{eqn:C2mat_m=0},\ref{eqn:C2mat_m=0_124})
\tabularnewline
$\{12,3\}$, $\{12,4\}$ & \cmark & 6D & & $3$ & (\ref{eqn:C1mat_m!=0},\ref{eqn:C1mat_m!=0_124}) & & $3$ & $\!\!\!$(\ref{eqn:C2mat_m!=0},\ref{eqn:C2mat_m!=0_124})
\end{tabular}\end{ruledtabular} 
\label{table:Cherns} 
\end{table}

To evaluate the implications of the above equation for the various Haldane models, we first extract the point-group matrices $M_g$ for each symmetry generator of a given $\{p,q\}$ lattice using computational group theory methods (Appendix~\ref{app:derive-M-matrices}).
Given those matrices, we then simplify the matrix $C_{(1)}$ via Eq.~(\ref{eqn:C1-constraint}).
The resulting matrices $C_{(1)}$ for each hyperbolic Haldane model considered are listed in Appendix~\ref{app:complete-C1-matrices}. 
The (significantly reduced) information about the number of linearly independent coordinate Chern numbers for each Haldane model is summarized in Table~\ref{table:Cherns}.

We next discuss symmetry constraints on second Chern numbers $C_{(2),ijk\ell}$ defined on 4D coordinate subtori spanned by momenta $(k^i,k^j,k^k,k^\ell)$. 
In analogy with the previous case, the exchange of any pair of momentum coordinates results in a reversed orientation on the 4D subtorus.
Therefore, $C_{(2),ijk\ell}$ flips sign under the exchange of any two indices, and $C_{(2)}$ is best interpreted as a fully skew-symmetric fourth-rank tensor. 
As for the number of independently defined components $C_{(2),ijk\ell}$, for models with 4D BZ there is a unique 4D subtorus (the whole BZ), whereas for a 6D BZ there are $\binom{6}{4}=\binom{6}{2}=15$ choices specified by the two momentum coordinates that do not appear among $(k^i,k^j,k^k,k^\ell)$.
In both cases, the reduction in independent components corresponds to the construction of the Hodge dual tensor $\mathscr{C}_{(2)} = \star C_{(2)}$~\cite{Fecko:2006} with components
\begin{equation}
\mathscr{C}_{(2)}^{k\cdots \ell} = \frac{1}{4!}C_{(2),i\cdots j}\epsilon^{i\cdots jk\cdots \ell}, \label{eqn:C2-Hodge-main}
\end{equation}
where $\epsilon^{i_1\cdots i_{\rm n}}$ (the number of superscript indices matches the BZ dimension) is the completely skew-symmetric Levi-Civita symbol, and a summation over repeated indices on the right-hand side is implicitly assumed. 
The dual $\mathscr{C}_{(2)}$ is a single number for a 4D BZ and a skew-symmetric matrix with components $\mathscr{C}_{(2)}^{ij}$ for a 6D BZ.

Similar to the case of first Chern numbers, point-group symmetries may reduce the number of linearly independent second Chern numbers. 
Let us first consider models with 4D BZ before discussing the ones with 6D BZ.
We derive in Appendix~\ref{app:Chern-second} that the sole independent second Chern number in a 4D BZ is constrained by
\begin{equation}
\mathscr{C}_{(2)} = (\det M_g) \mathscr{C}_{(2)} \label{eqn:C2-constraint}
\end{equation}
in the presence of a (unitary or antiunitary) symmetry $g$. 
In particular, the presence of any symmetry with $\det M_g = -1$ implies that the second Chern number must vanish.
By checking the point-group matrices listed in Appendix~\ref{app:derive-M-matrices}, we find that $\det M_g = +1$ for all symmetry generators of the hyperbolic Haldane models with 4D BZ ($\{8,3\}$, $\{6,4\}$, $\{8,4\}$, and $\{10,5\}$).
This implies that these models are compatible with a nonzero value of~$\mathscr{C}_{(2)}$.

For models with 6D BZ, we derive in Appendix~\ref{app:Chern-second} that
\begin{equation}
\mathscr{C}_{(2)} = (\det M_g) M_g \mathscr{C}_{(2)} M_g^\top \label{eqn:C2-constraint-6D}
\end{equation}
in the presence of a (unitary or antiunitary) symmetry $g$, where $\mathscr{C}_{(2)}$ is now a skew-symmetric $6\times 6$ matrix. 
The form of Eq.~(\ref{eqn:C2-constraint-6D}) crucially differs from that of Eq.~(\ref{eqn:C1-constraint}) in where the transpose ($^\top$) is located.
Given the point-group matrices, which are derived in \textsc{GAP} \cite{GAP4} and presented in Appendix~\ref{app:derive-M-matrices}, we again evaluate the consequences of Eq.~(\ref{eqn:C2-constraint-6D}) for hyperbolic Haldane models. 
We list the resulting symmetry-constrained matrices of second Chern numbers for the respective models in Appendix~\ref{app:complete-C2-matrices}, and indicate the number of linearly independent second Chern numbers in the last column of Table~\ref{table:Cherns} .

\subsection{Chern numbers and phase diagrams}\label{sec:Chern-phases}

\subsubsection{First Chern numbers}

We compute the first Chern numbers in the observed gapped regions of Fig.~\ref{fig:dos} using an efficient numerical method adapted from Ref.~\onlinecite{Fukui2005} (see Appendix~\ref{sec:numerical_C1} for details). 
In the 4D/6D hyperbolic BZ, we explicitly compute the coordinate Chern numbers $C_{ij}$ on six/fifteen coordinate subtori spanned by momenta $(k^i,k^j)$ and verify their relationships obtained by the symmetry analysis above. 
The remaining independent first Chern numbers are listed alongside the DOS plots in Fig.~\ref{fig:dos}. 
In cases where there is only one independent first Chern number, the momentum-space Chern numbers agree with the real-space Chern numbers of the corresponding \{$p,q$\} flake models.
The HBT DOS of \{$7,3$\} additionally reveals two small gapped regions with $C_{13}=\pm 2$ indistinguishable in the bulk DOS of the flake model.

The low-DOS regions in models \{$6,4$\}, \{$8,4$\}, and \{$10,5$\} are not fully gapped, so here we expect each $C_{ij}$ to depend on the choice of the fixed momentum components $k^{h\neq i,j}$, which can be thought of as tuning parameters. % driving the topological phase transition where the gap closes. 
Nevertheless, we notice that the $C_{ij}$'s of model \{$6,4$\} at half-filling are independent of the fixed momenta, indicating an accidental band-touching region where no topological phase transition occurs. 
The band-touching points are located at $(\pi,\pi,0,\pi)$, $(\pi,0,\pi,\pi)$, and $(0,\pi,\pi,0)$ for $(t_2,m,\phi)=(0.5,0,\pi/2)$. This semimetallic phase turns into a Chern insulator of topological charge $1$ upon the introduction of an infinitesimal sublattice mass $m$. 
The absence of sublattice mass in the experimental realization of the \{$6,4$\} Haldane model in Ref.~\onlinecite{Zhang:2022} leads us to hypothesize that the topological gap observed at half-filling is a finite-size effect.

\begin{figure} 
\includegraphics[width=\linewidth]{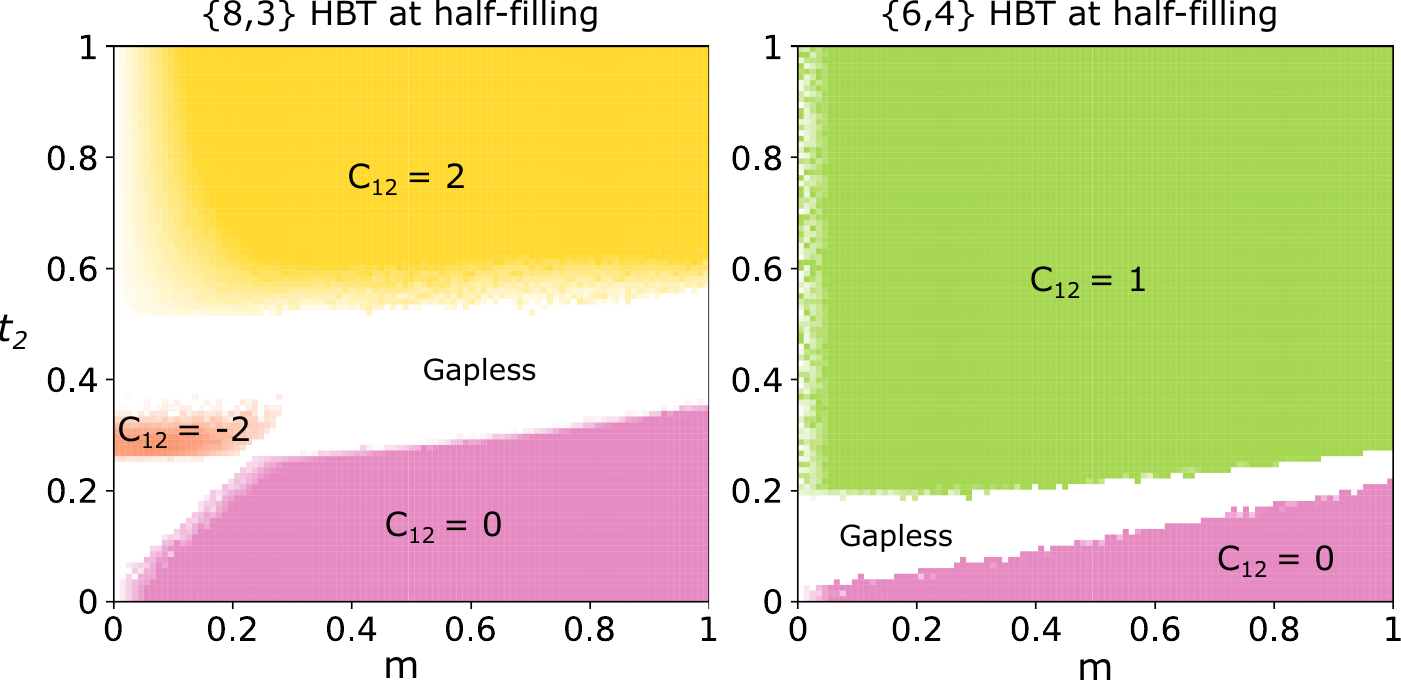}
\caption{Phase diagrams of hyperbolic Haldane models  $\{8,3\}$ and $\{6,4\}$ based on the first Chern number $C_{12}$ on the BZ subtorus spanned by momenta $(k^1,k^2)$. 
We compute $C_{12}$ at various  fixed momenta $k^{i\neq 1,2}$. 
If multiple distinct values are obtained, it can be inferred that the system is gapless. On the other hand, a unique result does not guarantee a gapped phase.
We superimpose the numerically estimated gap size $\Delta$ over the phase diagram, which is marked by white with linearly varying opacity (solid at $\Delta=0$ and transparent at $\Delta=0.1$). 
This allows us to distinguish semimetals with trivial gapless regions from gapped phases. 
The hyperbolic Haldane models exhibit an extended gapless phase, which, in the cases of $\{8,3\}$ and $\{6,4\}$, separates the trivial and nontrivial gapped phases.  \label{fig:pd}} \end{figure}

\subsubsection{Phase diagrams}

In the bipartite lattices, a nonzero sublattice mass $m$ opens a trivial gap at half-filling when $t_2=0$. 
Turning on the complex second-neighbor hoppings can induce a topological phase transition into a Chern insulator, as seen in the $\{6,3\}$ Haldane model~\cite{Haldane:1988}. 
We investigated the interplay between $m$ and $t_2$ in hyperbolic Haldane models through their phase diagrams over parameter space $(t_2,m)\in [0,1]\times[0,1]$, generated from the first Chern numbers. 
The phase diagrams of models $\{8,3\}$ and $\{6,4\}$  are shown in Fig.~\ref{fig:pd}. They reveal extensive gapless regions (unlike in the $\{6,3\}$ honeycomb lattice~\cite{Haldane:1988}) and Chern-insulating phases. 
Specifically, a point on the phase diagram is gapless (colored white) if the coordinate Chern number $C_{12}$
varies over a random sampling of the fixed momentum components $k^{i\neq 1,2}$, indicating a topological phase transition somewhere in the BZ. 
A unique $C_{12}$ indicates either a gapped phase or a gapless phase with an accidental band-touching region. 
Therefore we superimpose the numerically estimated gap size $\Delta$ over the phase diagram, which is marked by white with linearly varying opacity -- solid at $\Delta=0$ and transparent at $\Delta=0.1$.

Interestingly, $\{8,3\}$ exhibits gapped phases with Chern number $\pm 2$, albeit the gaps there are small (less than 5\% of the bandwidth). We also find that the $\{6,4\}$ model is a semimetal at $m=0$ and $t_2 \gtrsim 0.2$. The phase diagrams of models $\{8,4\}$, $\{10,5\}$, and $\{12,3\}$ (not shown) contain a trivially gapped region and a gapless region; the latter remains gapless even with $t_2$ increased to 10.

\subsubsection{Second Chern numbers} 

The 4D/6D hyperbolic band structures can be further characterized by higher-dimensional topological invariants. We computed the second Chern numbers using the approach of Ref.~\onlinecite{Mochol-Grzelak2018} (see Appendix~\ref{sec:numerical_C2} for details) and found $\mathscr{C}_{(2)}=0$ in all observed gaps and on all 4D subtori. We conclude that the hyperbolic Haldane models generally do not exhibit nontrivial second Chern numbers. 

\section{Flux patterns and \texorpdfstring{\\}{} magnetic hyperbolic space groups}\label{sec:flux-patterns}

\begin{figure*}
\includegraphics[width=0.9\linewidth]{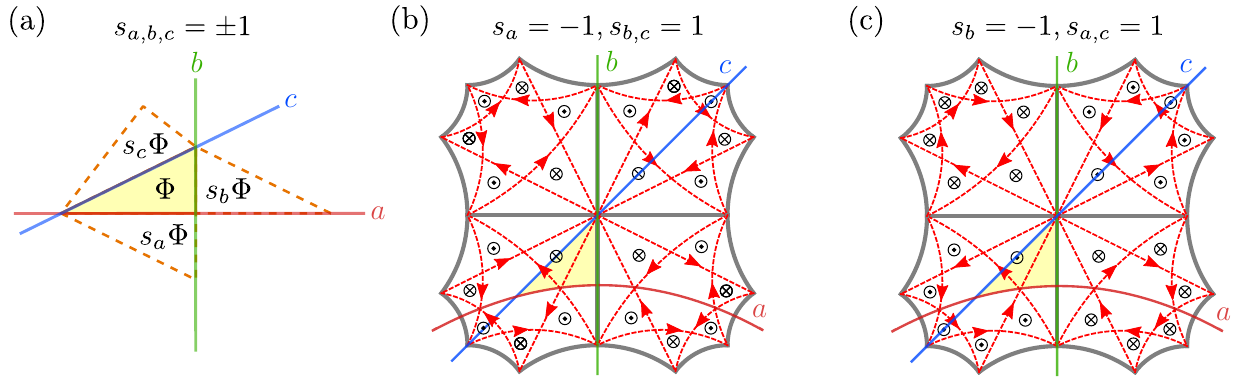}
\caption{Magnetic flux patterns in generalized Haldane models. 
(a)~If the magnetic flux through a chosen Schwarz triangle (yellow) is $\Phi$, then the flux through the three adjacent Schwarz triangles of opposite orientation (white) is $s_j \Phi$, where $j\in\{a,b,c\}$ are the three reflections in Fig.~\ref{fig:Schwarz-notation}, which generate the hyperbolic space group $\Delta(2,q,p)$, and $s_j = +1$ ($-1$) if reflection $j$ is composed (is not composed) with time reversal $\mcT$.
(b,c)~Magnetic flux patterns through four adjacent hexagons of the $\{6,4\}$ lattice, which result in particular instances of the generalized Haldane model, with their signs $s_{a,b,c}\in\{\pm 1\}$ indicated at the top of each panel.
Both illustrated flux configurations exhibit a reflection symmetry, which enforces a vanishing real-space Chern number; yet, these symmetries permit non-vanishing Chern numbers in the hyperbolic momentum space.}
\label{fig:flux_config} 
\end{figure*}

While Sec.~\ref{eqn:Chern-theory} contains a complete discussion of symmetry constraints on momentum-space Chern numbers for hyperbolic Haldane models, the presented symmetry-centered techniques are even more powerful as they can also be applied to study models symmetric under other magnetic symmetry groups on $\{p,q\}$ lattices. 
In fact, we have discussed in Sec.~\ref{sec:main-crystallography} that when constructing the Haldane models, one replaces the generators $a,b,c$ of the symmetry group $\Delta(2,q,p)$ by $a\mcT,b\mcT,c\mcT$, resulting in a particular type-III magnetic~\cite{Bradley:1972,Litvin:2013} hyperbolic space-group symmetry.\footnote{Starting with a non-magnetic space group $\mathsf{G}$ and its index-two subgroup $\mathsf{H}$, a type-III magnetic space group is constructed as $\mathsf{M}_\textrm{III}=\mathsf{H}\cup\mcT(\mathsf{G}\backslash\mathsf{H})$ where $\mcT$ is the time-reversal operator.} 
A simple and straightforward generalization is offered by composing only \emph{some} of the three symmetry generators with time reversal symmetry, which results in Haldane-like models with modified magnetic flux patterns.
By properly adapting our analysis, we find that some of these generalized models facilitate unusual topological features where symmetry enforces the real-space Chern number to vanish while allowing for non-vanishing Chern numbers in the momentum space.
For simplicity, in this section we set the sublattice mass $m$ to zero.

To characterize these generalized flux patterns, we introduce three $\mathbb{Z}_2$-valued quantities $s_{a,b,c}\in\{\pm 1\}$ as follows:
\begin{itemize}
\item $s_j = +1$ if the generator $j\in\{a,b,c\}$ in the presentation of the model's symmetry group is composed with time reversal $\mcT$; therefore, the reflection through edge $j$ is antiunitary [$\varsigma_{j\mcT} = -1$ in Eq.~(\ref{eqn:C1-constraint})].
\item $s_j = -1$ if the generator $j\in\{a,b,c\}$ in the presentation of the model's symmetry group is \emph{not} composed with $\mcT$, so that the reflection through edge $j$ is unitary [$\varsigma_{j} = +1$ in Eq.~(\ref{eqn:C1-constraint})].
\end{itemize} 
In both cases, if $\Phi$ is the magnetic flux through a selected Schwarz triangle [yellow in Fig.~\ref{fig:flux_config}(a)], the flux through the Schwarz triangle related by $j$ is given by $s_j\Phi$. 
With our definitions, it always holds that ``$\varsigma = -s$''.

Note that not all combinations of $s_{a,b,c}=\pm 1$ are available on all $\{p,q\}$ lattices. 
In particular, the flux through a Schwarz triangle related by $\tfrac{2\pi}{q}$-rotation around the vertex (rotation $Q=bc$ in Fig.~\ref{fig:Schwarz-notation}) equals $s_b s_c \Phi$. Therefore, consistency requires that $s_b s_c = +1$ if $q$ is odd.
Similarly, the flux through a Schwarz triangle related by $\tfrac{2\pi}{p}$-rotation around the $p$-sided polygon (rotation $P=ca$ in Fig.~\ref{fig:Schwarz-notation}) equals $s_c s_a \Phi$, and consistency requires that $s_c s_a = +1$ if $p$ is odd.
For simplicity, we here explicitly consider the application to the $\{6,4\}$ lattice, for which both $p$ and $q$ are even. Then, the two above restrictions do not arise, and all $2^3= 8$ combinations of $s_{a,b,c}=\pm 1$ are allowed.\footnote{In addition, all eight options are compatible also with the translation group given in Table~\ref{table:unit-cells}. To reveal this, note that in Appendix~\ref{app:6-4-symmetry} we express a representative element of the translation group as $\gamma_1 = (QP^{-1})^2=(bcac)^2$. Since this product contains an even number of each generator $a,b,c$, it follows that any pair of Schwarz triangles related by $\gamma_1$ carry the same magnetic flux irrespective of signs $s_{a,b,c}$, in accordance with the translation symmetry.} 
Haldane models, whose flux pattern is illustrated in Fig.~\ref{fig:hoppingdiagram}, correspond to the choice $s_{a,b,c}=1$.
Two of the generalized Haldane models that carry a modifed flux pattern, with an indicated orientation of magnetic field through parts of four adjacent hexagons of the $\{6,4\}$ lattice, are illustrated in Fig.~\ref{fig:flux_config}(b,c).

It is generally understood that reflection symmetry, which flips orientation in a 2D system, enforces the real-space Chern number $C_r(\mu)$ to vanish.
This property relates to the fact that the Chern number measures a chiral response, and that chirality is flipped under  reflections.
In our model, setting either of $s_{a,b,c}$ to $-1$ implies the presence of the corresponding reflection symmetry, thus enforcing $C_r(\mu) = 0$ at all chemical potentials $\mu$. 
Note that in 2D \emph{Euclidean} lattices, the sole momentum-space Chern number is equal to the real-space Chern number~\cite{Kitaev:2006}. Therefore, reflections in Euclidean lattices also imply vanishing of the momentum-space Chern number.
However, in hyperbolic lattices, the correspondence between real-space and momentum-space Chern numbers is presently not known. 
Therefore, it is not a priori clear whether reflection symmetry on a hyperbolic lattice also implies vanishing of momentum-space Chern numbers $C_{(1),ij}$ on the subtori of the higher-dimensional hyperbolic BZ. 

The question is resolved by analyzing the implications of Eq.~(\ref{eqn:C1-constraint}) with adjusted choices of $\varsigma_g$. Our analysis, detailed in Appendix~\ref{app:6-4-other}, suggests that in hyperbolic lattices it is possible to have non-vanishing Chern numbers $C_{(1),ij}$ on some 2D subtori in momentum space \emph{even in the presence of reflection symmetry}, in sharp contrast with the Euclidean case.
Specifically, the choices  $(s_a,s_b,s_c)$ being either $(+,-,+)$, $(+,-,-)$ or  $(-,+,+)$ [of which the first and the last are illustrated in Fig.~\ref{fig:flux_config}(c) resp.~(b)] possess orientation-reversing reflection symmetries while admitting one independent first Chern number in momentum space, with the explicit form of the Chern-number matrices shown in Eqs.~(\ref{eqn:C64+-+}--\ref{eqn:C64-++}) in Appendix~\ref{app:6-4-other}.
We anticipate models with such magnetic hyperbolic space-group symmetry to provide a fruitful ground for investigations of the bulk-boundary correspondence associated with momentum-space Chern numbers in the higher-dimensional BZs, since their responses and spectra should not be contaminated by chiral edge states originating from a real-space Chern number.

\section{Conclusions and outlook}\label{sec:conclusion}

In summary, we have generalized the Haldane honeycomb-lattice model to a wide range of regular \{$p,q$\} hyperbolic lattices and conducted a comprehensive analysis of its symmetry and topology. 
In Sec.~\ref{sec:real-space}, we numerically constructed the tight-binding Haldane models on hyperbolic flakes embedded in the Poincar\'{e} disk. 
Given that the boundary of a hyperbolic flake constitutes a substantial portion of the system, we devised and computed the bulk DOS to extract the bulk physics. 
By plotting the bulk DOS as a function of the time-reversal-breaking magnetic flux $\phi$, we observed that certain models exhibit bulk gaps characterized by first Chern number $1$, as indicated by the quantized plateaus in the real-space Chern number computed at all chemical potentials. 
Other models remain gapless, apart from the trivial gap at half-filling if the sublattice mass is nonzero.

To shed light on the underlying cause of the selective gap opening in only certain \{$p,q$\} models, we computed the Gaussian curvature of the hyperbolic lattices and observed a trend where only lattices with weaker curvature give rise to topologically nontrivial Haldane models. 
In particular, Fig.~\ref{fig:Cr-trend} suggests a curvature-driven phase transition from the Chern-insulating phase to the gapless phase.
However, it remains to be studied why the formation of energy gaps appears more unlikely for tight-binding models in strongly curved hyperbolic lattices.
Notably, a similar curvature-induced spectral transition in hyperbolic lattices was reported for the Hofstadter spectra~\cite{Stegmaier:2021}. 
In that work, the spectral transition was related to attributes of classical trajectories of charged particles, which in 2D hyperbolic space are open in weak magnetic fields~\cite{Comtet:1987}, whereas they always constitute closed cyclotron orbits in 2D Euclidean space. 
It is tempting to ponder an adaptation of this argument for Chern insulators; however, that would require a meaningful formulation of cyclotron orbits in the absence of externally applied magnetic fields. In addition, the present scenario is complicated by the following aspect: while usually one considers topological phase transitions at one particular gap, the \{$p,q$\} Haldane models studied here exhibit qualitatively different band structures for various choices of $p$, and their topological gaps arise at various filling fractions. We leave a more careful study of this curvature-driven transition to future investigations.

In Sec.~\ref{sec:k-space}, we provided a detailed formulation of hyperbolic crystallography and band theory, which can be straightforwardly generalized to 
all regular \{$p,q$\} lattices beyond those explicitly considered in this work. 
In particular, we present a previously missing construction of hyperbolic point groups and the hyperbolic crystallography of lattices \{$6,4$\} and \{$12,3$\}, which do not fall into any of the infinite families of \{$p,q$\} lattices discussed in Ref.~\onlinecite{Boettcher:2022}. 
Having obtained the hyperbolic Bravais lattices and unit cells, we employ the Bloch ansatz to construct Bloch Hamiltonians $H(\boldsymbol{k})$ with momentum $\boldsymbol{k}$ residing in the 4D and 6D hyperbolic BZs. 
The DOS computed from the Bloch Hamiltonians agrees with the bulk DOS of finite-sized models in the flake geometry, except for lattices with strong curvatures. 
We therefore anticipate that the contribution of non-Abelian states to the spectrum of finite hyperbolic flakes becomes increasingly prevalent for strongly curved $\{p,q\}$ lattices, although a careful examination of this question remains open for future studies.

To investigate the hyperbolic band topology, we derived constraints on the multiplet of first Chern numbers arising from the high-dimensional BZs by considering the hyperbolic point-group symmetry.
Subsequently, we numerically computed the first and second Chern numbers on various 2D and 4D subtori in the BZs. 
We found that the numerically computed first Chern numbers satisfy the analytically derived symmetry constraints. 
Furthermore, in models where only one independent first Chern number remains after the symmetry constraints, the momentum-space and real-space first Chern numbers were observed to agree. 
The physical meaning of having more than one independent first Chern number in a 2D lattice requires future investigations. 
We also demonstrated that the Bloch Hamiltonians facilitate an efficient computation of the 
phase diagrams, resulting in the presented plots of the first Chern numbers as a function of second-neighbor hopping $t_2$ and sublattice mass $m$, and revealing nontrivial phases with Chern number $2$. 

In addition, we analytically derived the anticipated constraints from hyperbolic point-group symmetry on the multiplet of second Chern numbers. 
However, the second Chern numbers are trivial for all the presently considered hyperbolic Haldane models; therefore, we were not able to illuminate these constraints with explicit numerical data.
It would be of much theoretical interest to devise a general mechanism, extending the generation of $\bs{k}$-dependent Haldane mass term via second-neighbor hopping in the presence of staggered magnetic field~\cite{Haldane:1988}, that would produce tight-binding models with non-trivial second (rather than first) Chern numbers on a range of hyperbolic $\{p,q\}$ lattices.
Let us remark that the derived relations between Chern numbers, encompassed by Eqs.~(\ref{eqn:C1-constraint}--\ref{eqn:C2-constraint-6D}) are, in principle, also applicable to any system with space group symmetry acting in a four- (or higher-) dimensional momentum space. 
Such topological Bloch Hamiltonians are not restricted to hyperbolic $\{p,q\}$ lattices, but also arise in certain quasicrystalline and superlattice models~\cite{Lohse:2018,Petrides:2018,Su:2020,Koshino:2022} and in models with synthetic dimensions~\cite{Zilberberg:2018,Weisbrich:2021,Chen:2021}. 
A four-dimensional topological lattice can also be directly realized as an electric-circuit network~\cite{Wang:2020}.

In Sec.~\ref{sec:flux-patterns}, we introduced the notion of \emph{magnetic} hyperbolic space groups, in analogy with the magnetic space groups of Euclidean lattices~\cite{Bradley:1972}. 
In particular, while the Haldane models are naturally captured by certain type-III~\cite{Litvin:2013} magnetic groups, simple modifications of the magnetic flux pattern over the $\{p,q\}$ lattice result in models symmetric under other type-III magnetic groups.
Curiously, while reflection symmetry in 2D Euclidean lattices implies a vanishing of the real-space and momentum-space Chern numbers, which on a flat lattice are equal to each other~\cite{Kitaev:2006}, we identified concrete magnetic symmetries on the $\{6,4\}$ lattice where reflection symmetry appears to be compatible with a non-vanishing first Chern number on 2D subtori of the 4D hyperbolic BZ. 
We anticipate such models to provide a fruitful playground for disentangling the bulk-boundary correspondence associated with momentum-space Chern numbers from that of the real-space Chern number.

Experimentally, while hyperbolic tight-binding models with only nearest-neighbor hopping are straightforward to implement using existing techniques of circuit QED and topolectrical circuitry, the complex-valued second-neighbor hopping in the hyperbolic Haldane models requires additional engineering. 
A complex-phase coupling between circuit nodes has been achieved in topolectrical circuits in Refs.~\onlinecite{Zhang:2022,Chen2023}. 
In particular, the authors of Ref.~\onlinecite{Zhang:2022} fabricated a circuit realization of the \{$6,4$\} hyperbolic Haldane model and observed chiral edge modes robust against back-scattering, which according to the bulk-boundary correspondence, is equivalent to a measurement of the bulk Chern topology. 
Going forward, hyperbolic Haldane models may be realizable by genuinely quantum-mechanical platforms, such as circuit QED equipped with complex-phase coupling and superconducting qubits~\cite{Kollar:2019}. With the rapidly developing techniques of optical tweezer arrays~\cite{Kaufman2021,Spar2022}, which provide exceptional control and versatility in lattice engineering, realizations of hyperbolic Haldane models in optical lattices may also be possible in the foreseeable future.

Finally, we briefly comment on the thermodynamic limit, which has been the subject of very recent works~\cite{Lux:2022,Lux:2023,Mosseri2023}. 
As shown in these studies, this limit can be approached by considering a suitably converging sequence of finite periodic clusters. 
From the HBT point of view, non-Abelian Bloch states belonging to increasingly high-dimensional IRs of the translation group are expected to play an increasingly important role as this sequence progresses: an IR of dimension $d$ appears $d$ times in the spectrum~\cite{Maciejko:2022}, and the allowed values of $d$ grow with the system size.
Abelian HBT can in fact be used to systematically explore non-Abelian Bloch states, by applying it to unit cells larger than the fundamental cell~\cite{Lenggenhager:2023}. 
In this regard, it would also be interesting to study how the number of independent first and second Chern numbers in momentum space (listed for the considered Haldane models in Table~\ref{table:Cherns}) changes with the choice of the unit cell, and whether such considerations shed light on a prospective exact correspondence between real-space and momentum-space Chern numbers. 
We also note that while of mathematical interest, the thermodynamic limit is not directly relevant to real experiments such as those of Refs.~\onlinecite{Kollar:2019,Lenggenhager:2021,Zhang:2022,Zhang:2023}, which contain at most a few hundred sites. 
In fact, the finite size was exploited in Ref.~\cite{Zhang:2023} to directly implement periodic boundary conditions in the laboratory, leading to the experimental realization of both Abelian and non-Abelian Bloch states.

\vspace{0.5cm}
\acknowledgements %\noindent 

We would like to thank S.~Dey, R.~Mosseri, T.~Neupert, E.~Prodan, S.~Rayan, G.~Shankar, A.~Stegmaier, C.~Sun, R.~Thomale, T.~Tummuru, and D.~Urwyler for valuable discussions.
A.~C. was supported by the Avadh Bhatia Fellowship at the University of Alberta. A.~C.~and  I.~B.~acknowledge support from the University of Alberta startup fund UOFAB Startup Boettcher. I.~B.~acknowledges funding from the Natural Sciences and Engineering Research Council of Canada (NSERC) Discovery Grants RGPIN-2021-02534 and DGECR2021-00043.
J.~M.~was supported by NSERC Discovery Grants \#RGPIN-2020-06999 and \#RGPAS-2020-00064; the Canada Research Chair (CRC) Program; the Government of Alberta's Major Innovation Fund (MIF); the Tri-Agency New Frontiers in Research Fund (NFRF, Exploration Stream); and the Pacific Institute for the Mathematical Sciences (PIMS) Collaborative Research Group program.
Y.~G.~acknowledges support by grant No.~204254 by the Swiss National Science Foundation (SNSF).
P.~M.~L.~and T.~B.~were supported by the Ambizione grant No.~185806 by SNSF.

\FloatBarrier
\clearpage

\appendix

\begin{figure*}
\includegraphics[width=0.7\linewidth]{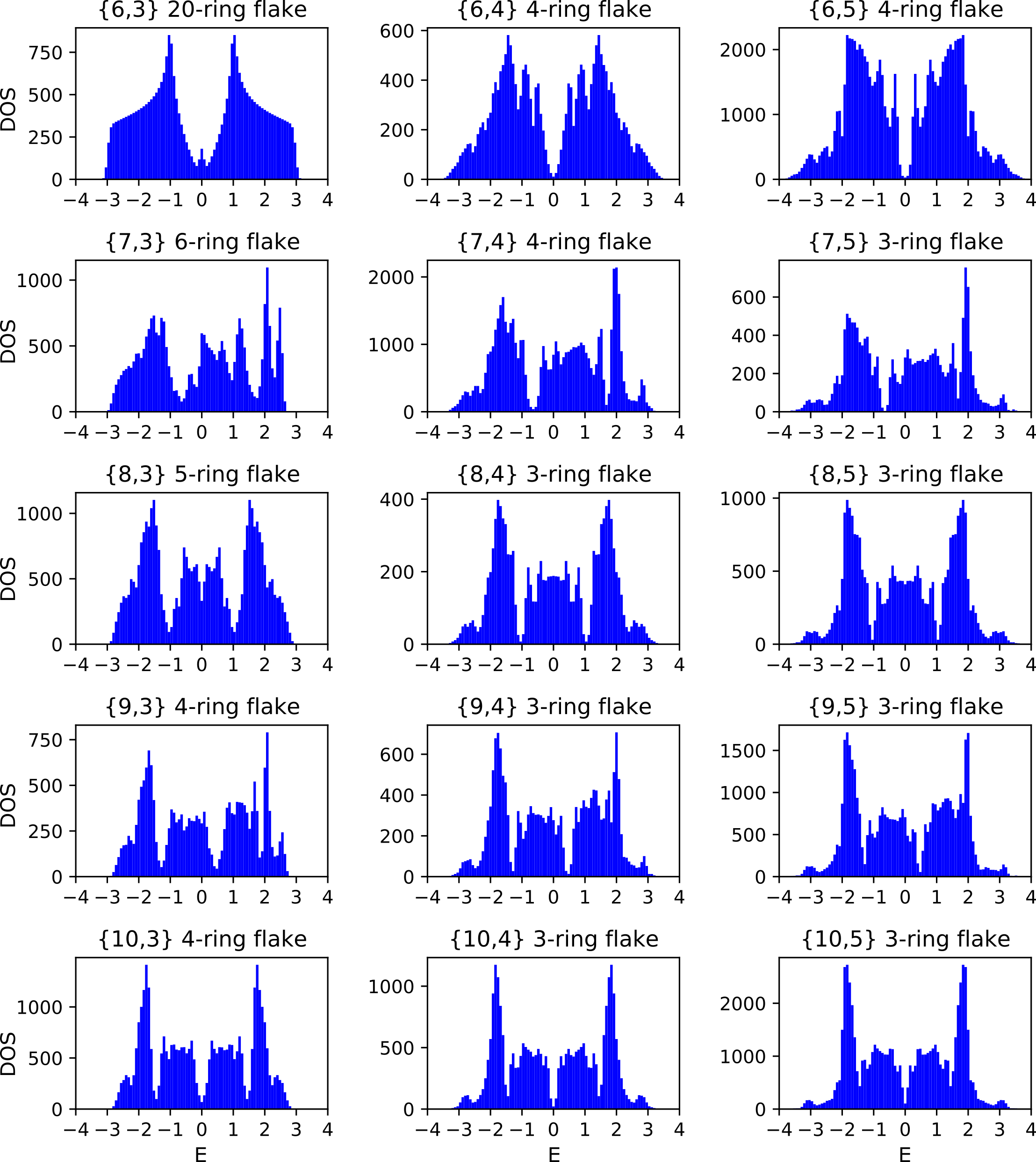}
\caption{\textbf{DOS for nearest-neighbor-hopping models.} The DOS of tight-binding models constructed on various \{$p,q$\} flakes share similar qualitative features in the location of high-DOS and low-DOS regions among models with the same $p$. Here we show the total DOS of \{$p,q$\} models with only nearest-neighbor hopping $t_1=-1$. Such universality is also observed in hyperbolic Hofstadter butterflies~\cite{Stegmaier:2021} and the DOS of hyperbolic Haldane models (Fig.~\ref{fig:dos} in the main text).}
\label{supp_fig:Hnn-dos}
\end{figure*}

\section{Hyperbolic Haldane models on flakes} \label{app:flake_Hh}

In the following, we describe numerical algorithms for constructing (1) the adjacency matrix $\mathcal{A}$ of any finite $\{p,q\}$ lattice, (2) the Poincar\'{e} disk coordinates of the lattice sites, (3) the diagonal matrix $\mathcal{M}_{S}$ for assigning a staggered on-site potential, such that $\mathcal{M}_{S,ii}=1$ for $i\in$ sublattice $A$ and $\mathcal{M}_{S,ii}=-1$ for $i\in$ sublattice $B$, and (4) the “arrow” matrix $\mathcal{M}_{H}$, defined as $\mathcal{M}_{H,ij}=1$ for all second-neighbor pairs $(i,j)$ such that the arrow going from $i$ to $j$ is clockwise within the polygon containing $i$ and $j$ ($\mathcal{M}_{H,ij}=0$ otherwise). The single-particle Hamiltonian of the Haldane model on a finite flake with $N$ sites in the position basis is then given by the $N\times N$ matrix 
\begin{equation}
    \mathcal{H}_H(m,t_{1},t_{2},\phi)=t_{1}\mathcal{A}+t_{2}(e^{\text{i\ensuremath{\phi}}}\mathcal{M}_{H}+e^{\text{-i\ensuremath{\phi}}}\mathcal{M}_{H}^\mathsf{T})+m\mathcal{M}_{S}. \label{eq:haldane_model}
\end{equation}
All Hamiltonians generated in this work are available in the Supplementary Code and Data~\cite{Chen:2023:SDC}.

\subsection{\texorpdfstring{$\{p,q\}$}{(p,q)} adjacency matrices} 

There are two methods for numerically ``growing'' finite-sized $\{p,q\}$ lattices: edge inflation and vertex inflation~\cite{Jahn2020}. The two procedures are equivalent when $q=3$. Vertex inflation results in a smoother boundary for $q\ge4$, so it is the method of choice in this work. In the following we first describe the numerical algorithm for edge inflation because it serves as a basis for the algorithm of vertex inflation.

\subsubsection{Edge inflation}

The adjacency matrix of a finite-sized $\{p,q\}$ lattice is an $N\times N$ matrix $\mathcal{A}$, where $N$ is the number of existing sites and $\mathcal{A}_{ij}=\mathcal{A}_{ji}=1$ if and only if site $i$ and $j$ are nearest neighbors; otherwise $\mathcal{A}_{ij}=\mathcal{A}_{ji}=0$. Every time we create a site, we add a row and a column to $\mathcal{A}$; to create an edge between sites $i$ and $j$, we add $1$ to $\mathcal{A}_{ij}$ and $\mathcal{A}_{ji}$. We start with a $p$-polygon, or tile, at the center of the Poincar\'{e} disk, and then iteratively attach \textit{rings} of tiles to the \textit{open edges}, which are the edges that belong to only one tile. Specifically, to attach the $(n+1)^{\text{th}}$ ring, we reflect the tiles in the $n^{\text{th}}$ ring across the open edges. After attaching each ring, some sites end up with more than $q$ neighbors. There are only two possible scenarios: site $z$ has either $q+1$ or $q+2$ neighbors. To remove the extra neighbors, we take two adjacent tiles containing site $z$ in the $(n+1)^{\text{th}}$ ring, and then either merge their adjacent, open edges into one edge (reducing one neighbor in the former scenario) or merge the two tiles into one tile (reducing two neighbors in the latter scenario). We perform the merging procedure after attaching each ring.

\subsubsection{Vertex inflation}

Each iteration in the above procedure leaves many \textit{open vertices}, which are the vertices with less than $q$ neighbors. On the other hand, vertex inflation demands every $(n+1)^{\text{th}}$ ring to contain all the tiles required to fill all the open vertices in the $n^{\text{th}}$ ring, resulting in a lattice without any open vertices in the bulk (inner rings). To construct the $(n+1)^{\text{th}}$ ring by vertex inflation, first we identify the list $L$ of open vertices as all those contained in the boundary of the $n^{\text{th}}$ ring,\footnote{It is insufficient to let open vertices be the ones with less than $q$ neighbors, since an open vertex can have $q$ neighbors but not $q$ tiles.} and then perform the following steps in a while-loop: \begin{enumerate} \item For every open edge\footnote{The list of open edges is constantly updated throughout the script.} $e_{i}$ that contains some vertex in $L$, identify the tile containing $e_{i}$ and invert it over $e_{i}$. \item Merge any duplicated tiles by the same merging procedure used in edge inflation. \item Remove any filled vertices from $L$. \end{enumerate} The while-loop ends when $L$ is empty, and the $(n+1)^{\text{th}}$ ring is completed. One can attach arbitrarily many rings, limited only by computational resources.

The DOS distributions of various hyperbolic flakes generated by vertex inflation are shown in Fig.~\ref{supp_fig:Hnn-dos}.

\subsection{Poincar\'{e} disk  coordinates} 

While the adjacency matrix alone is often sufficient for constructing various tight-binding models and computing energy spectra, it is useful to assign Poincar\'{e} disk coordinates
to the sites. In this work we need them for identifying the correct second-nearest neighbors according to the hyperbolic distance and computing the real-space
Chern numbers. The site coordinates of the central polygon are given
by 
\begin{equation} \label{eq:app_zn}
z_{n}=r_{0}e^{\text{i}2\pi n/p},\,n=1,\ldots,p,
\end{equation}
where 
\begin{equation}\label{eq:app_r0}
r_{0}=\sqrt{\frac{\cos(\pi/p+\pi/q)}{\cos(\pi/p-\pi/q)}}.
\end{equation}
To invert a site $z=z_{x}+\text{i}z_{y}$ across an edge $\{u=u_{x}+\text{i}u_{y},v=v_{x}+\text{i}v_{y}\}$,
we first determine whether $\{u,v\}$ is a straight segment (i.e.,
along the radial direction) or a circular arc. If $u_{x}v_{y}-u_{y}v_{x}=0$,
$\{u,v\}$ is along the radial direction, and the inverted point is
given by 
\begin{equation}
z'=\frac{\left(z_{x}(a^{2}-b^{2})-2abz_{y}\right)+\text{i}\left(-z_{y}(a^{2}-b^{2})-2abz_{x}\right)}{a^{2}+b^{2}}
\end{equation}
where $a=u_{x}-v_{x}$ and $b=v_{y}-u_{y}$. If $u_{x}v_{y}-u_{y}v_{x}\neq0$,
$\{u,v\}$ is a circular arc whose orthocenter is given by 
\begin{equation}
C_{\{u,v\}}=\frac{\left(\alpha v_{y}+u_{y}^{2}v_{y}-\beta u_{y}\right)+\text{i}\left(-\alpha v_{x}-u_{y}^{2}v_{x}+\beta u_{x}\right)}{2(u_{x}v_{y}-u_{y}v_{x})}=C_{x}+\text{i}C_{y}
\end{equation}
where $\alpha=1+u_{x}^{2}$ and $\beta=1+v_{x}^{2}+v_{y}^{2}$. The
orthoradius is 
\begin{equation}
R=\sqrt{C_{x}^{2}+C_{y}^{2}-1}.
\end{equation}
The inverted point is then given by
\begin{equation}
z'=C_{x}+\text{i}C_{y}+R^{2}\frac{(z_{x}-C_{x})+\text{i}(z_{y}-C_{y})}{(z_{x}-C_{x})^{2}+(z_{y}-C_{y})^{2}}.
\end{equation}
Note that the merging procedure described earlier is equivalent to removing
sites at the same coordinates. One can alternatively eliminate extra
neighbors by identifying sites with the same coordinates. However
this method is more computationally expensive and prone to numerical
rounding error for large systems in which near-boundary sites are
very close together.

\subsection{Sublattice mass}

We introduce a staggered on-site potential,
known as the sublattice mass in the case of graphene, on bipartite lattices.
Identification of the sublattices can be done using graph theory. The
degree matrix, defined as
\begin{equation}
\mathcal{D}=\text{diag}\left(\underset{i}{\sum}\mathcal{A}_{ij}\right),
\end{equation}
is a diagonal matrix whose diagonal element $\mathcal{D}_{jj}$ specifies
the number of nearest neighbors of site $j$. On the other hand, the
square of the adjacency matrix, $(\mathcal{A}^{2})_{ij}$, gives the number of length-2
hops between sites $i$ and $j$. Since a particle can trivially hop
two steps back to itself by passing through a nearest neighbor, the matrix $\mathcal{S}$
of neighbors that are two hops away is the difference
\begin{equation}\label{matrixS}
\mathcal{S}=\mathcal{A}^{2}-\mathcal{D}.
\end{equation}
It is a matrix of 0's and 1's because there is a unique two-hop path between
any two second-nearest neighbors. $\mathcal{S}$ has two connected
components---corresponding to two sublattices---which can be identified
using graph-theoretic subroutines available in Python's~\textsc{scipy}
library and \textsc{Mathematica}. An alternative way to identify the two sublattices is to compute the eigenvector of the largest eigenvalue of $\mathcal{S}$. This eigenvector is anti-ferromagnetic in the sense that it is positive on one sublattice and negative on the other. We then define the diagonal matrix $\mathcal{M}_{S}$ such that  $\mathcal{M}_{S,ii}=1$ if site $i$ is in the first sublattice and $\mathcal{M}_{S,ii}=-1$ if site $i$ is in the second sublattice.

\subsection{Haldane hopping} We want to introduce complex hopping terms $e^{\pm \imi \phi}$ between all second-nearest neighbors. For $\{p,3\}$ lattices, the second-neighbor pairs are readily encoded in the matrix $\mathcal{S}$. However, for lattices with $q>3$, not all sites which are two hops away are second-nearest neighbors. In a $\{p,q\}$ graph, each site has $q(q-1)$ sites that are two hops away. However if we consider the geometry of the lattice, a site $z_{0}$ is shared among $q$ polygons, and within each polygon, $z_{0}$ has two second-nearest neighbors, resulting in a total of $2q$ second neighbors. Since $q(q-1)>2q$ for $q>3$, we must check the hyperbolic distance $d(z_{i},z_{j})$ between sites $i$ and $j$, given by
\begin{equation}
d(z_{i},z_{j})=\mathcal{R}\ \text{arcosh}\left(1+\frac{2|z_{i}-z_{j}|^{2}}{(1-|z_{i}|^{2})(1-|z_{j}|^{2})}\right),
\end{equation}
where $\mathcal{R}$ is the curvature radius, to see if they are second neighbors.  Specifically, for sites $i$ and $j$ such that $\mathcal{S}_{ij}=1$, we reset $\mathcal{S}_{ij}=0$ if $d(z_{i},z_{j})\neq d(r_{0},r_{0}e^{\text{i}4\pi/p})$.

Next, we assign ``arrows" between all pairs of second neighbors which represent the positive hopping direction. The arrows must be consistently clockwise within each polygon. Consequently, the arrows at each site go in-and-out in an alternating fashion (as one can see in Fig. \ref{fig:hoppingdiagram} by focusing on one particular site), and we will use this observation in the following arrow-assigning algorithm. We start with a site in the central polygon $z_{1}=r_{0}e^{\text{i}2\pi/p}$, and define the first arrow to go from  $z_{3}=r_{0}e^{\text{i}6\pi/p}$ to $z_{1}$, denoted by $\overrightarrow{z_{3}z_{1}}$. The second neighbors of $z_{1}$ are $z_{i}$ such that $\mathcal{S}_{1i}=1$. We identify their relative positions from one another using graph theory. In particular, two second neighbors $\alpha$ and $\beta$ of $z_{1}$ are ``next to'' each other among all second neighbors of $z_{1}$ if and only if $\alpha$ and $\beta$ are either 2 hops or $(p-4)$ hops away (without passing through $z_{1}$). Knowing their relative positions and the initializing arrow $\overrightarrow{z_{3}z_{1}}$, we can then assign all the second-neighbor arrows of $z_{1}$ in an in-and-out fashion. Subsequently we repeat the above procedure over all sites, starting with sites with at least one known second-neighbor arrow. Eventually all the arrows will be assigned consistently within a sublattice (including the boundary region where sites do not have all $2q$ second neighbors). If the lattice has a single sublattice, then the arrow assignment is over. If the lattice is bipartite, we run another arrow-assigning algorithm with the initializing site $z_{2}=r_{0}e^{\text{i}4\pi/p}$ and arrow $\overrightarrow{z_{4}z_{2}}$ going from  $z_{4}=r_{0}e^{\text{i}8\pi/p}$ to $z_{2}$. All arrows are then recorded in a matrix $\mathcal{M}_H$ such that  $\mathcal{M}_{H,ij}=1$ if there exists an arrow $\overrightarrow{z_{i}z_{j}}$ and $\mathcal{M}_{H,ij}=0$ otherwise.

\begin{figure*}[p]
\includegraphics[width=0.9\linewidth]{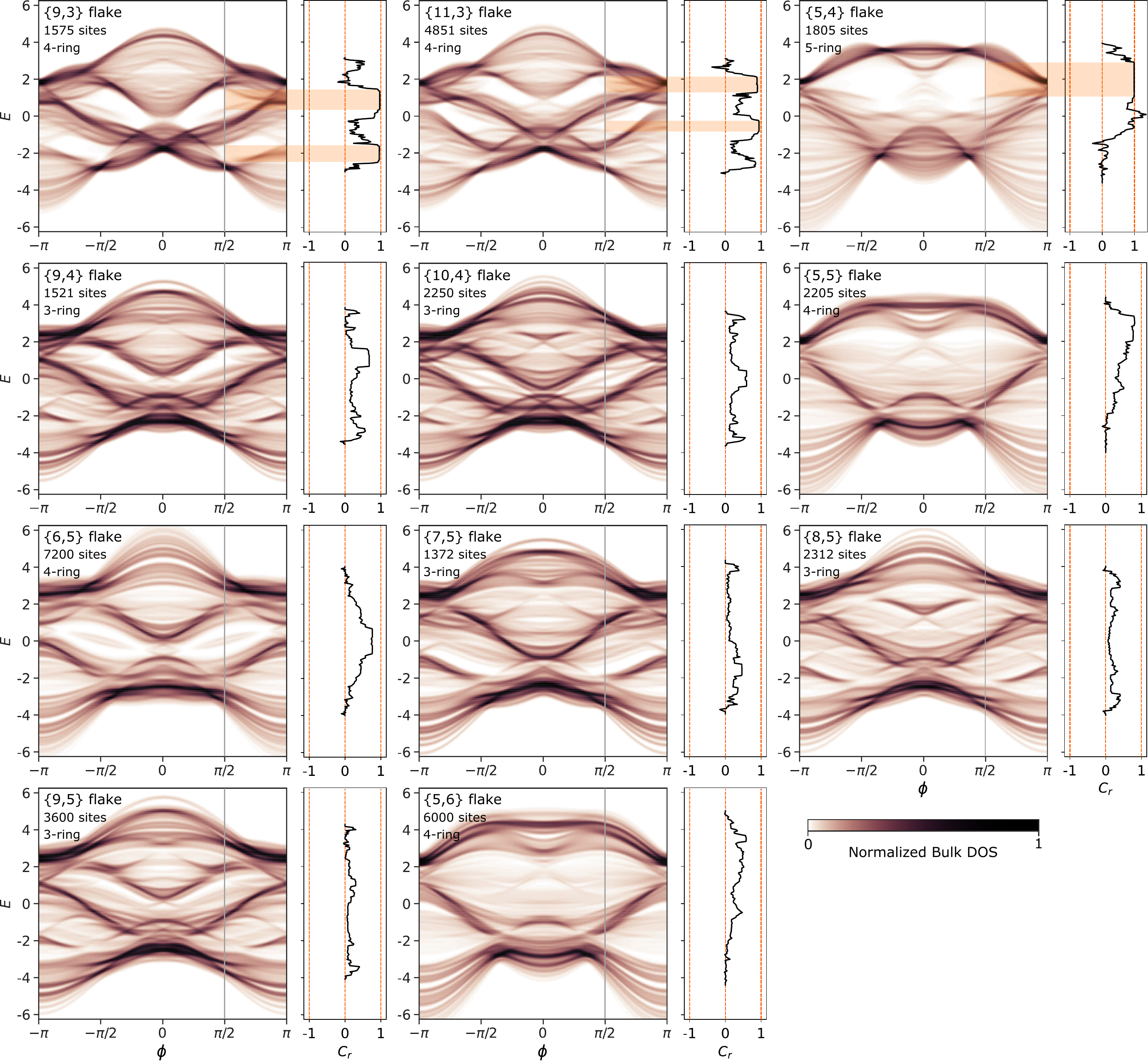}
\caption{\textbf{DOS of real-space hyperbolic Haldane models.}
In addition to Fig.~\ref{fig:dos} in the main text, the normalized bulk-DOS $\rho_{\text{bulk}}(E,\phi)$ of various $\{p,q\}$ hyperbolic Haldane models on a flake are shown here with $(t_2,m)$=$(0.5,0)$. They are ordered by increasing magnitude of Gaussian curvature (see Fig.~\ref{fig:Cr-trend}). White pockets indicate potentially gapped regions. The real-space Chern number $C_{r}(\mu)$ is computed at $\phi=\pi/2$, with nonzero quantized $C_{r}$ plateaus observed in the models with weaker curvature. Orange highlights are added to emphasize that the quantized $C_{r}$ plateaus coincide with the white pockets in $\rho_{\text{bulk}}(E,\phi)$.} 
\label{supp_fig:flake-dos}
\end{figure*}

In Figs.~\ref{fig:dos} and \ref{supp_fig:flake-dos}, we show the bulk-DOS of the Haldane model in Eq.~\eqref{eq:haldane_model} implemented on various hyperbolic flakes.

\makeatletter\onecolumngrid@push\makeatother
\clearpage
\makeatletter\onecolumngrid@pop\makeatother

\begin{figure*}
\includegraphics[width=0.9\linewidth]{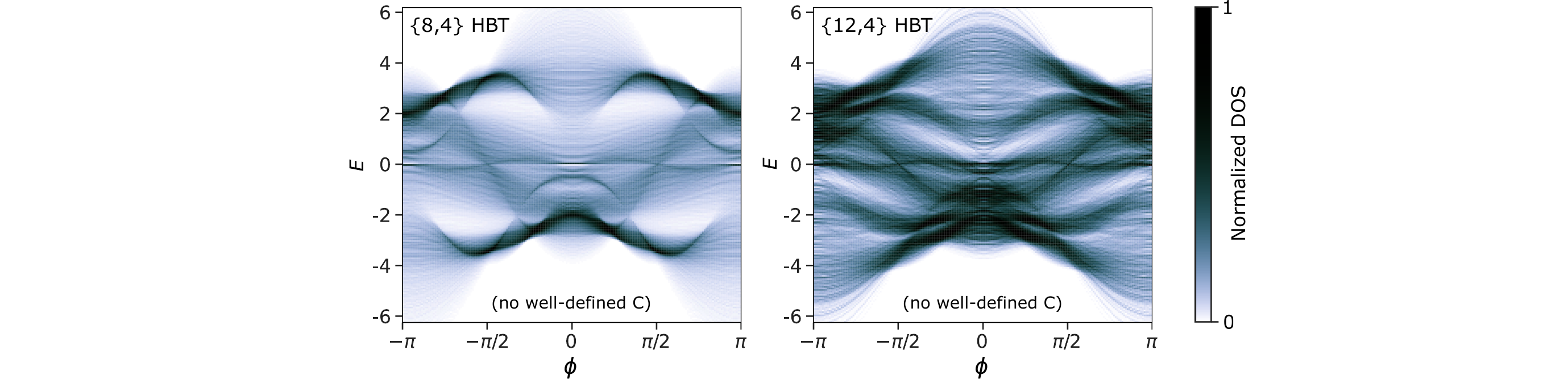}
\caption{The HBT DOS of Haldane models \{$8,4$\} and \{$12,4$\} are plotted here with parameters $(t_2,m)=(0.5,0)$. Neither show gaps induced by Haldane hoppings.}
\label{supp_fig:hbt-dos}
\end{figure*}

\section{Computation of momentum-space Chern numbers}\label{app:comput-k-Cherns}

\subsection{Hyperbolic Bloch Hamiltonians}

A brief description of our construction of hyperbolic Bloch Hamiltonians is included in Sec.~\ref{sec:HBT-DOS} of the main text. 
Here, we provide the explicit Hamiltonian matrices for three selected ones, namely $\{6,4\}$, $\{8,4\}$, and $\{10,5\}$. All Hamiltonians are available in the Supplementary Code and Data~\cite{Chen:2023:SDC}.

\begin{widetext}
For the $\{6,4\}$ lattice, there are $6$ sites per hyperbolic unit cell. 
We label the six sites by $1\leq j \leq 6$ according to the translation $\gamma_j$ appearing at the nearest unit cell edge in Fig.~\ref{fig:unit-cells}. 
With this labelling, and with the sign convention for the staggered fluxes shown in Fig.~\ref{fig:hoppingdiagram}, the hyperbolic Bloch Hamiltonian $\mcH_{\textrm{H}}^{\{6,4\}}(k^1,k^2,k^3,k^4;m,t_1,t_2,\phi)$ is given by the $6\times 6$ matrix
{\small \begin{equation}
\left(
\begin{array}{cccccc}
m   &
t_1(1{+}\mathrm{E}_1) &
\!\!\!\!\!\!\!\!\!t_2[\Phi(1{+}\mathrm{E}_{12}){+}\bar{\Phi}(\mathrm{E}_1{+}\mathrm{E}_2)]\!\!\!\!\!\!\!\!\! & 
0   &
\!\!\!\!\!\!\!\!\!t_2[\Phi(\mathrm{E}_{13}{+}\mathrm{E}_{24}){+}\bar{\Phi}(1{+}\mathrm{E}_{1234})]\!\!\!\!\!\!\!\!\! &
t_1(1{+}\mathrm{E}_{24}) \\
t_1(1{+}\mathrm{E}_{\bar{1}})    &
-m  &
t_1(1{+}\mathrm{E}_2) &
\!\!\!\!\!\!\!\!\!t_2[\Phi(1{+}\mathrm{E}_{23}){+}\bar{\Phi}(\mathrm{E}_2{+}\mathrm{E}_3)]\!\!\!\!\!\!\!\!\! &
0 &
\!\!\!\!\!\!\!\!\!t_2[\Phi(\mathrm{E}_{\bar{1}}{+}\mathrm{E}_{24}){+}\bar{\Phi}(1{+}\mathrm{E}_{\bar{1}24})] \!\!\!\! \\
t_2[\Phi(\mathrm{E}_{\bar{1}}{+}\mathrm{E}_{\bar{2}}){+}\bar{\Phi}(1{+}\mathrm{E}_{\bar{1}\bar{2}})]\!\!\!\!\!\!\!\!\! &
t_1(1+\mathrm{E}_{\bar{2}}) &
m   &
t_1(1{+}\mathrm{E}_3) &
\!\!\!\!\!\!\!\!\!t_2[\Phi(1{+}\mathrm{E}_{34}){+}\bar{\Phi}(\mathrm{E}_3{+}\mathrm{E}_4)]\!\!\!\!\!\!\!\!\! &
0 \\
0 &
\!\!\!\!\!\!\!\!\!t_2[\Phi(\mathrm{E}_{\bar{2}}{+}\mathrm{E}_{\bar{3}}){+}\bar{\Phi}(1{+}\mathrm{E}_{\bar{2}\bar{3}})]\!\!\!\!\!\!\!\!\! &
t_1(1{+}\mathrm{E}_{\bar{3}}) &
-m &
t_1(1{+}\mathrm{E}_{4}) &
\!\!\!\!\!\!\!\!\!t_2[\Phi(1{+}\mathrm{E}_{\bar{1}\bar{3}4}){+}\bar{\Phi}(\mathrm{E}_{\bar{1}\bar{3}}{+}\mathrm{E}_4)] \!\!\!\! \\
\!\!\! t_2[{\Phi}(1{+}\mathrm{E}_{\bar{1}\bar{2}\bar{3}\bar{4}}){+}\bar{\Phi}(\mathrm{E}_{\bar{1}\bar{3}}{+}\mathrm{E}_{\bar{2}\bar{4}})]\!\!\!\!\!\!\!\!\! &
0 &
\!\!\!\!\!\!\!\!\!t_2[\Phi(\mathrm{E}_{\bar{3}}{+}\mathrm{E}_{\bar{4}}){+}\bar{\Phi}(1{+}\mathrm{E}_{\bar{3}\bar{4}})]\!\!\!\!\!\!\!\!\! &
t_1(1{+}\mathrm{E}_{\bar{4}}) &
m &
t_1(1{+}\mathrm{E}_{\bar{1}\bar{3}}) \\
t_1(1{+}\mathrm{E}_{\bar{2}\bar{4}}) &
\!\!\!\!\!\!\!\!\!t_2[{\Phi}(1{+}\mathrm{E}_{1\bar{2}\bar{4}}){+}\bar{\Phi}(\mathrm{E}_{1}{+}\mathrm{E}_{\bar{2}\bar{4}})]\!\!\!\!\!\!\!\!\! &
0   &
\!\!\!\!\!\!\!\!\!t_2[{\Phi}(\mathrm{E}_{13}{+}\mathrm{E}_{\bar{4}}){+}\bar{\Phi}(1{+}\mathrm{E}_{13\bar{4}})]\!\!\!\!\!\!\!\!\! & 
t_1(1{+}\mathrm{E}_{13}) &
-m
\end{array}
\right),   \label{eqn:fat-6-4-Hamiltonian}
\end{equation}}
where we introduce the short-hand notations $\Phi=\e^{\imi\phi}$ and $\mathrm{E}_{ijk\ldots}=\e^{\imi k^i + \imi k^j + \imi k^k + \ldots}$, and the bar in $\mathrm{E}_{\bar{j}} = \mathrm{e}^{-\imi k^j}$ and in $\bar{\Phi}=\e^{-\imi\phi}$ encodes a minus sign inside the exponent.
Let us emphasize that in Eq.~(\ref{eqn:fat-6-4-Hamiltonian}) we have already replaced the redundant momenta $k^5\mapsto -k^1 -k^3$ and $k^6\mapsto -k^2 - k^4$ as implied by the relators in Table.~\ref{table:unit-cells}.

For the $\{8,4\}$ lattice, there are $4$ sites per hyperbolic unit cell,
which we label by $1 \leq j \leq 4$ according to the translation $\gamma_j$ appearing at the corresponding edge in Fig.~\ref{fig:unit-cells}.
With this labelling, and with the usual sign conventions, the hyperbolic Bloch Hamiltonian $\mcH_{\textrm{H}}^{\{8,4\}}(k^1,k^2,k^3,k^4;m,t_1,t_2,\phi)$ is given by the $4\times 4$ matrix
{\small
\begin{equation}
\left(\begin{array}{cccc}
m &
t_1(1{+}\mathrm{E}_{1\bar{2}}) &
\!\!\!\!\!\!\!\!\!\!\!\!\!\!\!\!\!\!\!\!\!t_2[\Phi(1{+}\mathrm{E}_{\bar{4}}{+}\mathrm{E}_{1\bar{3}}{+}\mathrm{E}_{1\bar{3}4}){+}\bar{\Phi}(\mathrm{E}_1{+}\mathrm{E}_{\bar{3}}{+}\mathrm{E}_{1\bar{2}}{+}\mathrm{E}_{2\bar{3}})]\!\!\!\!\!\!\!\!\!\!\!\!\!\!\!\!\!\!\!\!\! &
t_1(\mathrm{E}_1{+}\mathrm{E}_{\bar{4}}) \\
t_1[1{+}\mathrm{E}_{\bar{1}2}] &
-m &
t_1(1{+}\mathrm{E}_{2\bar{3}}) &
\!\!\!\!\!\!\!\!\!\!\!\!\!\!\!\!\!\!\!\!\!t_2[\Phi(1{+}\mathrm{E}_1{+}\mathrm{E}_{2\bar{4}}{+}\mathrm{E}_{\bar{1}2\bar{4}}){+}\bar{\Phi}(\mathrm{E}_{2}{+}\mathrm{E}_{\bar{4}}{+}\mathrm{E}_{2\bar{3}}{+}\mathrm{E}_{3\bar{4}})]\!\!\!\! \\
\!\!\! t_2[{\Phi}(\mathrm{E}_{\bar{1}}{+}\mathrm{E}_{3}{+}\mathrm{E}_{\bar{1}2}{+}\mathrm{E}_{\bar{2}3}){+}\bar{\Phi}(1{+}\mathrm{E}_{4}{+}\mathrm{E}_{\bar{1}3}{+}\mathrm{E}_{\bar{1}3\bar{4}})] \!\!\!\!\!\!\!\!\!\!\!\!\!\!\!\!\!\!\!\!\! &
t_1(1{+}\mathrm{E}_{\bar{2}3}) &
m &
t_1(1{+}\mathrm{E}_{3\bar{4}}) \\
 t_1(\mathrm{E}_{\bar{1}}{+}\mathrm{E}_{4}) &
\!\!\!\!\!\!\!\!\!\!\!\!\!\!\!\!\!\!\!\!\! t_2[{\Phi}(\mathrm{E}_{\bar{2}}{+}\mathrm{E}_{4}{+}\mathrm{E}_{\bar{2}3}{+}\mathrm{E}_{\bar{3}4}){+}\bar{\Phi}(1{+}\mathrm{E}_{\bar{1}}{+}\mathrm{E}_{\bar{2}4}{+}\mathrm{E}_{1\bar{2}4})]\!\!\!\!\!\!\!\!\!\!\!\!\!\!\!\!\!\!\!\!\! &
t_1(1{+}\mathrm{E}_{\bar{3}4}) &
-m
\end{array}\right),
\label{eqn:fat-8-4-Hamiltonian}
\end{equation}}
where we used the same short-hand notations as in Eq.~(\ref{eqn:fat-6-4-Hamiltonian}).
\end{widetext}

For the $\{10,5\}$ lattice, there are $2$ sites per hyperbolic unit
cell as shown in Fig.~\ref{fig:unit-cells}, which we label by $j=1,2$ such that $j=1$ corresponds to the site closer to the positive horizontal axis. 
With this labelling, and with the usual sign conventions, the
hyperbolic Bloch Hamiltonian $\mcH_{\textrm{H}}^{\{10,5\}}(k^{1},k^{2},k^{3},k^{4};m,t_{1},t_{2},\phi)$
is given by the $2\times2$ matrix 
\begin{equation}
\left(\begin{array}{cc}
t_{2}\alpha(\boldsymbol{k},\phi)+m & t_{1}\beta(\boldsymbol{k})\\
t_{1}\beta^{*}(\boldsymbol{k}) & t_{2}\alpha(\boldsymbol{k},-\phi)-m
\end{array}\right)
\end{equation}
where 
\begin{equation}
\alpha(\boldsymbol{k},\phi)=\Phi\mathrm{E}_{\bar{1}2}+\Phi\mathrm{E}_{2\bar{3}}+\Phi\mathrm{E}_{\bar{3}4}+\Phi\mathrm{E}_{1\bar{2}3}+\Phi\mathrm{E}_{\bar{2}3\bar{4}}+\text{c.c.}
\end{equation}
\begin{center}
and 
\end{center}
\begin{equation}
\beta(\boldsymbol{k})=1+\mathrm{E}_{1\bar{2}}+\mathrm{E}_{2\bar{3}}+\mathrm{E}_{1\bar{3}4}+\mathrm{E}_{2\bar{3}\bar{3}4}.
\end{equation}

The hyperbolic Bloch Hamiltonians for all seven $\{p,q\}$ lattices listed on page~\pageref{table:unit-cells} can be accessed through the Supplementary Code and Data~\cite{Chen:2023:SDC}, with the corresponding DOS shown in Figs.~\ref{fig:dos} and \ref{supp_fig:hbt-dos}.

\subsection{First Chern number}\label{sec:numerical_C1}
We follow Ref.~\onlinecite{Fukui2005} for the numerical computation of the first Chern number. Here we recount the key steps of this method. A Python implementation is available in the Supplementary Code and Data~\cite{Chen:2023:SDC}.

In a 2D continuum BZ, the  first Chern number of the $ b ^\textrm{th}$ band on the $(k^i,k^j)$-subtorus of BZ is given by 
\begin{equation}
C_{(1),ij}^{b}=\frac{1}{2\pi \imi}\int_{T^{2}}d^{2}\text{\ensuremath{\boldsymbol{k}\;F_{ij}^{b}(\boldsymbol{k})}}, \label{eq:cn}
\end{equation}
where $F_{ij}^{b}(\boldsymbol{k})=\partial_{i}A_{j}^{b}(\boldsymbol{k})-\partial_{j}A_{i}^{b}(\boldsymbol{k})$ is the Berry curvature with $A_{i}^{b}(\boldsymbol{k})=\langle b(\boldsymbol{k})|\partial_{i}|b(\boldsymbol{k})\rangle$ the Berry connection, and $|b(\boldsymbol{k})\rangle$ are the eigenstates in the $ b ^\textrm{th}$ band. The 2D torus is denoted by $T^2$. Integration over a compact surface, in this case the BZ torus, must give an integer. 

In a numerical computation, the Brillouin zone $T^{2}$ is divided into a fine grid of momentum points $\boldsymbol{k}$ with spacing $\ell$.  Equation \eqref{eq:cn} is then 
\begin{equation}
C_{(1),ij}^{b}=\frac{1}{2\pi \imi} \lim_{\ell\to 0}\ \sum_{\textbf{k}}
\ell^{2}F_{ij}^{b}(\boldsymbol{k}). \label{eq:cn_discrete}
\end{equation}
In the limit $\ell\text{\ensuremath{\rightarrow0}}$, the total flux of Berry curvature $F_{ij}^{b}(\boldsymbol{k})$ through each square of the momentum grid is equivalent to the Berry phase accumulated by a particle parallel-transported counterclockwise around the edge of the square. We have
\begin{equation}
\ell^{2}F_{ij}^{b}(\boldsymbol{k})=\ln(U_{i}(\boldsymbol{k})U_{j}(\boldsymbol{k}{+}\ell\hat{i})U_{i}(\boldsymbol{k}{+}\ell\hat{j})^{-1}U_{j}(\boldsymbol{k})^{-1}),
\label{eq:berryphase}
\end{equation}
where the $\mathsf{U}(1)$ link variable is defined as
\begin{equation}
U_{i}(\boldsymbol{k})\equiv\langle b(\boldsymbol{k})|b(\boldsymbol{k}+\ell\hat{i})\rangle. \label{eq:link}
\end{equation}
(Note that in the limit $\ell \rightarrow 0$, the $\mathsf{U}(1)$ links are unitary. Numerical computation uses finite $\ell$, so the $\mathsf{U}(1)$ links  should be ``unitarized'' by dividing them by their complex modulus.)
Equation~\eqref{eq:berryphase} is gauge invariant and summing it over all discretized $\boldsymbol{k}$ always gives an integer~\cite{Fukui2005}. It is also computationally efficient as the squares do not need to be infinitesimal as long as the numerical algorithm ensures that the Berry phase on each square (extracted only modulo $2\pi$ with the logarithm in Eq.~(\ref{eq:berryphase})), 
consistently lies in the branch $(-\pi,\pi]$ throughout the torus.

For our analysis, we need to compute the total Chern number for a multiplet of $n$ filled bands 
\begin{equation}
\Psi(\boldsymbol{k})= \Bigl(|1(\boldsymbol{k})\rangle,|2(\boldsymbol{k})\rangle,\dots,|n(\boldsymbol{k})\rangle\Bigr).    
\end{equation}
If they are mutually non-degenerate, one simply adds up the Chern number from each band. In the case of degeneracy, the Chern number is 
\begin{equation} 
C_{(1),ij}^{\Psi}=\frac{1}{2\pi \imi}\int_{T^{2}}d^{2}\text{\ensuremath{\boldsymbol{k}\;\tr\left[F_{ij}^{\Psi}(\boldsymbol{k})\right]}}, 
\end{equation} 
where $F_{ij}^{\Psi}(\boldsymbol{k})=\ensuremath{\partial_{i}A_{j}^{\Psi}(\boldsymbol{k})-\partial_{j}A_{i}^{\Psi}(\boldsymbol{k})}+\imi[A_{i}^{\Psi}(\boldsymbol{k}),A_{j}^{\Psi}(\boldsymbol{k})]$ is the $n\times n$ Berry-Wilczek-Zee (BWZ) Berry curvature of the filled bands, with BWZ connection $\left(A_{i}^{\Psi}(\boldsymbol{k})\right)^{ab}=\langle a(\boldsymbol{k})|\partial_{i}|b(\boldsymbol{k})\rangle$~~\cite{Berry:1984,Wilczek:1984}. 
(Note that Appendix~\ref{app:Chern-matrixforms} introduces the BWZ connection and curvature using the language of matrix-valued differential forms, which is not adopted in the present Appendix~\ref{app:comput-k-Cherns}.). 
The discretized integral is 
\begin{equation} C_{(1),ij}^{\Psi}=\frac{1}{2\pi \imi}\lim_{\ell \to 0}\ \sum_{\textbf{k}}\ell^{2}\;\tr\left[F_{ij}^{\Psi}(\boldsymbol{k})\right],
\end{equation} 
where 
\begin{equation} F_{ij}^{\Psi}(\boldsymbol{k})\ell^{2}=\ln(U_{i}^{\Psi}(\boldsymbol{k})U_{j}^{\Psi}(\boldsymbol{k}{+}\ell\hat{i})U_{i}^{\Psi}(\boldsymbol{k}{+}\ell\hat{j})^{-1}U_{j}^{\Psi}(\boldsymbol{k})^{-1}) \label{eq:discrete_F_Psi}
\end{equation} 
with $\mathsf{U}(1)$ link tensors 
\begin{equation} U_{i}^{\Psi}(\boldsymbol{k})\equiv\Psi^{\dagger}(\boldsymbol{k})\Psi(\boldsymbol{k}+\ell\hat{i}).   \label{eq:discrete_U_Psi}
\end{equation}
Using the identity $\tr\ln M=\ln\det M$ for any matrix $M$,  $C_{(1),ij}^{\Psi}$ can be easily computed from Eqs.~(\ref{eq:cn_discrete}--\ref{eq:link}) by replacing the $\mathsf{U}(1)$ link variable in Eq.~\eqref{eq:link} with
\begin{equation}
U_{i}(\boldsymbol{k})\equiv\det[\Psi^{\dagger}(\boldsymbol{k})\Psi(\boldsymbol{k}+\ell\hat{i})].
\end{equation}
$\,$

\subsection{Second Chern number}\label{sec:numerical_C2}
We follow Ref.~\onlinecite{Mochol-Grzelak2018} for the numerical computation of the second Chern number, which is a generalization of the aforementioned method for the first Chern number. A Python implementation is available in the Supplementary Code and Data~\cite{Chen:2023:SDC}.

In a 4D BZ spanned by $(k^{1},k^{2},k^{3},k^{4})$, the second Chern
number of a multiplet of $n$ filled bands $\Psi(\boldsymbol{k})=(|1(\boldsymbol{k})\rangle,|2(\boldsymbol{k})\rangle,...,|n(\boldsymbol{k})\rangle)$
is given by
\begin{widetext}
\begin{equation}
C_{(2),1234}^{\Psi}=\frac{1}{4\pi^{2}}\int_{\mathrm{BZ}}d^{4}\boldsymbol{k}\tr\left[F_{12}^{\Psi}(\boldsymbol{k})F_{34}^{\Psi}(\boldsymbol{k})+F_{41}^{\Psi}(\boldsymbol{k})F_{32}^{\Psi}(\boldsymbol{k})+F_{31}^{\Psi}(\boldsymbol{k})F_{24}^{\Psi}(\boldsymbol{k})\right].
\end{equation}
We numerically compute the discretized integral according to
\begin{equation}
C_{(2),1234}^{\Psi}=\frac{1}{4\pi^{2}}\underset{\ell\rightarrow0}{\lim}\underset{\boldsymbol{k}}{\sum}\tr\left[F_{12}^{\Psi}(\boldsymbol{k})F_{34}^{\Psi}(\boldsymbol{k})+F_{41}^{\Psi}(\boldsymbol{k})F_{32}^{\Psi}(\boldsymbol{k})+F_{31}^{\Psi}(\boldsymbol{k})F_{24}^{\Psi}(\boldsymbol{k})\right]\ell^{4},
\end{equation}
\end{widetext}
where the discretized BWZ Berry curvature $F_{ij}^{\Psi}(\boldsymbol{k})$ is computed
from $\mathsf{U}(1)$ link tensors via Eqs.~(\ref{eq:discrete_F_Psi}) and~(\ref{eq:discrete_U_Psi}). Note that we perform
the additional step of unitarizing the $\mathsf{U}(1)$ link tensors via singular
value decomposition. Namely, for each link tensor $U_{i}^{\Psi}(\boldsymbol{k})$,
we first factorize
\begin{equation}
U_{i}^{\Psi}(\boldsymbol{k})=WSV^{*},
\end{equation}
where $W$ and $V$ are complex unitary matrices and $S$ is a diagonal
matrix with non-negative real numbers on the diagonal. Then we use
the unitarized $\widetilde{U}_{i}^{\Psi}(\boldsymbol{k})\equiv WV^{*}$
to compute $F_{ij}^{\Psi}(\boldsymbol{k})$.

\section{Hyperbolic band theory} 
\label{sec:64-12-HBT}

In this paper, we investigate the momentum-space topology of selected hyperbolic Haldane models on $\{p,q\}$ lattices. 
For a complete analysis of tight-binding models in HBT, one must first determine for each lattice (1)~the correct translation group, (2)~the structure of momentum space, and (3)~the correct point-group symmetries.
The translation group and momentum space for five out of the seven lattices considered (Fig.~\ref{fig:unit-cells}) has been clarified in previous works~\cite{Maciejko:2021,Boettcher:2022,Cheng:2022,Bzdusek:2022}. 
In this Appendix we elucidate the translation group and momentum-space structure for the remaining two lattices, $\{6,4\}$ and $\{12,3\}$, which have featured in recent works~\cite{Zhang:2022,Tao:2022,Liu:2022b}. 
Additionally, we present a comprehensive analysis of point-group symmetries in both real space and momentum space for all seven lattices, which so far had only been carried out for the $\{8,8\}$ and $\{8,3\}$ lattices in Refs.~\onlinecite{Maciejko:2021,Urwyler:2022}. 
Finally, we also outline how HBT can be constructed for an arbitrary $\{p,q\}$ lattice, a task whose completion we leave for a future work~\cite{Lenggenhager:2023}. 

This Appendix is subdivided into four subsections. First, in Appendix~\ref{app:crystallography} we review general notions of hyperbolic crystallography, providing some additional details that were omitted in Sec.~\ref{sec:main-crystallography} of the main text. Next, in Appendix~\ref{app:6-4-symmetry} we study the $\{6,4\}$ lattice in detail, showing that its BZ is 4D. 
In Appendix~\ref{app:12-3-symmetry}, we repeat the same exercise for the $\{12,3\}$ lattice to reveal the construction of its 6D~BZ.
Finally, in Appendix~\ref{sec:hyper-PGs} we discuss hyperbolic point groups for all the $\{p,q\}$ lattices listed in Fig.~\ref{fig:unit-cells}, with the key information summarized in Table~\ref{table:PG-summary}.

\subsection{Hyperbolic crystallography}\label{app:crystallography}

We begin by supplementing Sec.~\ref{sec:main-crystallography} with additional remarks. 
In the discussion we extensively utilize the notation ($a,b,c,P,Q,R,\mcT$) for the group elements  as summarized in Fig.~\ref{fig:Schwarz-notation}.
The triangle group $\Delta(2,q,p)$, to be interpreted as the space group of the $\{p,q\}$ lattice, is specified in terms of the reflections $a,b,c$ in Fig.~\ref{fig:Schwarz-notation} by the presentation
\begin{equation}
\Delta(2,q,p) = \left<a,b,c\,\big|\,a^2,b^2,c^2,(ab)^2,(bc)^q,(ca)^p\right>,\label{eqn:Delta-def}
\end{equation}
where all the compositions (called \emph{relators}) listed to the right of the vertical line are set to equal the identity element. 
Our symmetry analysis necessitates borrowing results from Riemann surface theory~\cite{Conder:2007}, in which one is primarily interested in orientation-preserving isometries of hyperbolic space. 
In this context, one considers the index-2 subgroup of orientation-preserving symmetries in $\Delta(2,q,p)$, known as the \emph{von Dyck group} (or \emph{proper triangle group}), defined by the presentation
\begin{equation}
\mathsf{D}(2,q,p) = \left< P,Q,R\,\big| \, R^2, Q^q, P^p, RQP\right>,\label{eqn:D-2pq}
\end{equation}
where 
\begin{equation}
R=ab,\qquad Q=bc,\qquad P=ca \label{eqn:mirror2rot}
\end{equation} 
are rotations by $\pi$, $2\pi/q$, and $2\pi/p$ around the three respective corners of the Schwarz triangle (see footnote~\ref{eqn:foot-group-action}). The relator $RQP=(ab)(bc)(ca)=1$ follows trivially from the relators listed in Eq.~(\ref{eqn:Delta-def}).
As a fundamental domain of the von Dyck group, we can take any triangle consisting of two adjacent Schwarz triangles; this larger domain is conventionally called the \emph{fundamental triangle}.

In the main text we have defined the translation group\footnote{More broadly, by a hyperbolic translation group we mean any strictly hyperbolic subgroup of $\mathsf{PSU}(1,1)$, the full group of orientation-preserving isometries of the hyperbolic plane. Being strictly hyperbolic is a stronger notion than being torsion-free. The group $\mathsf{T}(2,q,p)$ as defined in the text can be shown to also be strictly hyperbolic.} $\mathsf{T}(2,q,p)$ as the smallest-index, i.e., ``largest'' subgroup of $\Delta(2,q,p)$ that is torsion-free, normal, and orientation-preserving. This last condition implies that we can instead look for the smallest-index torsion-free normal subgroup of $\mathsf{D}(2,q,p)$, as it contains as subgroups all the orientation-preserving subgroups of $\Delta(2,q,p)$. The proper point group
\begin{equation}
    \mathsf{P}_{\!\mathsf{s}}(2,q,p) = \mathsf{D}(2,q,p)/\mathsf{T}(2,q,p) = \mathsf{P}(2,q,p)/\mathbb{Z}_2 \label{eqn:PS-group}
\end{equation}
captures the residual orientation-preserving symmetries that act on the unit cell,
\begin{equation}
\textrm{unit cell} = \mathbb{D}/\mathsf{T}(2,q,p).\label{eqn:unit-cell-def} 
\end{equation}
The unit cell as defined by Eq.~(\ref{eqn:unit-cell-def}) has the geometry and topology of a Riemann surface of genus $\mathfrak{g}\geq 2$. 

Quotients of hyperbolic von Dyck groups which act on a Riemann surface with genus $2\leq\mathfrak{g}\leq 101$ have been tabulated by Conder~\cite{Conder:2007}.
In particular, Conder's table provides the presentation of every such quotient group in terms of cosets\footnote{Given a normal subgroup $\mathsf{T}\triangleleft\mathsf{D}$ and an element $g\in\mathsf{D}$, the left (right) coset $\breve{g} \subset \mathsf{D}$ is a collection of elements in $\mathsf{D}$ constructed as $\breve{g} = g\mathsf{T} :=\{g \gamma \,|\, \gamma\in\mathsf{T}\}$ (resp.~as $\breve{g}=\mathsf{T}g:=\{\gamma g \,|\, \gamma\in\mathsf{T}\})$. Since $\mathsf{T}$ is a \emph{normal} subgroup, left and right cosets coincide, i.e., $\breve{g} = g\mathsf{T} = \mathsf{T}g$. The coset of the identity is $\breve{1}=\mathsf{T}$.} $\breve{P},\breve{Q},\breve{R}$ of rotations $P,Q,R$ in the von Dyck group.\footnote{Contrary to our introduction of the breve ($\,\breve{}\,$) symbol, Ref.~\onlinecite{Conder:2007} does not symbolically distinguish elements of the von Dyck group from elements of the hyperbolic point group. In addition, the following relabelling is necessary to translate between our notation and theirs: $\{R,Q,P\} \leftrightarrow \{x,y,z\}$.} 
Knowing the presentation of the quotient (proper point group) $\mathsf{P}_{\!\mathsf{s}}(2,q,p)$, in combination with the presentation of the von Dyck group $\mathsf{D}(2,q,p)$ in Eq.~(\ref{eqn:D-2pq}), we are able to invert the relation~(\ref{eqn:PS-group}) to uniquely identify the translation group $\mathsf{T}(2,q,p)$. 
Subsequently, by considering the action of $\mathsf{T}(2,q,p)$ on the Schwarz triangles, we utilize Eq.~(\ref{eqn:unit-cell-def}) to identify the unit cell of the hyperbolic translation group as well as its appropriate identification of edges.
This analysis is carried out in detail for the $\{6,4\}$ lattice in Appendix~\ref{app:6-4-symmetry} and for the $\{12,3\}$ lattice in Appendix~\ref{app:12-3-symmetry}.
For convenience, we list in Table~\ref{table:p-g-genie} all the $\{p,q\}$ lattices which per the above construction result in a unit cell of genus $2\leq \mathfrak{g} \leq 10$, corresponding to a $2\mathfrak{g}$-dimensional BZ.

\begin{table}[t]
\begin{threeparttable}
\caption{
Overview of hyperbolic $\{p,q\}$ lattices with $p\geq q$ for which the construction in Appendix~\ref{app:crystallography} yields a compactified unit cell of genus $2\leq \mathfrak{g} \leq 10$; these are associated with a $2\mathfrak{g}$-dimensional BZ. The dual lattices with $p \leftrightarrow q$ are not explicitly shown.
Lattices whose unit cell was identified in Ref.~\onlinecite{Boettcher:2022} are indicated with a superscript whose meaning is explained in the notes. [Note: the family $\{2(2\mathfrak{g}+1),3\}$ is not reproduced by our method, presumably because the translation group suggested in Ref.~\onlinecite{Boettcher:2022} is not a normal subgroup of $\Delta(2,3,2(2\mathfrak{g}+1))$.]
\label{table:p-g-genie} 
}
\begin{ruledtabular}
\begin{tabular}{clc}
genus &
$\{p,q\}$ lattices & $\qquad$
\tabularnewline
\hline 
$\mathfrak{g} = 2$ &
$\{8,3\}$\tnote{a}, $\{6,4\}$, $\{8,4\}$\tnote{b}, $\{10,5\}$\tnote{c}, $\{6,6\}$, $\{8,8\}$\tnote{d}
\\[0.13cm] %\tabularnewline
$\mathfrak{g} = 3$ &
$\{7,3\}$\tnote{a}, $\{12,3\}$, $\{12,4\}$\tnote{b}, $\{14,7\}$\tnote{c}, $\{12,12\}$\tnote{d}
\\[0.13cm] %\tabularnewline
\multirow{2}{*}{$\mathfrak{g} = 4$} &
$\{5,4\}$, $\{10,4\}$, $\{16,4\}$\tnote{b}, $\{5,5\}$, $\{12,6\}$, $\{18,9\}$\tnote{c},
\tabularnewline
&  $\{10,10\}$, $\{16,16\}$\tnote{d}
\\[0.1cm] %\tabularnewline
$\mathfrak{g} = 5$ &
$\{10,3\}$, $\{20,4\}$\tnote{b}, $\{15,6\}$, $\{22,11\}$\tnote{c}, $\{20,20\}$\tnote{d}
\\[0.13cm] %\tabularnewline
\multirow{2}{*}{$\mathfrak{g} = 6$} &
$\{9,4\}$, $\{14,4\}$, $\{24,4\}$\tnote{b}, $\{8,6\}$, $\{9,9\}$, $\{15,10\}$
\tabularnewline
& $\{26,13\}$\tnote{c}, $\{14,14\}$, $\{24,24\}$\tnote{d}
\\[0.13cm] %\tabularnewline
$\mathfrak{g} = 7$ &
$\{28,4\}$\tnote{b}, $\{9,6\}$, $\{21,6\}$, $\{7,7\}$, $\{30,15\}$\tnote{c}, $\{28,28\}$\tnote{d}
\\[0.13cm] %\tabularnewline
\multirow{2}{*}{$\mathfrak{g} = 8$} &
$\{18,4\}$, $\{32,4\}$\tnote{b}, $\{10,6\}$, $\{24,6\}$, $\{12,8\}$, $\{20,10\}$,
\tabularnewline
& $\{34,17\}$\tnote{c}, $\{18,18\}$, $\{32,32\}$\tnote{d}
\\[0.13cm] %\tabularnewline
$\mathfrak{g} = 9$ &
$\{36,4\}$\tnote{b}, $\{6,5\}$, $\{24,8\}$, $\{21,14\}$, $\{38,19\}$\tnote{c}, $\{36,36\}$\tnote{d}
\\[0.13cm] %\tabularnewline
\multirow{2}{*}{$\mathfrak{g} = 10$} &
$\{9,3\}$, $\{15,3\}$, $\{18,3\}$, $\{24,3\}$, $\{7,4\}$, $\{22,4\}$, $\{40,4\}$\tnote{b},
\tabularnewline
 & $\{30,6\}$, $\{24,12\}$, $\{42,21\}$\tnote{c}, $\{22,22\}$, $\{40,40\}$\tnote{d}
\end{tabular}\end{ruledtabular} 
\begin{tablenotes}[flushleft]
\item[a] One of the exceptional cases identified by Ref.~\onlinecite{Boettcher:2022}.
\item[b] Member of the $\{4\mathfrak{g},4\}$ family.
\item[c] Member of the $\{2(2\mathfrak{g}+1),2\mathfrak{g}+1\}$ family.
\item[d] Member of the $\{4\mathfrak{g},4\mathfrak{g}\}$ family.
\end{tablenotes}
\end{threeparttable}
\end{table}

\begin{figure*}
\includegraphics[width=\linewidth]{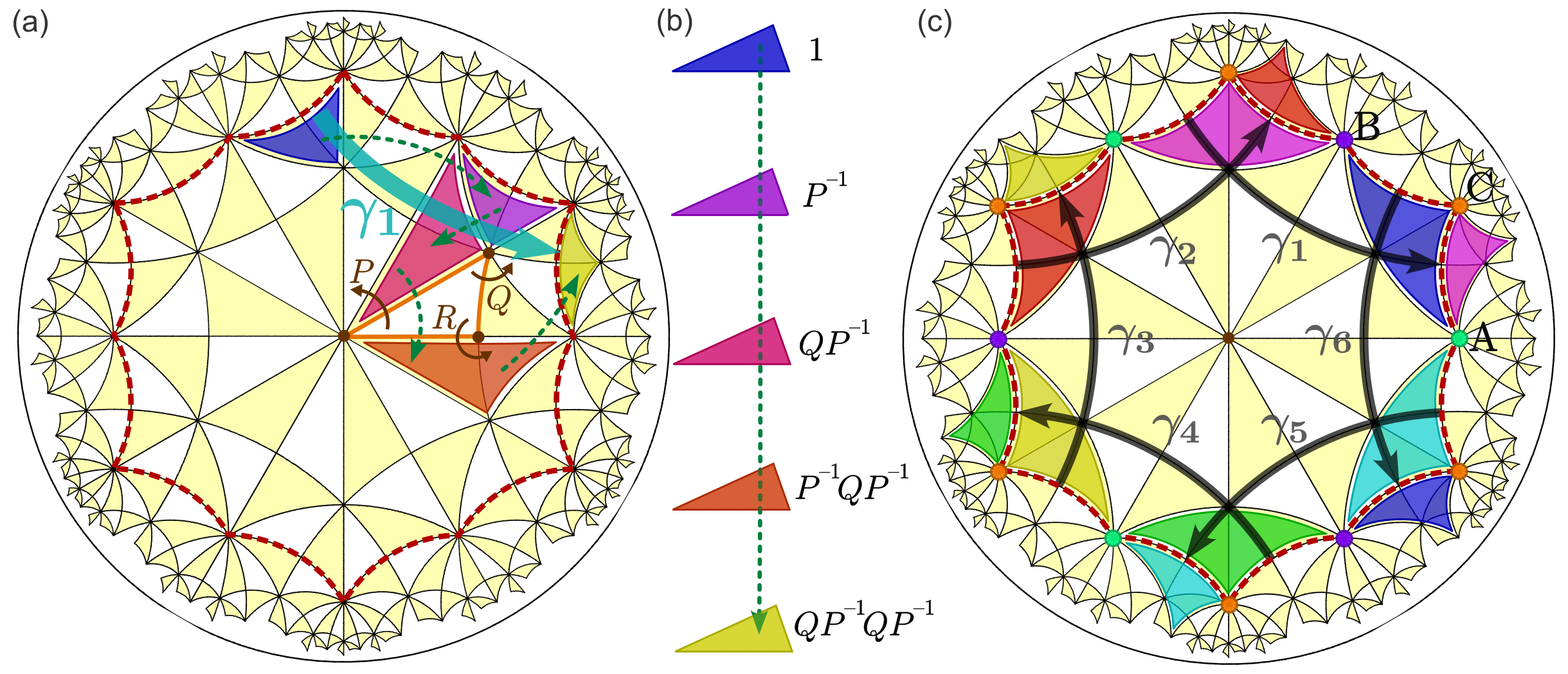}
\caption{
Identification of the hyperbolic translation group of the $\{6,4\}$ lattice (see Appendices~\ref{app:6-4-symmetry} and~\ref{eqn:PG-6-4}).}
\label{fig:6-4-cell}
\end{figure*}

To obtain the symmetry group of the hyperbolic Haldane model in the absence of a sublattice mass $\pm m$, all orientation-reversing elements of $\Delta(2,q,p)$ must be composed with time reversal $\mcT$. 
The resulting type-III~\cite{Bradley:1972,Litvin:2013} 
magnetic hyperbolic space group is thus obtained simply by replacing in Eq.~(\ref{eqn:Delta-def}) the generators $a,b,c$ by $\mcT a,\mcT b ,\mcT c$, namely
\begin{eqnarray}
\mathsf{M}(2,q,p)&=&\big<\mcT a,\mcT b,\mcT c\,\big|\,  \\
&\phantom{=}&  \,\big|\, (\mcT a)^2,(\mcT b)^2,(\mcT c)^2,(ab)^2,(bc)^q,(ca)^p\big>,\nonumber \label{eqn:magnetic-def}
\end{eqnarray}
or, equivalently,
\begin{equation}
\mathsf{M}(2,q,p)=\mathsf{D}(2,q,p){\,\cup\,} \mcT \left[\Delta(2,q,p) \backslash \mathsf{D}(2,q,p)\right],
\end{equation}
where we have treated the groups as sets.
Note that time reversal commutes with space-group symmetries, therefore we have simplified $\mcT a \mcT b = ab$ in Eq.~(\ref{eqn:magnetic-def}).\footnote{To be more precise, time reversal commutes with reflections in spinless systems (i.e., $\mcT^2 = 1$ and $\mcT a = a \mcT$) and anticommutes with reflections in spinful systems (i.e., $\mcT^2 = -1$ and $\mcT a = - a \mcT$). One finds that $\mcT a \mcT b = ab$ in both cases.}
The fundamental domain of $\mathsf{M}(2,q,p)$ is again the Schwarz triangle. The subgroup of orientation-preserving elements of $\mathsf{M}(2,q,p)$ is again the von Dyck group (\ref{eqn:D-2pq}), implying that $\mathsf{M}(2,q,p)$ and $\Delta(2,q,p)$ are characterized by the same translation group.

In the presence of a sublattice mass ($m \neq 0$), which is allowed in Haldane models with even $p$, the symmetry group is generated by elements $\mcT b$ and $\mcT c$ together with $(ca)^2=:P^2$, where the last operation is a counter-clockwise rotation by $4\pi/p$ around the face of the underlying $\{p,q\}$-lattice.
We label the resulting type-III magnetic hyperbolic space group as $\widetilde{\mathsf{M}}(2,q,p)$ to distinguish it from Eq.~(\ref{eqn:magnetic-def}). 
It is uniquely specified by its presentation
\begin{eqnarray}
\widetilde{\mathsf{M}}(2,q,p)&=&\big<P^2,\mcT b,\mcT c\,\big|\,  \\
&\phantom{=}&  \,\big|\, (\mcT b)^2,(\mcT c)^2,(bc)^q,(P^2)^{p/2},P^2(\mcT c) P^2(\mcT c)\big>.\nonumber \label{eqn:magnetic-def-2}
\end{eqnarray}
This is an index-2 subgroup of $\mathsf{M}(2,q,p)$, and its fundamental domain consists of a pair of Schwarz triangles meeting along their common edge associated with the broken reflection $a$.
The identification of the translation group in the presence of a non-vanishing sublattice mass $m$ in the Haldane Hamiltonian~(\ref{eq:Hh}) requires some care.
We are unaware of a general proof showing that $\mathsf{T}(2,q,p)$ as constructed above should always be a subgroup of $\widetilde{\mathsf{M}}(2,q,p)$; however, this is indeed the case for the seven $\{p,q\}$ lattices displayed in Fig.~\ref{fig:unit-cells}. 

In summary, given any of the seven $\{p,q\}$ lattices in Fig.~\ref{fig:unit-cells}, one can study hyperbolic Haldane models on them (with or without sublattice mass) by adopting the hyperbolic translation group $\mathsf{T}(2,q,p)$ derived from the von Dyck group $\mathsf{D}(2,q,p)$ using the steps outlined above.

\subsection{Symmetry and momentum space of \texorpdfstring{$\{6,4\}$}{(6,4)} lattice}\label{app:6-4-symmetry}

We proceed by applying the general prescription from Appendix~\ref{app:crystallography} to identify the translation group and construct momentum space for the $\{6,4\}$ lattice.
In Ref.~\onlinecite{Conder:2007}, the quotient of the von Dyck group $\mathsf{D}(2,4,6)$ that acts on a Riemann surface with the smallest possible genus $\mathfrak{g}=2$ is listed as ``$\textrm{T}{2.2}$'' [which we relabel here as $\mathsf{P_{\!\mathsf{s}}}(2,4,6)$].
This is stated to be a finite group of order $24$, specified by the presentation 
\begin{equation}
\mathsf{P_{\!\mathsf{s}}}(2,4,6) = \left< \breve{P},\breve{Q},\breve{R}\,\big| \, \breve{R}^2, \breve{Q}^4, \breve{P}^6, \breve{R}\breve{Q}\breve{P}, (\breve{Q}\breve{P}^{-1})^2\right>.\label{eqn:P64-presentation}
\end{equation}
By comparing the presentation in Eq.~(\ref{eqn:P64-presentation}) to that in Eq.~(\ref{eqn:D-2pq}), we observe one additional relator $(\breve{Q}\breve{P}^{-1})^2=\breve{1}$ in the point group.
Therefore, $(QP^{-1})^2 \in \mathsf{D}(2,4,6)$ must be an element of the translation subgroup $\mathsf{T}(2,4,6)$. We denote it~as~${(QP^{-1})^2=:\gamma_1}$. 

We approach the identification of $\mathsf{T}(2,4,6)$ geometrically, while leaving a systematic algorithmization for future work~\cite{Lenggenhager:2023}.
In particular, our strategy for finding the hyperbolic unit cell is to follow the trajectory of selected Schwarz triangles in the hyperbolic plane under the action of $(QP^{-1})^2=QP^{-1}QP^{-1}$.
To that end, we define the generators $P,Q,R$ in Eq.~(\ref{eqn:D-2pq}) with respect to the corners of the positively-oriented Schwarz triangle outlined with orange edges in Fig.~\ref{fig:6-4-cell}(a) (see Fig.~\ref{fig:Schwarz-notation} in the main text).
To study the action of $(QP^{-1})^2=QP^{-1}QP^{-1}$ on the triangular tessellation, it is convenient to take as the initial Schwarz triangle for the trajectory the one displayed in blue in Fig.~\ref{fig:6-4-cell}(a), and we act on it successively with rotations $P^{-1}$, $Q$, $P^{-1}$, and $Q$ (dashed green arrows); then the initial (blue) Schwarz triangle will follow a trajectory [summarized in Fig.~\ref{fig:6-4-cell}(b)] that passes through triangles displayed in purple/red/orange until reaching the Schwarz triangle displayed in dark yellow. 
The net motion along $QP^{-1}QP^{-1}=:\gamma_1$ [turquoise arrow in Fig.~\ref{fig:6-4-cell}(a)] corresponds to a nontrivial element of the translation group $\mathsf{T}(2,4,6)$, meaning that either the initial (blue) or the final (dark yellow) triangle lies in the fundamental domain, \emph{but not both}.

\begin{figure*}
\includegraphics[width=\linewidth]{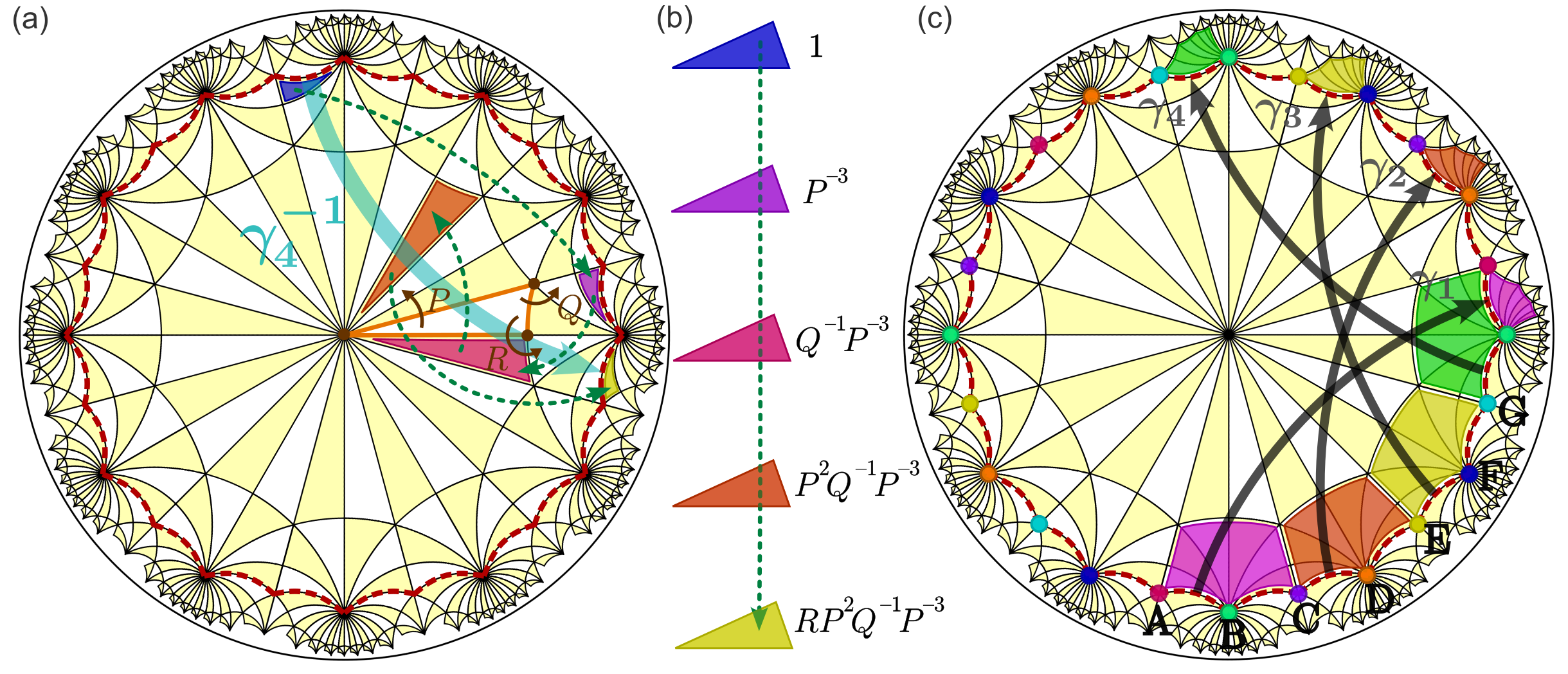}
\caption{
Identification of the hyperbolic translation group of the $\{12,3\}$ lattice (see Appendices~\ref{app:12-3-symmetry} and~\ref{eqn:PG-12-3}). 
For clarity, only the first four translation generators $\gamma_{1},\gamma_2,\gamma_3,\gamma_3$, out of the total of twelve, are explicitly shown in the right panel.}
\label{fig:12-3-cell}
\end{figure*}

Next, we construct six elements of $\mathsf{T}(2,4,6)$ as
\begin{equation}
\gamma_j := P^{(j-1)}\gamma_1 P^{-(j-1)},\hspace{5mm}j=1,\ldots,6.   \label{eqn:6-4-gammas} 
\end{equation}
Indeed, since $\breve{\gamma}_1 = \breve{1}$ inside the quotient $\mathsf{P}_{\!\mathsf{s}}(2,4,6)$, one also has $\breve{\gamma}_j = \breve{P}^{(j-1)}\breve{\gamma}_1 \breve{P}^{-(j-1)} = \breve{P}^{(j-1)} \breve{P}^{-(j-1)} = \breve{1}$, meaning that $\{\gamma_j\}_{j=1}^6$ are indeed elements of $\mathsf{T}(2,4,6)$.
Since the defining relation in Eq.~(\ref{eqn:6-4-gammas}) corresponds to conjugation with rotations around the center of the disk [see~Fig.~\ref{fig:6-4-cell}(a)], we easily identify the action of all $\{\gamma_j\}_{j=1}^6$ on the hyperbolic plane.
In particular, for each translation $\gamma_j$ we show a pair of triangular regions which are related through that translation [pairs of matching colors in Fig.~\ref{fig:6-4-cell}(c)].
Since only one member of each colored triangle-pair in Fig.~\ref{fig:6-4-cell}(c) can lie within the fundamental domain of $\mathsf{T}(2,4,6)$, a simple choice for the hyperbolic unit cell becomes apparent: the dashed red, semi-regular dodecagon outlined in Fig.~\ref{fig:6-4-cell}(c). 
This unit cell and its pairwise identification of edges were also derived using algebraic geometry methods in Ref.~\onlinecite{Kuusalo:1995}. The resulting compactified surface is a genus-2 Accola-Maclachlan surface~\cite{accola1968,maclachlan1969}, corresponding to the complex algebraic curve $w^2=z^6-1$. 
The dodecagonal unit cell contains $24$ fundamental triangles (i.e., $48$ Schwarz triangles), thus matching the order of $\mathsf{P}_{\!\mathsf{s}}(2,4,6)$ as specified in Ref.~\onlinecite{Conder:2007}.

We next need to derive a presentation of the translation group $\mathsf{T}(2,4,6)$ in terms of generators and relators. 
Since the six translations in Eq.~(\ref{eqn:6-4-gammas}) relate all twelve edges of the dodecagonal unit cell in Fig.~\ref{fig:6-4-cell}(c), they comprise a sufficient set of generators. 
To identify the relators, we follow the trajectory of \emph{corners} of the unit cell under suitable combinations of $\{\gamma_j\}_{j=1}^6$.
We find that the dodecagonal unit cell is demarcated by three distinct corners (by ``distinct'' we mean modulo translations; i.e., $\delta\mathfrak{v}=2$ in terms of footnote~\ref{foot:corner-Euler}), which we label $\textrm{A}$, $\textrm{B}$, and $\textrm{C}$.
We observe that corner $\mathrm{A}$ [green dots in Fig.~\ref{fig:6-4-cell}(c)] is brought to its initial position under sequential translations by $\gamma_5$, $\gamma_3$, and $\gamma_1$, i.e., $\mathrm{A} = \gamma_{1}\gamma_{3}\gamma_{5}\mathrm{A}$.
Thus we deduce the first relator, namely $\gamma_1\gamma_3\gamma_5 = 1$. 
By similarly following the trajectory of the remaining corners $\mathrm{B}$ and $\mathrm{C}$ [purple resp.~orange dots in Fig.~\ref{fig:6-4-cell}(c)], we complete the set of relators with $\gamma_2\gamma_4\gamma_6 = 1$ and $\gamma_1\gamma_2\gamma_3\gamma_4\gamma_5\gamma_6=1$. 
Therefore, we obtain
\begin{equation}
\mathsf{T}(2{,}4{,}6)=\left<\gamma_1,...,\gamma_6\,\big|\,\gamma_1\gamma_3\gamma_5, \gamma_2\gamma_4\gamma_6,\gamma_1\gamma_2\gamma_3\gamma_4\gamma_5\gamma_6\right>  \label{eqn:6-4-translations}
\end{equation}
as the presentation of the hyperbolic translation group of the $\{6,4\}$ lattice. 
This matches the information listed in Table~\ref{table:unit-cells} of the main text.

Finally, to construct HBT, we study constraints on the map $\rho_{\bs{k}}: \gamma_j \mapsto \e^{\imi k^j}$ to be a 1D IR. The first two relators in Eq.~(\ref{eqn:6-4-translations}) respectively imply that
\begin{equation}
k^5 = -k^1 -k^3\quad\textrm{and}\quad k^6 = -k^2 - k^4,\label{eqn:6-4-momenta}
\end{equation}
which turn out to be automatically consistent with the third relator. 
We thus conclude that the BZ of 1D IRs for the $\{6,4\}$ lattice is spanned by four momenta $k^{1,\ldots,4}\in[-\pi,\pi]$, corresponding to a dodecagonal unit cell and six twisted boundary conditions $\rho_{\bs{k}}: \gamma_j \mapsto \e^{\imi k^j}$ shown in Fig.~\ref{fig:6-4-cell}(c) and with phases $k^5$ and $k^6$ fixed by Eq.~(\ref{eqn:6-4-momenta}).

\subsection{Symmetry and momentum space of \texorpdfstring{$\{12,3\}$}{(12,3)} lattice}\label{app:12-3-symmetry}

We now repeat the exercise of the previous section for the $\{12,3\}$ lattice.
In Ref.~\onlinecite{Conder:2007}, the quotient of the von Dyck group $\mathsf{D}(2,3,12)$ that acts on a Riemann surface with the smallest possible genus $\mathfrak{g}=3$ is listed as ``$\textrm{T}{3.3}$'' [which we relabel here as $\mathsf{P_{\!\mathsf{s}}}(2,3,12)$].
This is stated to be a finite group of order $48$ with the presentation 
\begin{eqnarray}
\mathsf{P_{\!\mathsf{s}}}(2,3,12) &=& \big< \breve{P},\breve{Q},\breve{R}\,\big|\, \label{eqn:P-12-3-presentation} \\
&\phantom{=}& \,\big|\, \breve{R}^2, \breve{Q}^3, \breve{R}\breve{Q}\breve{P}, \breve{R}\breve{P}^2\breve{Q}^{-1}\breve{P}^{-3}\big>.\nonumber 
\end{eqnarray}
By comparing Eqs.~(\ref{eqn:P-12-3-presentation}) and~(\ref{eqn:D-2pq}), we recognize one additional relator $\breve{R}\breve{P}^2\breve{Q}^{-1}\breve{P}^{-3}=\breve{1}$ in the point group.
Therefore, ${R}{P}^2{Q}^{-1}{P}^{-3} \in \mathsf{D}(2,3,12)$ must be an element of the translation group $\mathsf{T}(2,3,12)$. We denote it as ${R}{P}^2{Q}^{-1}{P}^{-3}=:\gamma_{4}^{-1}$. 

Similar to the discussion in Appendix~\ref{app:6-4-symmetry}, we define the generators $P,Q,R$ in Eq.~(\ref{eqn:D-2pq}) with respect to the corners of the positively-oriented Schwarz triangle positioned above the positive horizontal axis [outlined with orange edges in Fig.~\ref{fig:12-3-cell}(a)].
For concreteness, we depart from the Schwarz triangle displayed in blue, and we act on it successively with rotations $P^{-3}$, $Q^{-1}$, $P^2$, and $R$ (dashed green arrows); then the initial (blue) Schwarz triangle will follow a trajectory [summarized in Fig.~\ref{fig:12-3-cell}(b)] that passes through triangles displayed in purple/red/orange until reaching the Schwarz triangle displayed in dark yellow. 
The net motion along ${R}{P}^2{Q}^{-1}{P}^{-3}=:\gamma_4^{-1}$ [turquoise arrow in Fig.~\ref{fig:6-4-cell}(a)] corresponds to a nontrivial element of the translation group $\mathsf{T}(2,3,12)$, meaning that either the initial (blue) or the final (dark yellow) triangle lies in the fundamental domain, but not both.

We now construct twelve elements of $\mathsf{T}(2,3,12)$ via
\begin{equation}
\gamma_j^{-1} := P^{(j-4)}\gamma_4^{-1} P^{-(j-4)},\hspace{5mm}j=1,\ldots,12.   \label{eqn:3-12-gammas} 
\end{equation}
Indeed, knowing that $\breve{\gamma}_4^{-1} = \breve{1}$ inside the quotient $\mathsf{P}_{\!\mathsf{s}}(2,3,12)$, one has $\breve{\gamma}_j^{-1} = \breve{P}^{(j-4)}\breve{\gamma}_4^{-1} \breve{P}^{-(j-4)} = \breve{P}^{(j-4)} \breve{P}^{-(j-4)} = \breve{1}$, meaning that $\gamma_j$ in Eq.~(\ref{eqn:3-12-gammas}) are indeed elements of $\mathsf{T}(2,3,12)$.
Since the defining relation in Eq.~(\ref{eqn:3-12-gammas}) corresponds to conjugation with rotations around the center of the disk [cf.~Fig.~\ref{fig:12-3-cell}(a)], we easily identify the action of all $\{\gamma_j\}_{j=1}^{12}$ on the hyperbolic plane.
In particular, for each translation $\gamma_j$ we show a pair of pentagonal regions which are related through that translation [pairs of matching colors in Fig.~\ref{fig:12-3-cell}(c)].
Since only one member of each colored pentagon-pair in Fig.~\ref{fig:12-3-cell}(c) can lie within the fundamental domain of $\mathsf{T}(2,3,12)$, a simple choice for the hyperbolic unit cell becomes apparent: the dashed red $24$-gon outlined in Fig.~\ref{fig:12-3-cell}(c). Upon identification of the edges, the unit cell becomes a $\mathfrak{g}=3$ Riemann surface known as the $\textrm{M}(3)$ surface~\cite{schmutz1993}.
Counting reveals that there are $48$ fundamental triangles (i.e., $96$ Schwarz triangles) inside the constructed unit cell, which matches the order of $\mathsf{P}_{\!\mathsf{s}}(2,3,12)$ listed in Ref.~\onlinecite{Conder:2007}.

We next derive a presentation of the translation group $\mathsf{T}(2,3,12)$ in terms of $\{\gamma_j\}_{j=1}^{12}$ and relators. 
To identify the relators, we follow the trajectory of corners of the unit cell under suitable combinations of the translation generators.
We find that the unit cell is demarcated by seven distinct corners (i.e., $\delta\mathfrak{v}=6$ in terms of footnote~\ref{foot:corner-Euler}), which we label by letters from $\textrm{A}$ to $\textrm{G}$.
We observe that corner $\mathrm{A}$ [red dots in Fig.~\ref{fig:12-3-cell}(c)] is brought to its initial position under sequential translations by $\gamma_1$, $\gamma_5$, and $\gamma_9$, i.e., $\mathrm{A} = \gamma_{9}\gamma_{5}\gamma_{1}\mathrm{A}$.
Thus we deduce the first relator, namely $\gamma_9\gamma_5\gamma_1 = 1$. 
By similarly following the trajectory of the remaining corners [dots of matching colors in Fig.~\ref{fig:12-3-cell}(c)], we complete the set of relators with
\begin{eqnarray}
\textrm{A} \quad & \Rightarrow & \quad \gamma_9\gamma_5\gamma_1 = 1, \nonumber \\
\textrm{C} \quad & \Rightarrow &  \quad \gamma_{10}\gamma_6\gamma_2 = 1, \nonumber \\
\textrm{E} \quad & \Rightarrow & \quad \gamma_{11}\gamma_7\gamma_3 = 1, \nonumber \\
\textrm{G} \quad & \Rightarrow & \quad \gamma_{12}\gamma_8\gamma_4 = 1,  \\
\textrm{B} \quad & \Rightarrow & \quad \gamma_{10}\gamma_7\gamma_4\gamma_1 = 1, \nonumber \\
\textrm{D} \quad & \Rightarrow & \quad \gamma_{11}\gamma_8\gamma_5\gamma_2 = 1, \nonumber \\
\textrm{F} \quad & \Rightarrow & \quad \gamma_{12}\gamma_9\gamma_6\gamma_3 = 1, \nonumber 
\end{eqnarray}
as listed in Table~\ref{table:unit-cells} of the main text. 
Finally, from the above relators we derive constraints on the momenta parameterizing 1D IR $\rho_{\bs{k}}: \gamma_j \mapsto \e^{\imi k^j}$, which we also list in Table~\ref{table:unit-cells}. 
The constraints are explicitly solved as
\begin{eqnarray}
&k^7 = -k^1 + k^2 - k^4 + k^6,& \nonumber \\
&k^8 = -k^1 + k^3 - k^4 - k^5 + k^6,& \nonumber \\
&k^9 = -k^1 - k^5, \qquad k^{10} = -k^2 - k^6, & \\
&k^{11} = k^1 - k^2 - k^3 + k^4 - k^6,& \nonumber \\
&k^{12} = k^1 - k^3 + k^5 - k^6.& \nonumber
\end{eqnarray}

\subsection{Hyperbolic point groups}\label{sec:hyper-PGs}

\begin{table}[t]
\begin{threeparttable}
\caption{
Proper ($\mathsf{P}_{\!\mathsf{s}}$) and full ($\mathsf{P}$)  point groups characterizing the symmetry of hyperbolic unit cells for the seven $\{p,q\}$ lattices considered. The data for $\mathsf{P}_{\!\mathsf{s}}(2,q,p)$ are taken from Ref.~\onlinecite{Wolfart:2005}, while those for $\mathsf{P}(2,q,p)$ were extracted using GAP following the analysis in Appendix~\ref{sec:hyper-PGs}. 
The last column gives the order (number of elements) of $\mathsf{P}(2,q,p)$, which equals twice the order of $\mathsf{P}_{\!\mathsf{s}}(2,q,p)$.
\label{table:PG-summary} 
}
\begin{ruledtabular}
\begin{tabular}{lccr}
$\{p,q\}$ & $\mathsf{P}_{\!\mathsf{s}}(2,q,p)$ & $\mathsf{P}(2,q,p)$ & $\abs{\mathsf{P}(2,q,p)}$
\tabularnewline \hline
$\{8,3\}$  & $\mathsf{GL}(2,\mathbb{Z}_3)$ & $\mathsf{GL}(2,\mathbb{Z}_3)\rtimes\mathbb{Z}_2$ & $96$ 
\tabularnewline
$\{6,4\}$  & $(\mathbb{Z}_3 \times \mathbb{Z}_2 \times \mathbb{Z}_2)\rtimes \mathbb{Z}_3$ & $ \mathsf{D}_4\times \mathsf{S}_3$ & $48$
\tabularnewline
$\{8,4\}$  & $\mathbb{Z}_8 \rtimes \mathbb{Z}_2$ & $ \mathbb{Z}_8 \rtimes (\mathbb{Z}_2\times \mathbb{Z}_2)$ & $32$
\tabularnewline
$\{10,5\}$  & $\mathbb{Z}_{10}$ & $ \mathsf{D}_{10}$ & $20$
\tabularnewline
$\{7,3\}$  & $\mathsf{PSL}(2,\mathbb{Z}_7)$ & $ \mathsf{PGL}(2,\mathbb{Z}_7)$ & $336$
\tabularnewline
$\{12,3\}$  & $\mathsf{SL}(2,\mathbb{Z}_3)\times \mathbb{Z}_4/\langle(-\mathbb{1},t^2)\rangle$ & $\mathsf{GL}(2,\mathbb{Z}_3)\rtimes \mathbb{Z}_2$  & $96$
\tabularnewline
$\{12,4\}$  & $\mathbb{Z}_4 \times \mathsf{S}_3$ & $\mathsf{D}_4 \times \mathsf{S}_3$  & $48$
\end{tabular}\end{ruledtabular} 
\begin{tablenotes}[flushleft]
\item[$\!\!$] Notation: $\mathbb{Z}_n$ is the cyclic group of $n$ elements, $\mathsf{D}_n=\left<a,b\,|\,a^n,b^2,b^{-1}aba\right>$ is the dihedral group of order $2n$; $\mathsf{S}_n$ is the permutation group of $n$ elements and has order $n!$, $\mathsf{GL}(n,\mathbb{F})$ is the group of invertible $n\times n$ matrices over field $\mathbb{F}$, $\mathsf{SL}(n,\mathbb{F})$ is the index-two subgroup of $\mathsf{GL}(n,\mathbb{F})$ that contains matrices of unit determinant, projective linear ($\mathsf{P}$) groups are obtained from the corresponding linear ($\mathsf{GL}$ or $\mathsf{SL}$) groups by identifying pairs of elements related by the negative identity matrix, and $\langle(-\mathbb{1},t^2)\rangle$ in the $\{12,3\}$ entry is the order-two group generated by $(-\mathbb{1},t^2)\in \mathsf{SL}(2,\mathbb{Z}_3)\times\mathbb{Z}_4$, where $t$ is the generator of $\mathbb{Z}_4$. The group homomorphism $\varphi:\mathbb{Z}_{2n}\to\textrm{Aut}(\mathsf{G})$ entering all listed semidirect products $\mathsf{G}\rtimes_\varphi \mathbb{Z}_{2n}$ (with $\varphi$ dropped in the table) corresponds to the parity of elements in~$\mathbb{Z}_{2n}$. 
\end{tablenotes}
\end{threeparttable}
\end{table}

In this section, we systematically go through the hyperbolic $\{p,q\}$ lattices listed in Fig.~\ref{fig:unit-cells}, and for each discuss the (full) point group $\mathsf{P}(2,q,p)$ and the proper point group $\mathsf{P}_{\!\mathsf{s}}(2,q,p)$. 
In particular, we discuss how the point-group relators listed by Conder~\cite{Conder:2007} can be reduced to certain standardized forms which are usually found in the mathematical literature, and that allow for easier derivations of the corresponding translation groups and hyperbolic unit cells.
The considerations that reveal the hyperbolic unit cells and the identification of their edges under hyperbolic translations are illustrated for the seven distinct lattices in Figs.~\ref{fig:6-4-cell}--\ref{fig:12-4-cell}.
The collected information on the hyperbolic point groups is summarized by Table~\ref{table:PG-summary}.

\subsubsection{Point group of \texorpdfstring{$\{8,3\}$}{(8,3)} lattice}\label{eqn:PG-8-3}

\begin{figure*}
\includegraphics[width=0.625\linewidth]{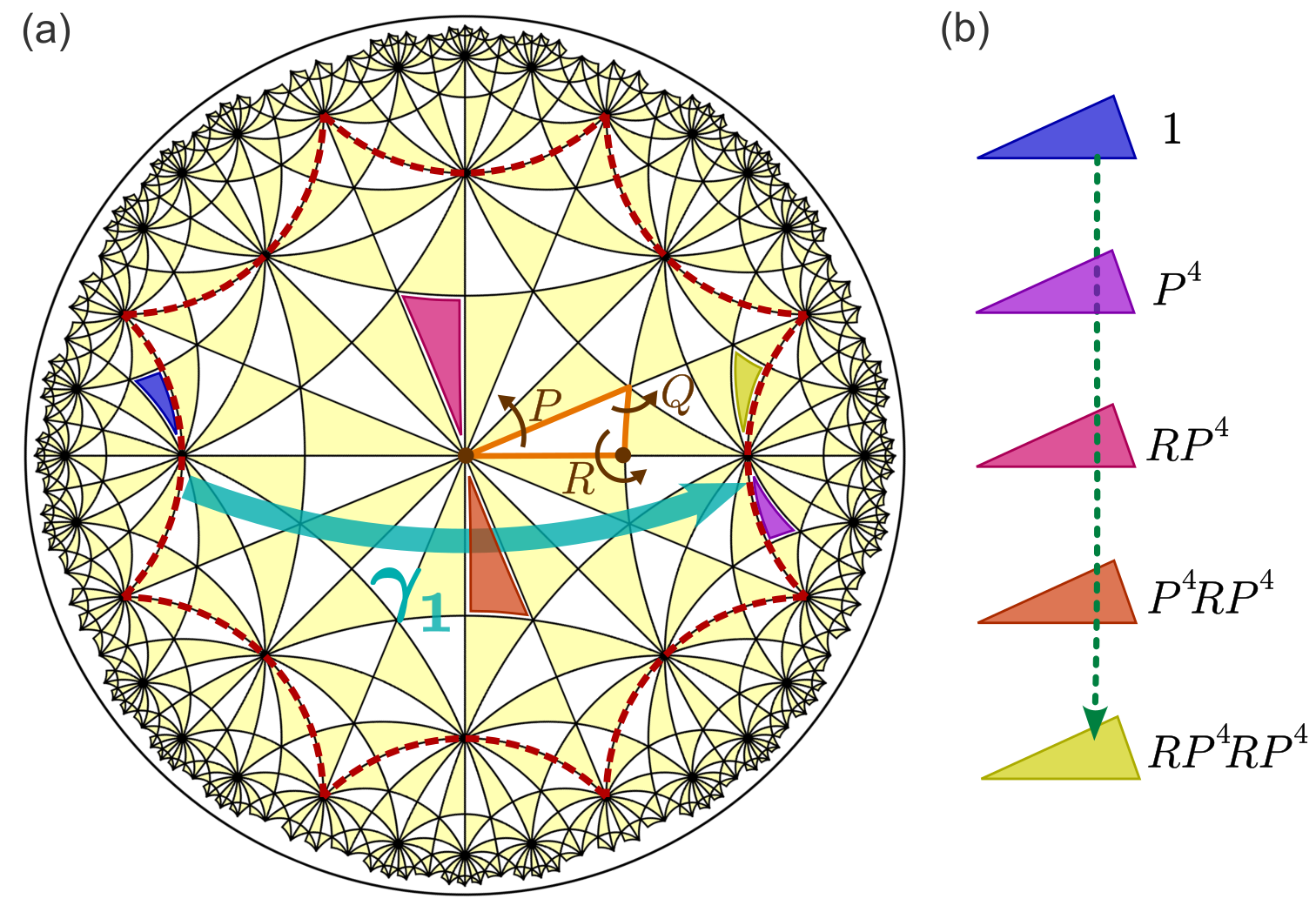}
\caption{Identification of the hyperbolic translation group of the $\{8,3\}$ lattice (see Appendix~\ref{eqn:PG-8-3}).}
\label{fig:8-3-cell}
\end{figure*}

The proper point group of the $\{8,3\}$ lattice is specified in Ref.~\onlinecite{Conder:2007} by the presentation
\begin{eqnarray}
\!\!\!\!\!\!\mathsf{P}_{\!\mathsf{s}}(2,3,8) \!=\! \left<\breve{P}{,}\breve{Q}{,}\breve{R}\,\big|\,\breve{R}^2{,}\breve{Q}^3{,}\breve{R}\breve{Q}\breve{P}{,}\breve{P}\breve{Q}\breve{R}\breve{P}\breve{Q}^{-1}\breve{P}^{-1}\breve{R}\breve{P}\right>.\,\,\, \label{eqn:PG-2-3-8}
\end{eqnarray}
In contrast, Ref.~\onlinecite{Kuusalo:1995} defines the same group by
\begin{equation}
\mathsf{P}_{\!\mathsf{s}}(2,3,8) = \left<\breve{P},\breve{Q},\breve{R}\,\big|\,\breve{R}^2,\breve{Q}^3,\breve{P}^8,\breve{R}\breve{Q}\breve{P},\breve{R}\breve{P}^4\breve{R}\breve{P}^{-4}\right>.     \label{eqn:PG-2-3-8-Kuusalo}
\end{equation}
We show how the more standard relators in Eq.~(\ref{eqn:PG-2-3-8-Kuusalo}) can be obtained from those listed in Eq.~(\ref{eqn:PG-2-3-8}), thus making the two definitions equivalent.

To begin, we obtain from $\breve{R}\breve{Q}\breve{P}=\breve{1}$ with $\breve{R}^2=\breve{1}$ that
\begin{equation}
\breve{Q}=\breve{R}\breve{P}^{-1}\qquad\textrm{and}\qquad\breve{Q}^{-1}=\breve{P}\breve{R}. \label{eqn:2-3-8-PQR}
\end{equation} 
We use these expressions to replace $\breve{Q}$ and its inverse in the long relator in Eq.~(\ref{eqn:PG-2-3-8}), resulting in
\begin{eqnarray}
\breve{1}&=&\breve{P}\breve{R}\breve{P}^{-1}\breve{R}\breve{P}^2\breve{R}\breve{P}^{-1}\breve{R}\breve{P} \nonumber \\
&=& \breve{P}^2 \left(\breve{P}^{-1}\breve{R}\breve{P}^{-1}\breve{R}\breve{P}^{-1}\right) \breve{P}^4 \left(\breve{P}^{-1}\breve{R}\breve{P}^{-1}\breve{R}\breve{P}^{-1}\right) \breve{P}^2.\label{eqn:2-3-8-relators-at-worst}
\end{eqnarray}
To simplify Eq.~(\ref{eqn:2-3-8-relators-at-worst}), note that 
\begin{equation}
\breve{1}=\breve{Q}^3 = \breve{R} \breve{P}^{-1}\breve{R}\breve{P}^{-1}\breve{R}\breve{P}^{-1}\quad\!\! \Rightarrow \!\!\quad  \breve{P}^{-1}\breve{R}\breve{P}^{-1}\breve{R}\breve{P}^{-1} = \breve{R},\label{eqn:2-3-8-Q3}
\end{equation}
so that Eq.~(\ref{eqn:2-3-8-relators-at-worst}) reduces to
\begin{eqnarray}
\breve{1} &=& \breve{P}^2 \breve{R} \breve{P}^4 \breve{R} \breve{P}^2 \nonumber \\
&=& \breve{R} \breve{P}^4 \breve{R} \breve{P}^4,\label{eqn:RP4RP4}
\end{eqnarray}
where the second line is obtained from the first one through conjugation with $\breve{P}^{-2}$.

Furthermore, note that 
\begin{equation}
\left(\breve{P}^{-1}\breve{R}\breve{P}\right)^2 = \breve{P}^{-1}\breve{R}\breve{P}^{-1}\breve{P}\breve{R}\breve{P} = \breve{P}^{-1} \breve{R}^2 \breve{P}=\breve{1},
\end{equation}
and that the long relator in Eq.~(\ref{eqn:PG-2-3-8}) implies $\breve{P}\breve{Q}\breve{R}\breve{P}\breve{Q}^{-1} = \breve{P}^{-1}\breve{R}\breve{P}$. 
By combining the two previous results, we have
\begin{eqnarray}
\breve{1} &=& \breve{P}\breve{Q}\breve{R}\breve{P}\breve{Q}^{-1}\breve{P}\breve{Q}\breve{R}\breve{P}\breve{Q}^{-1} \nonumber \\
&=& \breve{P}\breve{R}\breve{P}^{-1}\breve{R}\breve{P}^2\breve{R}\breve{P}\breve{R}\breve{P}^{-1}\breve{R}\breve{P}^2\breve{R},\label{eqn:2-3-8-long-and-stubborn}
\end{eqnarray}
where in the second step we used Eq.~(\ref{eqn:2-3-8-PQR}). 
From Eq.~(\ref{eqn:2-3-8-Q3}) we observe that $\breve{R}\breve{P}^{-1}\breve{R} = \breve{P}\breve{R}\breve{P}$, which we plug into Eq.~(\ref{eqn:2-3-8-long-and-stubborn}). 
This results in
\begin{eqnarray}
\breve{1} &=& \breve{P}\left(\breve{R}\breve{P}^{-1}\breve{R}\right)\breve{P}^2\breve{R}\breve{P}\left(\breve{R}\breve{P}^{-1}\breve{R}\right)\breve{P}^2\breve{R} \nonumber \\
&=&\breve{P}\left(\breve{P}\breve{R}\breve{P}\right)\breve{P}^2\breve{R}\breve{P}\left(\breve{P}\breve{R}\breve{P}\right)\breve{P}^2\breve{R} \nonumber \\
&=& \breve{P}^2 \breve{R} \breve{P}^3 \breve{R} \breve{P}^2 \breve{R} \breve{P}^3 \breve{R}.\label{eqn:2-3-8-powers}
\end{eqnarray}
Next, we extract from Eq.~(\ref{eqn:RP4RP4}) that \begin{equation}
\breve{R}\breve{P}^3 = \breve{P}^{-4}\breve{R}\breve{P}^{-1}\quad\textrm{and}\quad \breve{P}^3\breve{R} = \breve{P}^{-1}\breve{R}\breve{P}^{-4}.
\end{equation}
Substituting for these expression in Eq.~(\ref{eqn:2-3-8-powers}), we obtain
\begin{eqnarray}
\breve{1} 
&=& \breve{P}^2 \left(\breve{R} \breve{P}^3 \right) \breve{R} \breve{P}^2 \breve{R} \left(\breve{P}^3 \breve{R}\right) \nonumber \\
&=& \breve{P}^2 \left(\breve{P}^{-4} \breve{R} \breve{P}^{-1} \right) \breve{R} \breve{P}^2 \breve{R} \left(\breve{P}^{-1} \breve{R} \breve{P}^{-4}\right) \nonumber \\
&=&\breve{P}^{-2}\breve{R}\breve{P}^{-1}\breve{R}\breve{P}^2\breve{R}\breve{P}^{-1}\breve{R}\breve{P}^{-4}. 
\end{eqnarray}
Finally, we recognize in the previous line the known expression~(\ref{eqn:2-3-8-Q3}) for $\breve{R}$, leading to
\begin{eqnarray}
\breve{1} &=& \breve{P}^{-1}\left(\breve{P}^{-1} \breve{R} \breve{P}^{-1} \breve{R} \breve{P}^{-1} \right) \breve{P}^4 \left( \breve{P}^{-1} \breve{R} \breve{P}^{-1} \breve{R} \breve{P}^{-1} \right) \breve{P}^{-3} \nonumber \\
&=& \breve{P}^{-1} \breve{R} \breve{P}^4 \breve{R} \breve{P}^{-3} \nonumber \\
&=& \breve{R} \breve{P}^4 \breve{R} \breve{P}^{-4},\label{eqn:RP4RP-4}
\end{eqnarray}
where the last line is obtained from the previous one through conjugation with $\breve{P}$. 
By comparing Eqs.~(\ref{eqn:RP4RP4}) and~(\ref{eqn:RP4RP-4}), we find that $\breve{P}^8=1$. 
Note that the last two results are precisely the relators listed in Eq.~(\ref{eqn:PG-2-3-8-Kuusalo}).

The full point group is obtained through the usual replacement of rotation generators by reflections, as given by Eq.~(\ref{eqn:mirror2rot}), resulting in
\begin{eqnarray}
\!\!\!\mathsf{P}(2,3,8)&=&\left<\breve{a},\breve{b},\breve{c}\,\big|\,\breve{a}^2,\breve{b}^2,\breve{c}^2,(\breve{a}\breve{b})^2,\right. \nonumber \\
&\phantom{=}&\qquad \left. (\breve{b}\breve{c})^3,(\breve{c}\breve{a})^8,(\breve{a}\breve{b})(\breve{c}\breve{a})^4(\breve{a}\breve{b})(\breve{c}\breve{a})^{-4}\right>.\label{eqn:P-2-3-8-full}
\end{eqnarray}
Note that Eq.~(\ref{eqn:PG-2-3-8-Kuusalo}) differs from the corresponding von Dyck group in Eq.~(\ref{eqn:D-2pq}) by the additional relator $\breve{R}\breve{P}^4\breve{R}\breve{P}^{-4} = \breve{1}$. 
Therefore, $RP^4 RP^{-4} \in \mathsf{D}(2,3,8)$ is an element of the translation subgroup $\mathsf{T}(2,3,8)$, which we label in Fig.~\ref{fig:8-3-cell} as $\gamma_1$ .
We recognize from the sketch in Fig.~\ref{fig:8-3-cell} that the hyperbolic unit cell corresponds to the Bolza cell~\cite{Maciejko:2021,Urwyler:2021} and that translations relate antipodal edges of the Bolza cell.
Compactification of the unit cell edges results in the famous Bolza surface, with algebraic curve description $w^2=z^5-z$~\cite{bolza1887}.
We verify in GAP that the group defined by Eq.~(\ref{eqn:P-2-3-8-full}) is isomorphic to $\mathsf{GL}(2,\mathbb{Z}_3)\rtimes\mathbb{Z}_2$, which is the known automorphism group of the Bolza surface~\cite{bolza1887,CookThesis} of order $96$.
The proper point group (\ref{eqn:PG-2-3-8-Kuusalo}) is found to be isomorphic to $\mathsf{GL}(2,\mathbb{Z}_3)$.

\subsubsection{Point group of \texorpdfstring{$\{6,4\}$}{(6,4)} lattice}\label{eqn:PG-6-4}

We read from Ref.~\onlinecite{Conder:2007} that $\mathsf{P}_{\!\mathsf{s}}(2,4,6)$ is given by Eq.~(\ref{eqn:P64-presentation}), whereas Ref.~\onlinecite{Kuusalo:1995} provides the seemingly different presentation
\begin{eqnarray}
\mathsf{P}(2,4,6) &=& \left<\breve{a},\breve{b},\breve{c}\,\big|\,\breve{a}^2,\breve{b}^2,\breve{c}^2,(\breve{a}\breve{b})^2,\right. \nonumber \\
&\phantom{=}&\qquad \left. (\breve{b}\breve{c})^4,(\breve{c}\breve{a})^6,(\breve{b}\breve{c}\breve{a}\breve{c})^2\right>. \label{eqn:P-2-4-6-full}
\end{eqnarray}
These two presentations of $\mathsf{P}(2,4,6)$ are easily shown to be equivalent under the usual replacement of generators in Eq.~(\ref{eqn:mirror2rot}). In particular, for the nontrivial relator one finds using $\breve{Q}=\breve{b}\breve{c}$ and $\breve{P}^{-1} = \breve{a}^{-1}\breve{c}^{-1} = \breve{a}\breve{c}$ that
\begin{equation}
\breve{1}=(\breve{Q}\breve{P}^{-1})^2 = (\breve{b}\breve{c}\breve{a}\breve{c})^2.  
\end{equation}
We thus identify $(QP^{-1})^2 = (bcac)^2$ as an element of the translation subgroup $\mathsf{T}(2,4,6)\triangleleft\Delta(2,4,6)$, which is labelled as $\gamma_1$ in Fig.~\ref{fig:6-4-cell}. 

Using GAP, we find that $\mathsf{P}(2,4,6)$ is a non-Abelian finite group of order 48, isomorphic to $\mathsf{D}_4\times \mathsf{S}_3$. 
Here, $\mathsf{D}_4$ is the dihedral group with $8$ elements, which is the symmetry group of a square (in the mathematical literature the same group is sometimes denoted as $\mathsf{D}_8$), and $\mathsf{S}_3$ is the permutation group with $3!=6$ elements.
The proper point group is similarly found to be isomorphic to $(\mathbb{Z}_3 \times \mathbb{Z}_2 \times \mathbb{Z}_2)\rtimes \mathbb{Z}_3$.

\subsubsection{Point group of \texorpdfstring{$\{8,4\}$}{(8,4)} lattice}\label{eqn:PG-8-4}

\begin{figure*}
\includegraphics[width=0.625\linewidth]{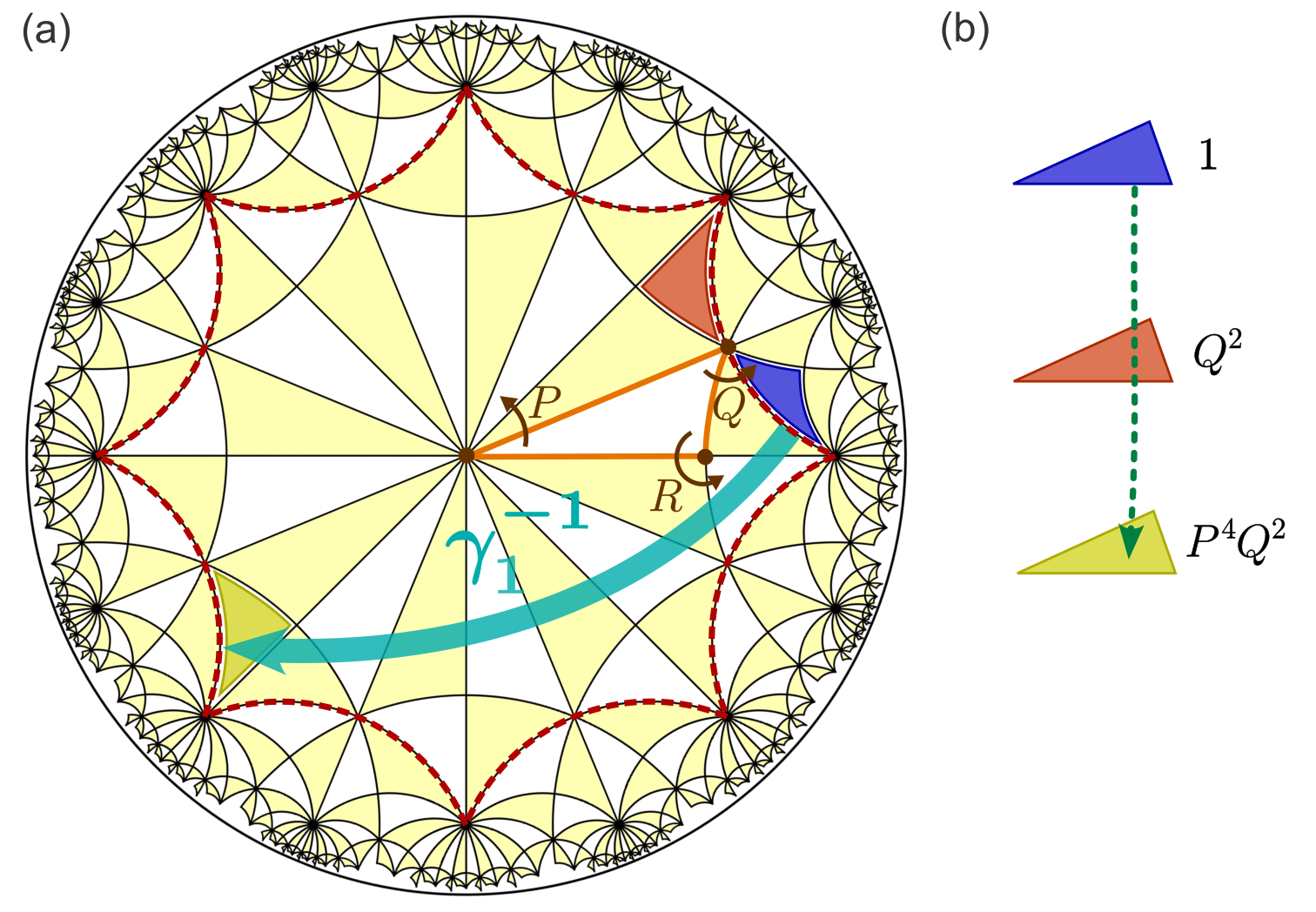}
\caption{Identification of the hyperbolic translation group of the $\{8,4\}$ lattice (see Appendix~\ref{eqn:PG-6-4}).}
\label{fig:8-4-cell}
\end{figure*}

From Ref.~\onlinecite{Conder:2007} we find that
\begin{eqnarray}
\mathsf{P}_{\!\mathsf{s}}(2,4,8) &=& \left<\breve{P},\breve{Q},\breve{R}\,\big|\,\breve{R}^2,\breve{Q}^4,\right.\nonumber \\
&\phantom{=}& \qquad \left. \breve{R}\breve{Q}\breve{P},(\breve{Q}\breve{P}^{-1})^2,\breve{P}^{-1}\breve{R}\breve{Q}^{-1}\breve{P}^{-2}\right>.  \label{eqn:PG-2-4-8}
\end{eqnarray}
We show how the above definition can be brought to a more standard form. First, 
we find from the long relator in Eq.~(\ref{eqn:PG-2-4-8}) that
\begin{eqnarray}
\breve{1} 
&=& \breve{P}^{-1}\breve{R}\breve{Q}^{-1}\breve{P}^{-2} \nonumber \\
&=& \breve{P}^{-1}\breve{Q}\breve{P}\breve{Q}^{-1}\breve{P}^{-2} \nonumber \\
&=& \breve{P}^{-1}\breve{Q}^2\breve{P}^{-3},
\label{eqn:2-4-8-manipulations}
\end{eqnarray}
where in the second line we used that $\breve{R}\breve{Q}\breve{P} = \breve{1}$ implies $\breve{R}=\breve{Q}\breve{P}$, and in the third line that $\breve{Q}\breve{P}^{-1}\breve{Q}\breve{P}^{-1} = \breve{1}$ implies $\breve{P}\breve{Q}^{-1}=\breve{Q}\breve{P}^{-1}$.
From the last line of Eq.~(\ref{eqn:2-4-8-manipulations}) we obtain
$\breve{Q}^2 = \breve{P}^4$. 
By squaring the last equality, we derive that $\breve{P}^8 = \breve{Q}^4$, which equals $\breve{1}$ according to the relators listed in Eq.~(\ref{eqn:PG-2-4-8}).
In addition, by conjugating Eq.~(\ref{eqn:2-4-8-manipulations}) with $\breve{P}^{-3}$ and using $\breve{P}^{-4}=\breve{P}^4$, we obtain the relator
\begin{equation}
\breve{P}^4\breve{Q}^2 = \breve{1}.    
\end{equation}
This allows us to bring the presentation of the proper point group to the more standard form
\begin{equation}
\mathsf{P}_{\!\mathsf{s}}(2,4,8) = \left<\breve{P},\breve{Q},\breve{R}\,\big|\,\breve{R}^2,\breve{Q}^4,\breve{P}^8, \breve{R}\breve{Q}\breve{P},\breve{P}^4\breve{Q}^2\right>.  \label{eqn:PG-2-4-8-modified}
\end{equation}
We recognize from the nontrivial relator that $P^4Q^2$ must be an element of the translation subgroup $\mathsf{T}(2,4,8)\triangleleft\mathsf{D}(2,4,8)$, which we label as $\gamma_1^{-1}$ in Fig.~\ref{fig:8-4-cell}. 
We recognize from the sketch in Fig.~\ref{fig:8-4-cell} that the hyperbolic unit cell is the Bolza cell in which antipodal edges are identified by translations.

By performing the substitution from rotation generators to reflections via Eq.~(\ref{eqn:mirror2rot}), we find that the point group is
\begin{eqnarray}
\mathsf{P}(2,4,8)&=&\left<\breve{a},\breve{b},\breve{c}\,\big|\,\breve{a}^2,\breve{b}^2,\breve{c}^2,(\breve{a}\breve{b})^2,\right. \nonumber \\
&\phantom{=}&\qquad \left. (\breve{b}\breve{c})^4,(\breve{c}\breve{a})^8,(\breve{c}\breve{a})^4(\breve{b}\breve{c})^{2}\right>.\label{eqn:P-2-4-8-full}
\end{eqnarray}
In GAP, we find that the point group as defined above is a non-Abelian finite group of order 32, isomorphic to $\mathbb{Z}_8\rtimes(\mathbb{Z}_2\times\mathbb{Z}_2)$. 
Similarly, the proper point group in Eq.~(\ref{eqn:PG-2-4-8}) is found to be isomorphic to $\mathbb{Z}_8 \rtimes \mathbb{Z}_2$.

\subsubsection{Point group of \texorpdfstring{$\{10,5\}$}{(10,5)} lattice}\label{eqn:PG-10-5}

\begin{figure*}
\includegraphics[width=0.625\linewidth]{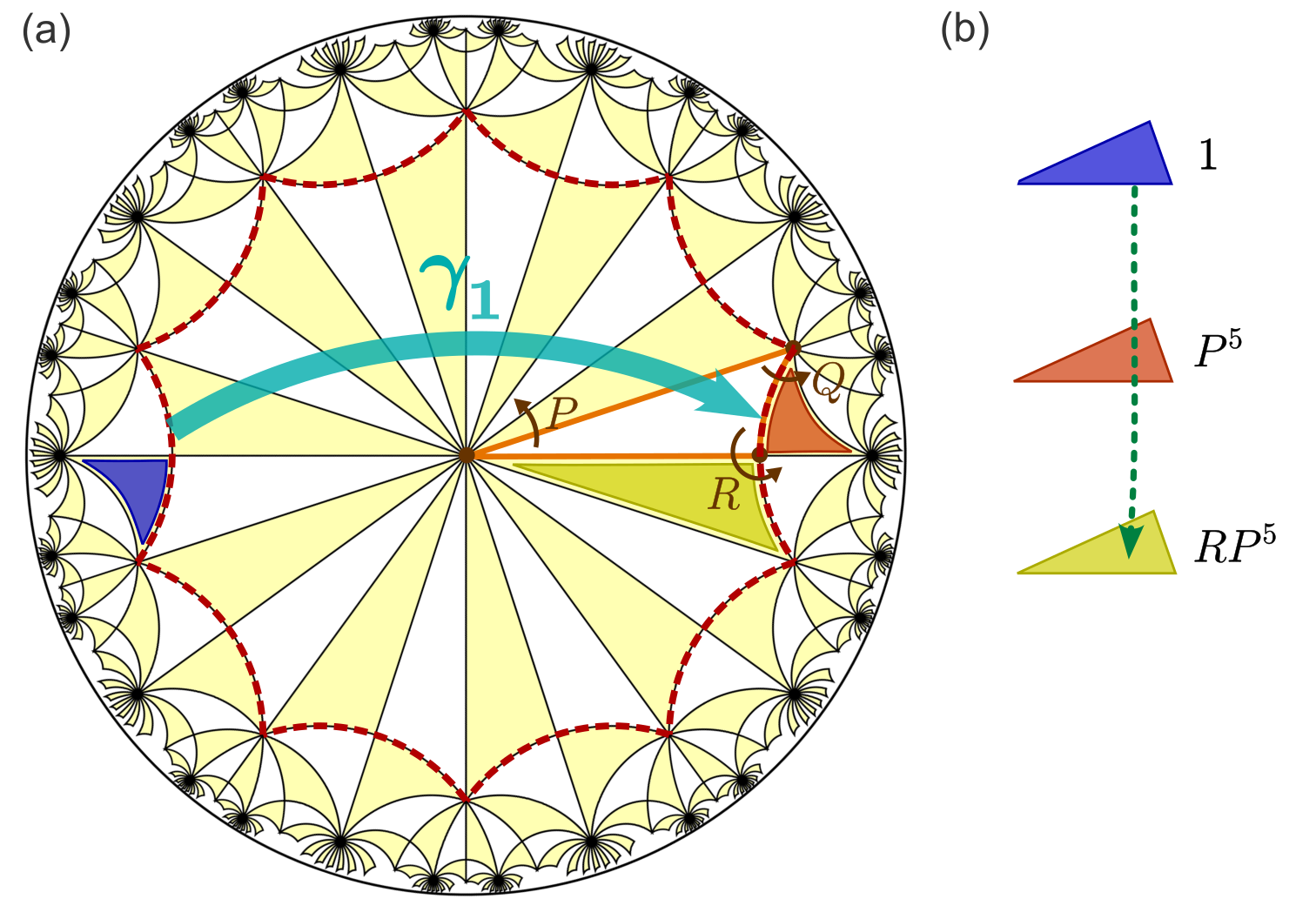}
\caption{Identification of the hyperbolic translation group of the $\{10,5\}$ lattice (see Appendix~\ref{eqn:PG-10-5}).}
\label{fig:10-5-cell}
\end{figure*}

The $\{10,5\}$ lattice has recently been experimentally implemented in the simulation of hyperbolic graphene in Ref.~\onlinecite{Chen2023}, where identification of antipodal edges of the unit cell has been adopted. 
We here confirm that this is indeed the choice imposed by the underlying translation group. 
To that end, we first read from Ref.~\onlinecite{Conder:2007} that
\begin{equation}
\mathsf{P}_{\!\mathsf{s}}(2,5,10) = \left<\breve{P},\breve{Q},\breve{R}\,\big|\,\breve{R}^2,\breve{Q}^5,\breve{R}\breve{Q}\breve{P},\breve{R}\breve{P}\breve{Q}\right>. \label{eqn:PG-2-5-10}
\end{equation}
Note that the pair of relators $\breve{R}\breve{Q}\breve{P} = \breve{1} = \breve{R}\breve{P}\breve{Q}$ imply the commutativity of generators $Q$ and $P$. 
Furthermore, by conjugating the first (second) of these relators with $\breve{R}^{-1}$ (with $\breve{Q}$), we obtain $\breve{Q}\breve{P}\breve{R} = \breve{1} = \breve{Q}\breve{R}\breve{P}$, which immediately implies the commutativity of generators $P$ and $R$.
In the same spirit one obtains $\breve{P}\breve{R}\breve{Q} = \breve{1} = \breve{P}\breve{Q}\breve{R}$, meaning that generators $\breve{R}$ and $\breve{Q}$ also commute. 
Since all three generators pairwise commute, we conclude that the proper point group $\mathsf{P}_{\!\mathsf{s}}(2,5,10)$ is Abelian.

To bring the presentation in Eq.~(\ref{eqn:PG-2-5-10}) into a more standard form, we first consider the fifth power of $\breve{R}\breve{Q}\breve{P}=\breve{1}$. 
Taking into account the commutativity of generators, we obtain
\begin{equation}
\breve{1}=(\breve{R}\breve{Q}\breve{P})^5 = \breve{R}^5\breve{Q}^5\breve{P}^5 = \breve{R}\breve{P}^5 \label{eqn:2-5-10-RP5}
\end{equation}
where in the last step we used that $\breve{R}^2=\breve{1}$ and $\breve{Q}^5=\breve{1}$. 
Further squaring Eq.~(\ref{eqn:2-5-10-RP5}) results in $\breve{1}=\breve{R}^2\breve{P}^{10}$, from which we deduce that $\breve{P}^{10} = \breve{1}$. Thus we obtain the more standard presentation of the proper point group as
\begin{equation}
\mathsf{P}_{\!\mathsf{s}}(2,5,10) = \left<\breve{P},\breve{Q},\breve{R}\,\big|\,\breve{R}^2,\breve{Q}^5,\breve{P}^{10},\breve{R}\breve{Q}\breve{P},\breve{R}\breve{P}^5\right>. \label{eqn:PG-2-5-10-simple}
\end{equation}
By comparing the previous equation to the presentation of the von Dyck group $\mathsf{D}(2,5,10)$ in Eq.~(\ref{eqn:D-2pq}), we recognize that $RP^5$ is an element of the translation subgroup $\mathsf{T}(2,5,10)$, which we label $\gamma_1$ in Fig.~\ref{fig:10-5-cell}. 
We also recognize from the same illustration that the unit cell is a regular decagon, and that translation symmetry relates its antipodal edges. The corresponding $\mathfrak{g}=2$ Riemann surface is a Wiman surface of type~I with algebraic curve description $w^2=z^5-1$~\cite{Kuusalo:1995}.

The full point group is obtained by the replacement in Eq.~(\ref{eqn:mirror2rot}). 
Note that in Eq.~(\ref{eqn:PG-2-5-10}) we can replace $\breve{R}\breve{P}\breve{Q}=(\breve{a}\breve{b})(\breve{c}\breve{a})(\breve{b}\breve{c}) = (\breve{a}\breve{b}\breve{c})^2$, suggesting the presentation
\begin{eqnarray}
\mathsf{P}(2,5,10)&=&\left<\breve{a},\breve{b},\breve{c}\,\big|\,\breve{a}^2,\breve{b}^2,\breve{c}^2,\right. \nonumber \\
&\phantom{=}&\qquad \left.(\breve{a}\breve{b})^2, (\breve{b}\breve{c})^5,(\breve{c}\breve{a})^{10},(\breve{a}\breve{b}\breve{c})^2\right>.\label{eqn:P-2-5-10-full}
\end{eqnarray}
Using GAP, we find that $\mathsf{P}(2,5,10)$ is a non-Abelian group of order $20$ isomorphic to $\mathsf{D}_{10}$.  
On the other hand $\mathsf{P}_{\!\mathsf{s}}(2,5,10)$ is an Abelian group of order $10$ isomorphic to $\mathbb{Z}_{10}$.

\subsubsection{Point group of \texorpdfstring{$\{7,3\}$}{(7,3)} lattice}\label{eqn:PG-7-3}

The point group symmetry of the $\{7,3\}$ unit cell is well known in mathematics~\cite{EightfoldWay}, since it corresponds to the largest amount of conformal symmetry possible for a genus-$3$ surface~\cite{Miranda:1995}.
Namely, the proper [full] point group is isomorphic to $\mathsf{PSL}(2,\mathbb{Z}_7)$ [$\mathsf{PGL}(2,\mathbb{Z}_7)$]. 
Therefore, we only include a basic discussion of the relators and comment on the identification of the unit cell edges.

\begin{figure*}
\includegraphics[width=0.72\linewidth]{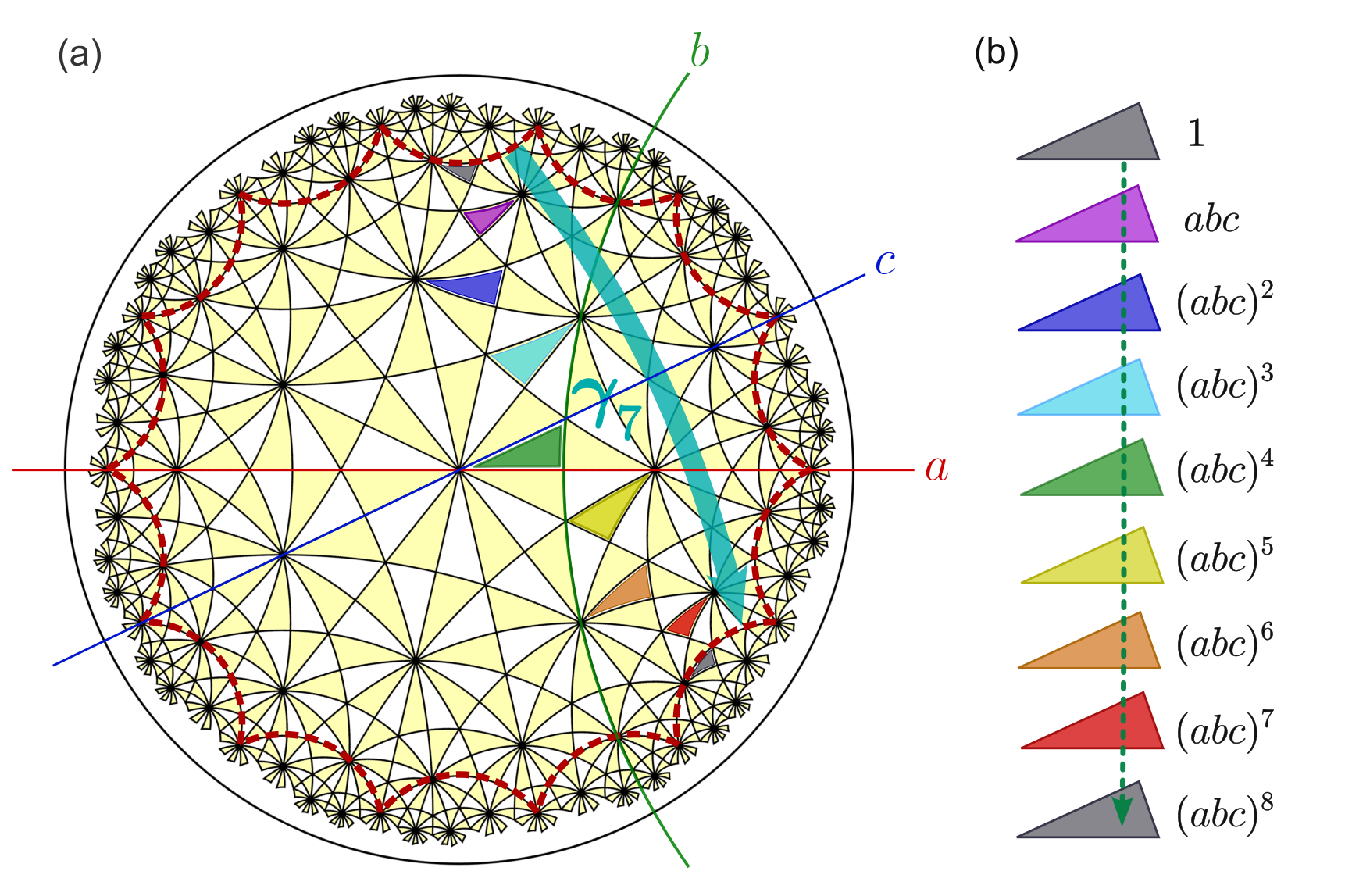}
\caption{Identification of the hyperbolic translation group of the $\{7,3\}$ lattice (see Appendix~\ref{eqn:PG-7-3}).}
\label{fig:7-3-cell}
\end{figure*}

We read from Ref.~\onlinecite{Conder:2007} that
\begin{eqnarray}
\mathsf{P}_{\!\mathsf{s}}(2,3,7) &=& \left<\breve{P},\breve{Q},\breve{R}\,\big|\,\breve{R}^2,\breve{Q}^3,\breve{R}\breve{Q}\breve{P},\breve{P}^{-3}\breve{R}\breve{Q}\breve{P}^{-3},\right. \nonumber \\
&\phantom{=}&\qquad \left.\breve{Q}\breve{R}\breve{P}\breve{Q}\breve{R}\breve{P}\breve{Q}^{-1}\breve{P}\breve{Q}^{-1}\breve{P}^{-1}\breve{R}\breve{P}\right>.  \label{eqn:PG-2-3-7}
\end{eqnarray}
We next perform the usual replacement (\ref{eqn:mirror2rot}) of the rotation generators by the reflection generators. 
Note that $\breve{1}=\breve{R}^2=\breve{a}\breve{b}\breve{a}\breve{b}$ in combination with $\breve{a}^2 = \breve{1} = \breve{b}^2$ implies that $\breve{a}$ and $\breve{b}$ commute.
Furthermore, $\breve{1}=\breve{Q}^3=\breve{b}\breve{c}\breve{b}\breve{c}\breve{b}\breve{c}$ implies that $\breve{c}\breve{b}=\breve{b}\breve{c}\breve{b}\breve{c}$.
With these rules in mind, we find
\begin{eqnarray}
\breve{1}
&=&\breve{Q}\breve{R}\breve{P}\breve{Q}\breve{R}\breve{P}\breve{Q}^{-1}\breve{P}\breve{Q}^{-1}\breve{P}^{-1}\breve{R}\breve{P} \nonumber \\
&=& [\breve{b}\breve{c}][\breve{a}\breve{b}][\breve{c}\breve{a}][\breve{b}\breve{c}][\breve{a}\breve{b}][\breve{c}\breve{a}][\breve{c}\breve{b}][\breve{c}\breve{a}][\breve{c}\breve{b}][\breve{a}\breve{c}][\breve{a}\breve{b}][\breve{c}\breve{a}]\qquad \nonumber \\
&=&\breve{b}\breve{c}(\breve{a}\breve{b}\breve{c})^3 \breve{a}\underline{\breve{c}\breve{b}}\breve{c}\breve{a}\breve{c}\underline{\breve{b}\breve{a}}\breve{c}\breve{a}\breve{b}\breve{c}\breve{a} \nonumber \\
&=&\breve{b}\breve{c}(\breve{a}\breve{b}\breve{c})^3 \breve{a}[\breve{b}\breve{c}\breve{b}\underline{\breve{c}]\breve{c}}\breve{a}\breve{c}[\breve{a}\breve{b}]\breve{c}\breve{a}\breve{b}\breve{c}\breve{a} \nonumber \nonumber \\
&=& \breve{b}\breve{c}(\breve{a}\breve{b}\breve{c})^3 \breve{a}\breve{b}\breve{c}\underline{\breve{b}\breve{a}}\breve{c}\breve{a}\breve{b}\breve{c}\breve{a}\breve{b}\breve{c}\breve{a} \nonumber \\
&=& \breve{b}\breve{c}(\breve{a}\breve{b}\breve{c})^3 \breve{a}\breve{b}\breve{c}[\breve{a}\breve{b}]\breve{c}\breve{a}\breve{b}\breve{c}\breve{a}\breve{b}\breve{c}\breve{a}  = \breve{b}\breve{c}(\breve{a}\breve{b}\breve{c})^7\breve{a}, \label{eqn:breves.breves-everywhere}
\end{eqnarray}
where the underlined terms in any expression indicate those combinations of generators $\breve{a},\breve{b},\breve{c}$ that are substituted by the bracketed terms or cancelled in the next line.
Conjugating Eq.~(\ref{eqn:breves.breves-everywhere}) with $\breve{a}$, we obtain $\breve{1}=(\breve{a}\breve{b}\breve{c})^8$. 
Indeed, the frequently given presentation of $\mathsf{P}(2,3,7)$ is~\cite{CookThesis}
\begin{eqnarray}
\mathsf{P}(2,3,7)&=&\left<\breve{a},\breve{b},\breve{c}\,\big|\,\breve{a}^2,\breve{b}^2,\breve{c}^2,\right. \nonumber \\
&\phantom{=}&\qquad \left.(\breve{a}\breve{b})^2, (\breve{b}\breve{c})^3,(\breve{c}\breve{a})^{7},(\breve{a}\breve{b}\breve{c})^8\right>.\label{eqn:P-2-3-7-full}
\end{eqnarray}
By comparing the previous equation to the presentation of the triangle group $\Delta(2,3,7)$ in Eq.~(\ref{eqn:Delta-def}), we recognize that endpoints of the ``eight-step geodesic'' $(abc)^8$ constitute an element of the translation subgroup $\mathsf{T}(2,3,7)$, which we label $\gamma_7$ in Fig.~\ref{fig:7-3-cell}. 
It is worth to emphasize that $abc\in\triangle(2,3,7)$ acts as a hyperbolic glide reflection, which reflects and translates the Schwarz triangles as illustrated in Fig.~\ref{fig:7-3-cell}. 
By following the underlying glide plane, we observe that the correct identification of the unit cell edges is not antipodal~\cite{Cheng:2022,Bzdusek:2022}. 
The corresponding compactified $\mathfrak{g}=3$ Riemann surface is the famous Klein quartic~\cite{klein1878,EightfoldWay}.

\subsubsection{Point group of \texorpdfstring{$\{12,3\}$}{(12,3)} lattice}\label{eqn:PG-12-3}

We read from Ref.~\onlinecite{Conder:2007} that $\mathsf{P_{\!\mathsf{s}}}(2,3,12)$ is given by Eq.~(\ref{eqn:P-12-3-presentation}).
To bring this presentation to the standard form, we need to use the relators in Eq.~(\ref{eqn:P-12-3-presentation}) to derive $\breve{P}^{12}=\breve{1}$. 
To that end, we first establish two results: first, that $\breve{Q}$ and $\breve{P}^3$ commute, and second, that $\breve{P}\breve{Q}\breve{P}=\breve{Q}^{-1}$.
To show that $\breve{Q}$ and $\breve{P}^3$ commute, we first find from the last relator in (\ref{eqn:P-12-3-presentation}) that $\breve{P}^3=\breve{R}\breve{P}^2\breve{Q}^{-1}$. Using $\breve{R}\breve{Q}\breve{P}=\breve{1} = \breve{R}^2$, we can replace $\breve{R}$ by $\breve{Q}\breve{P}$ in the previous equation, resulting in
\begin{equation}
\breve{P}^3=\breve{Q}\breve{P}^3\breve{Q}^{-1}\quad\Rightarrow\quad \breve{Q}\breve{P}^3=\breve{P}^3\breve{Q},
\end{equation}
which is the sought-after commutativity.
Next, we eliminate $\breve{R}$ in $\breve{R}\breve{Q}\breve{P}=\breve{R}\breve{P}^2\breve{Q}^{-1}\breve{P}^{-3}$, resulting in
\begin{eqnarray}
\breve{1} &=& \breve{Q}\breve{P}^4\breve{Q}\breve{P}^{-2} \label{eqn:2-3-12-early-stage}\\
&=& \breve{Q}\breve{P}\breve{P}^3\breve{Q}\breve{P}^{-2}   \nonumber \\
&=& \breve{Q} \breve{P}\breve{Q}\breve{P}^3\breve{P}^{-2} = \breve{Q} \breve{P}\breve{Q}\breve{P},
\end{eqnarray}
where the last line is equivalent to the sought relation $\breve{P}\breve{Q}\breve{P}=\breve{Q}^{-1}$.
With these two results established, and utilizing $\breve{Q}^3=\breve{1}$, we derive that
\begin{eqnarray}
\breve{P}^{12} &=& (\breve{Q}^{-1}\breve{P}^2\breve{Q}^{-1})^3 \nonumber \\
&=& \breve{Q}^{-1}\breve{P}^2\breve{Q}^{-2}\breve{P}^2\breve{Q}^{-2}\breve{P}^2\breve{Q}^{-1} \nonumber \\
&=& \breve{Q}^{2}\breve{P}^2\breve{Q}\breve{P}^2\breve{Q}\breve{P}^2\breve{Q}^{-1} = \breve{Q}^2\breve{P}(\breve{P}\breve{Q}\breve{P})^2\breve{P}\breve{Q}^{-1} \nonumber \\
&=&\breve{Q}^2\breve{P} \breve{Q}^{-2}\breve{P}\breve{Q}^{-1}
= \breve{Q}^2\breve{P} \breve{Q}\breve{P}\breve{Q}^{-1} \nonumber \\
&=& \breve{Q}^2 \breve{Q}^{-1}\breve{Q}^{-1} = \breve{1}
\end{eqnarray}
where in the first line we used Eq.~(\ref{eqn:2-3-12-early-stage}). 
Thus, we find that Eq.~(\ref{eqn:P-12-3-presentation}) differs from $\mathsf{D}(2,3,12)$ only by the additional relator $\breve{R}\breve{P}^2\breve{Q}^{-1}\breve{P}^{-3}=\breve{1}$. 
We recognize that $RP^2Q^{-1}P^{-3}$ is an element of the translation group $\mathsf{T}(2,3,12)$, as already analyzed in Appendix~\ref{app:12-3-symmetry} and in Fig.~\ref{fig:12-3-cell}. 

The full point group is specified~as 
\begin{eqnarray}
\mathsf{P}(2,3,12)&=&\left<\breve{a},\breve{b},\breve{c}\,\big|\,\breve{a}^2,\breve{b}^2,\breve{c}^2,(\breve{a}\breve{b})^2, (\breve{b}\breve{c})^3,\right. \nonumber \\
&\phantom{=}&\qquad \left.(\breve{c}\breve{a})^{12},(\breve{a}\breve{b})(\breve{c}\breve{a})^2(\breve{b}\breve{c})^{-1}(\breve{c}\breve{a})^{-3}\right>.\quad \label{eqn:P-2-3-12-full}
\end{eqnarray}
In GAP, we find that $\mathsf{P}(2,3,12)$ thus defined is a non-Abelian finite group of order~$96$. 
Interestingly, $\mathsf{P}(2,3,12)$ is isomorphic to $\mathsf{GL}(2,\mathbb{Z}_3)\rtimes\mathbb{Z}_2$, which is also the automorphism group of the Bolza surface. 
However, the compactified unit cell this group acts on here is a genus-$3$ Riemann surface~\cite{schmutz1993}.
In contrast, the proper point group $\mathsf{P}_{\!\mathsf{s}}(2,3,12)$ is isomorphic to\footnote{By $\langle(-\mathbb{1},t^2)\rangle$ we mean the group of order two generated by $(-\mathbb{1},t^2)\in\mathsf{SL}(2,\mathbb{Z}_3)\times\mathbb{Z}_4$, where $t$ is the generator of $\mathbb{Z}_4$.} $\mathsf{SL}(2,\mathbb{Z}_3)\times \mathbb{Z}_4/\langle(-\mathbb{1},t^2)\rangle$~\cite{Wolfart:2005} which differs from the proper automorphism group of the Bolza surface.

\subsubsection{Point group of \texorpdfstring{$\{12,4\}$}{(12,4)} lattice}\label{eqn:PG-12-4}

\begin{figure*}
\includegraphics[width=0.625\linewidth]{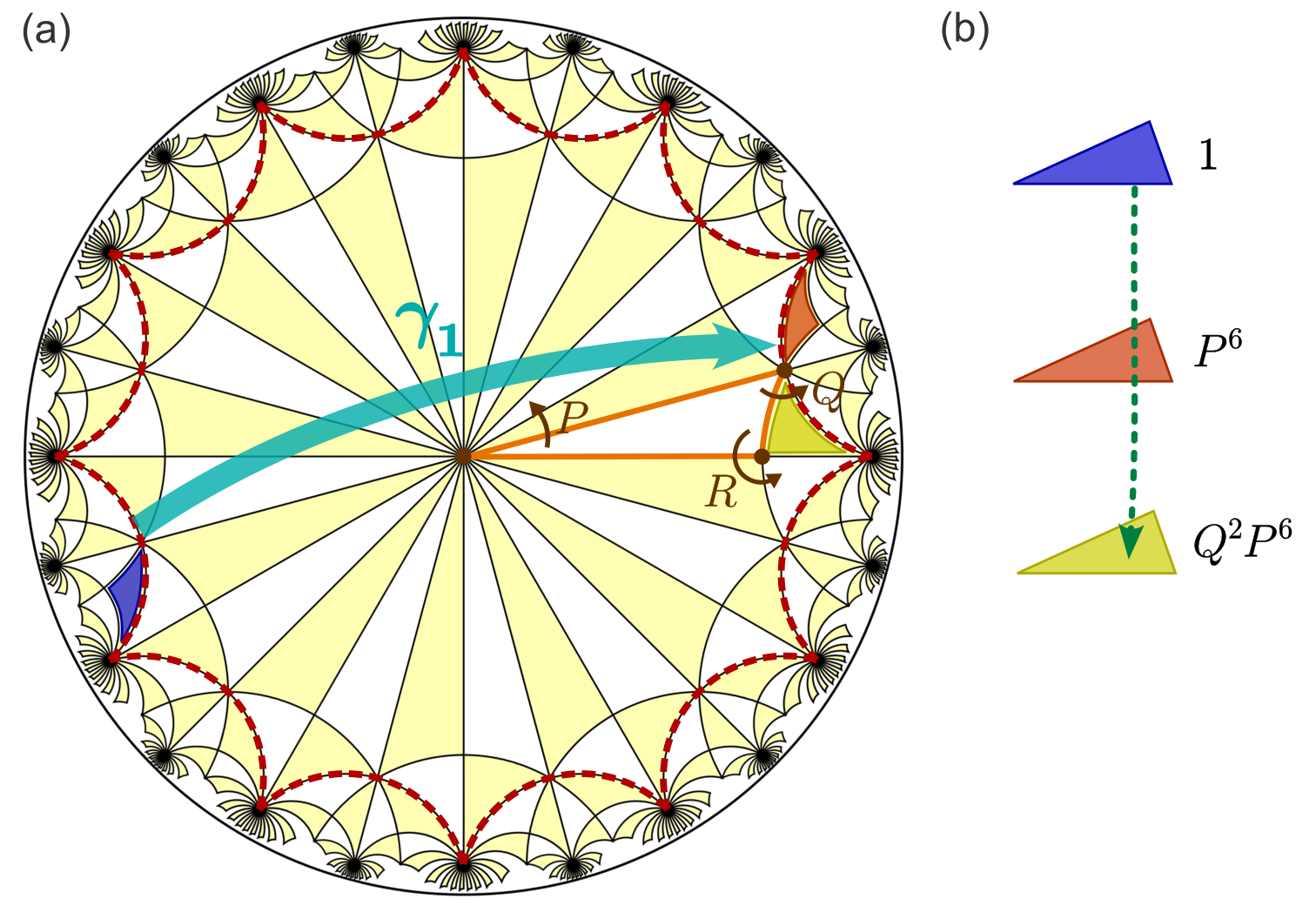}
\caption{Identification of the hyperbolic translation group of the $\{12,4\}$ lattice (see Appendix~\ref{eqn:PG-12-4}).}
\label{fig:12-4-cell}
\end{figure*}

From Ref.~\onlinecite{Conder:2007} we find that
\begin{eqnarray}
\mathsf{P}_{\!\mathsf{s}}(2,4,12) &=& \left<\breve{P},\breve{Q},\breve{R}\,\big|\,\breve{R}^2,\breve{Q}^4,\breve{R}\breve{Q}\breve{P},\right.\nonumber \\
&\phantom{=}& \qquad \left. (\breve{Q}\breve{P}^{-1})^2,\breve{P}^{-2}\breve{R}\breve{Q}^{-1}\breve{P}^{-3}\right>.  \label{eqn:PG-2-4-12}
\end{eqnarray}
To bring this presentation into the standard form, we need to show that $\breve{P}^{12}=\breve{1}$.
To achieve this, first note that $\breve{R}\breve{Q}\breve{P}=\breve{1}$ implies $\breve{R}=\breve{P}^{-1}\breve{Q}^{-1}$. Substituting this into the last relator in Eq.~(\ref{eqn:PG-2-4-12}) leads to
\begin{equation}
\breve{1} = \breve{P}^{-3}\breve{Q}^{-2}\breve{P}^{-3},
\end{equation}
which is equivalent to
\begin{equation}
\breve{P}^6=\breve{Q}^2.\label{eqn:PG-12-4-P6Q2}
\end{equation} 
Squaring the last equation results in to $\breve{P}^{12} = \breve{Q}^4 = \breve{1}$ where the last step corresponds to one of the relators in Eq.~(\ref{eqn:PG-2-4-12}). 

Note that we can take a modified version of Eq.~(\ref{eqn:PG-12-4-P6Q2}), namely $\breve{Q}^2\breve{P}^6=\breve{1}$, as the nontrivial relator that reduces the von Dyck group $\mathsf{D}(2,4,12)$ to the proper point group $\mathsf{P}_{\!\mathsf{s}}(2,4,12)$, i.e.
\begin{equation}
\mathsf{P}_{\!\mathsf{s}}(2,4,12) = \left<\breve{P},\breve{Q},\breve{R}\,\big|\,\breve{R}^2,\breve{Q}^4,\breve{P}^{12},\breve{R}\breve{Q}\breve{P},\breve{Q}^2\breve{P}^6\right>. \label{eqn:PG-2-4-12-proper}
\end{equation}
We therefore recognize that $Q^2 P^6 \in\mathsf{D}(2,4,12)$ is an element of the translation subgroup $\mathsf{T}(2,4,12)$, which we label $\gamma_1$ in Fig.~\ref{fig:12-4-cell}.
Using GAP we find that the proper point group as defined in Eq.~(\ref{eqn:PG-2-4-12-proper}) is isomorphic to $\mathbb{Z}_4\times \mathsf{S}_3$, where $\mathsf{S}_3$ is the permutation group of three elements.
The full point group corresponds to 
\begin{eqnarray}
\mathsf{P}(2,4,12) &=&\left<\breve{a},\breve{b},\breve{c}\,\big|\,\breve{a}^2,\breve{b}^2,\breve{c}^2,\right. \nonumber \\
&\phantom{=}&\qquad \left.(\breve{a}\breve{b})^2, (\breve{b}\breve{c})^4,(\breve{c}\breve{a})^{12},(\breve{b}\breve{c})^2(\breve{c}\breve{a})^6\right>,\;\;\label{eqn:PG-2-4-12-full}
\end{eqnarray}
which with GAP we find to be isomorphic to $\mathsf{D}_4\times \mathsf{S}_3$.

\section{Action of symmetries on hyperbolic momentum}\label{app:derive-M-matrices}

In this Appendix, we study the action of symmetry elements $g$ in $\Delta$ (or in any of the magnetic hyperbolic space groups derived from it) on the hyperbolic momentum $\bs{k}$. 
Our discussion formalizes and generalizes the discussion of $\{8,3\}$ lattices in the Supplementary Materials of Ref.~\onlinecite{Maciejko:2021}.
Specifically, we show that the action of $g$ amounts to linear transformations $g: \bs{k} \mapsto M_g \bs{k}$, and call $M_g \in \mathsf{GL}(2\mathfrak{g},\mathbb{Z})$ the point-group matrix of $g$. 
Our discussion is divided into two parts.
First, in Appendix~\ref{app:derive-M-matrices-subsec} we present the general considerations that reveal the general form of the transformation, and indicate how $M_g$ is computationally extracted. 
Then, in Appendix~\ref{app:list-M-matrices} we list the extracted matrices $M_g$ for the symmetry generators $a,b,c,P,Q,R$ of all seven hyperbolic $\{p,q\}$ lattices listed in Fig.~\ref{fig:unit-cells}.

\subsection{General considerations}\label{app:derive-M-matrices-subsec}

For simplicity we drop here the dependence of the groups $\Delta$, $\mathsf{T}$ and $\mathsf{P}$ on the choice of $\{p,q\}$ lattice.
To begin, we define 
\begin{equation}
\mathcal{S}_g\psi(z)=\psi(g^{-1}(z))\label{eqn:action-on-states}
\end{equation}
for any symmetry $g\in\Delta$.
The definition~(\ref{eqn:action-on-states}) ensures that $\mathcal{S}$ correctly behaves as a representation rather than an anti-representation, i.e., $\mathcal{S}_g\mathcal{S}_{g'}=\mathcal{S}_{gg'}$.
By definition of hyperbolic momenta, an Abelian Bloch state transforms as $\mathcal{S}_{\gamma}\psi(z)=\chi(\gamma)\psi(z)$ for translations $\gamma\in \mathsf{T}\triangleleft\Delta$,  where $\chi: \mathsf{T} \to \mathsf{U}(1)$ depends on the choice of momentum in the hypercubic Brillouin zone. 

We next consider a general symmetry $g\in\Delta$, so that we have $[\mathcal{S}_{{g}},\mcH]=0$.
Thus if $\psi(z)$ is an eigenstate of $\mcH$ with energy $E$ and certain momentum $\bs{k}$, then $\psi^{{g}}(z):= \mathcal{S}_{{g}}\psi(z)$ is also an eigenstate of $\mcH$ with the same energy.
Furthermore, $\psi^{{g}}(z)$ also obeys the Abelian Bloch theorem, but with a different crystal momentum $\bs{k}^g$ linearly related to the momentum $\bs{k}$ of the original state. 
To see this, consider any generator $\gamma_i$ of $\mathsf{T}$, and observe that
\begin{equation}\label{PGpsi}
\begin{aligned}
    \mathcal{S}_{\gamma_i}\psi^g(z)
    &=\mathcal{S}_g\mathcal{S}_g^{-1}\mathcal{S}_{\gamma_i}\mathcal{S}_g\psi(z) \\
    &=\mathcal{S}_g\mathcal{S}_{g^{-1}\gamma_i g}\psi(z)  \\
    &= \mathcal{S}_g \chi(g^{-1} \gamma_i g) \psi(z) \\
    &=\chi(g^{-1}\gamma_i g)\psi^g(z), 
\end{aligned}
\end{equation}
where in the third line we used that $g^{-1}\gamma_i g \in\mathsf{T}$ due to $\mathsf{T}$ being a \emph{normal} subgroup of $\Delta$, and in the last line we used that $\chi(g^{-1}\gamma_i g)$ is not operator-valued and can thus be commuted with $\mathcal{S}_g$. Furthermore, since $g^{-1}\gamma_i g\in \mathsf{T}$, this element can be expressed as a word in the generators $\gamma_j$.
Denote by $(M_g)^{i}_{\phantom{i}j}\in\mathbb{Z}$ the number of times $\gamma_j$ appears in this word, with negative entries corresponding to powers of the inverse; we then have 
\begin{equation}
\chi(g^{-1}\gamma_i g)=\e^{\imi (M_g)^{i}_{\phantom{i}j} k^j}   
\end{equation}
using $\chi(\gamma_j)=\e^{\imi k^j}$. 
From Eq.~(\ref{PGpsi}), if $\psi(z)$ is an Abelian Bloch state with momentum $\bs{k}$, then $\psi^g(z)$ is an Abelian Bloch state with momentum $\tilde{\bs{k}}$ given by
\begin{align}
    \tilde{\bs{k}}=M_g\bs{k},
\end{align}
where $M_g$ is a $2\mathfrak{g}$-dimensional representation of $\Delta$ over the field of integers, i.e., $M:\Delta \to \mathsf{GL}(2\mathfrak{g},\mathbb{Z})$. 
Note that $\det M_g = \pm 1$, which follows from the fact that both matrices $M_g$ and $M_{g^{-1}}$ have integer entries and therefore an integer determinant, combined with the fact that $(\det M_g)(\det M_{g^{-1}}) = 1$.

Let us emphasize that $M_{g_1} = M_{g_2}$ when $g_1$ and $g_2$ belong to the same coset of $\mathsf{T}$ in $\Delta$. 
Indeed, assuming that $g_2 = \gamma g_1$ with some $\gamma\in\mathsf{T}$, we find that the number $(M_{g_1})^i_{\phantom{i}j}$ of times that $\gamma_j$ appears in $g_1^{-1} \gamma_i g_1$ is clearly equal to the number $(M_{g_2})^i_{\phantom{i}j}$ of times that $\gamma_j$ appears in $g_2^{-1} \gamma_i g_2 = g_1^{-1} \gamma^{-1} \gamma_i \gamma g_2$, as the contributions from $\gamma$ and $\gamma^{-1}$ cancel.
Therefore, we actually deal with a map $M:\mathsf{P}\to \mathsf{GL}(2\mathfrak{g},\mathbb{Z})$.
For this reason, we call $M_g$ the \emph{point-group matrix of $g$}.
We automatize the extraction of matrices $M_g$ as defined above by writing a computer program in the computational group theory language GAP~\cite{GAP4,Chen:2023:SDC}. 

Finally, for anti-unitary elements, which take the form $g\mcT$ where $\mcT$ is time reversal, we modify Eq.~(\ref{eqn:action-on-states}) to
\begin{equation}
\mathcal{S}_{g\mcT}\psi(z)=\psi^*(g^{-1}(z))=: \psi^{g\mcT}(z)\label{eqn:action-on-states-AU}
\end{equation}
where the star indicates complex conjugation. 
Then Eq.~(\ref{PGpsi}) is modified to 
\begin{equation}\label{PGpsi-AU}
\begin{aligned}
    \mathcal{S}_{\gamma_i}\psi^{g\mcT}(z)
    &=\mathcal{S}_{g\mcT}\mathcal{S}_{(g\mcT^{-1})}\mathcal{S}_{\gamma_i}\mathcal{S}_{g\mcT}\psi(z) \\
    &=\mathcal{S}_{g\mcT}\mathcal{S}_{(g\mcT)^{-1}\gamma_i (g\mcT)}\psi(z) = \mathcal{S}_{g\mcT}\mathcal{S}_{g^{-1}\gamma_i g}\psi(z) \\
    &= \mathcal{S}_{g\mcT} \,\chi(g^{-1}\gamma_i g ) \psi(z) =\chi^*(g^{-1}\gamma_i g )\psi^{g\mcT}(z), 
\end{aligned}
\end{equation}
where in the third equality we used that time reversal commutes with all spatial transformations $g$, and complex conjugation in the last expression arises from commuting $\chi(g^{-1} \mcT g)$ with the antiunitary operator $\mathcal{S}_{g\mcT}$. 
Complex conjugation flips the sign of $\e^{\imi M_g\bs{k}}$, thus we find
\begin{equation}
M_{g \mcT} = -M_g,
\end{equation}
i.e., time reversal reverses all components of the hyperbolic momentum.

\subsection{Hyperbolic point-group matrices}\label{app:list-M-matrices}

Here we list the point-group matrices for the symmetry generators $a,b,c,P,Q,R$, computed for all seven $\{p,q\}$ lattices considered.
We verify that for each lattice, the point-group matrices correctly obey the group relations in Appendix~\ref{sec:hyper-PGs}.

\begin{widetext}

For the $\{8,3\}$ lattice:
\begin{align}
    M_a^{\{8,3\}}=\left(
    \begin{array}{cccc}
    1 & 0 & 0 & 0 \\
    0 & 0 & 0 & -1 \\
    0 & 0 & -1 & 0 \\
    0 & -1 & 0 & 0
    \end{array}
    \right),\hspace{5mm}
    M_b^{\{8,3\}}=\left(
    \begin{array}{cccc}
    -1 & 0 & 0 & 0 \\
    -1 & 1 & 0 & 0 \\
    0 & 1 & -1 & 1 \\
    1 & 0 & 0 & 1
    \end{array}
    \right),\hspace{5mm}
    M_c^{\{8,3\}}=\left(
    \begin{array}{cccc}
    0 & 1 & 0 & 0 \\
    1 & 0 & 0 & 0 \\
    0 & 0 & 0 & -1 \\
    0 & 0 & -1 & 0
    \end{array}
    \right),
\end{align}
\begin{align}
    M_P^{\{8,3\}}=\left(
    \begin{array}{cccc}
    0 & 0 & 0 & -1 \\
    1 & 0 & 0 & 0 \\
    0 & 1 & 0 & 0 \\
    0 & 0 & 1 & 0 \\
    \end{array}
    \right),\hspace{5mm}
    M_Q^{\{8,3\}}=\left(
    \begin{array}{cccc}
    0 & -1 & 0 & 0 \\
    1 & -1 & 0 & 0 \\
    1 & 0 & -1 & 1 \\
    0 & 1 & -1 & 0 \\
    \end{array}
    \right),\hspace{5mm}
    M_R^{\{8,3\}}=\left(
    \begin{array}{cccc}
    -1 & 0 & 0 & 0 \\
    -1 & 0 & 0 & -1 \\
    0 & -1 & 1 & -1 \\
    1 & -1 & 0 & 0 \\
    \end{array}
    \right).
\end{align} 

For the $\{6,4\}$ lattice:
\begin{align}
    M_a^{\{6,4\}}=\left(
    \begin{array}{cccc}
    1 & 0 & 1 & 0  \\
    0 & 0 & 0 & -1 \\
    0 & 0 & -1 & 0 \\
    0 & -1 & 0 & 0
    \end{array}
    \right),\hspace{5mm}
    M_b^{\{6,4\}}=\left(
    \begin{array}{cccc}
    -1 & 0 & 0 & 0  \\
    0 & 1 & 0 & 0 \\
    0 & 0 & -1 & 0 \\
    0 & 0 & 0 & 1
    \end{array}
    \right),\hspace{5mm}
    M_c^{\{6,4\}}=\left(
    \begin{array}{cccc}
    0 & 1 & 0 & 1  \\
    1 & 0 & 1 & 0 \\
    0 & 0 & 0 & -1 \\
    0 & 0 & -1 & 0
    \end{array}
    \right),
\end{align} 
\begin{align}
    M_P^{\{6,4\}}=\left(
    \begin{array}{cccc}
    0 & -1 & 0 & -1  \\
    1 & 0 & 0 & 0 \\
    0 & 1 & 0 & 0 \\
    0 & 0 & 1 & 0
    \end{array}
    \right),\hspace{5mm}
    M_Q^{\{6,4\}}=\left(
    \begin{array}{cccc}
    0 & -1 & 0 & -1  \\
    1 & 0 & 1 & 0 \\
    0 & 0 & 0 & 1 \\
    0 & 0 & -1 & 0
    \end{array}
    \right),\hspace{5mm}
    M_R^{\{6,4\}}=\left(
    \begin{array}{cccc}
    -1 & 0 & -1 & 0  \\
    0 & 0 & 0 & -1 \\
    0 & 0 & 1 & 0 \\
    0 & -1 & 0 & 0
    \end{array}
    \right).
\end{align} 

For the $\{8,4\}$ lattice:
\begin{align}
    M_a^{\{8,4\}}=\left(
    \begin{array}{cccc}
    0 & 0 & 0 & -1  \\
    0 & 0 & -1 & 0 \\
    0 & -1 & 0 & 0 \\
    -1 & 0 & 0 & 0
    \end{array}
    \right),\hspace{5mm}
    M_b^{\{8,4\}}=\left(
    \begin{array}{cccc}
    0 & 1 & -1 & 1  \\
    1 & 0 & -1 & 1 \\
    1 & -1 & 0 & 1 \\
    1 & -1 & 1 & 0
    \end{array}
    \right),\hspace{5mm}
    M_c^{\{8,4\}}=\left(
    \begin{array}{cccc}
    1 & 0 & 0 & 0  \\
    0 & 0 & 0 & -1 \\
    0 & 0 & -1 & 0 \\
    0 & -1 & 0 & 0
    \end{array}
    \right),
\end{align} 
\begin{align}
    M_P^{\{8,4\}}=\left(
    \begin{array}{cccc}
    0 & 0 & 0 & -1 \\
    1 & 0 & 0 & 0 \\
    0 & 1 & 0 & 0 \\
    0 & 0 & 1 & 0 \\
    \end{array}
    \right),\hspace{5mm}
    M_Q^{\{8,4\}}=\left(
    \begin{array}{cccc}
    0 & -1 & 1 & -1 \\
    1 & -1 & 1 & 0 \\
    1 & -1 & 0 & 1 \\
    1 & 0 & -1 & 1 \\
    \end{array}
    \right),\hspace{5mm}
    M_R^{\{8,4\}}=\left(
    \begin{array}{cccc}
    -1 & 1 & -1 & 0 \\
    -1 & 1 & 0 & -1 \\
    -1 & 0 & 1 & -1 \\
    0 & -1 & 1 & -1 \\
    \end{array}
    \right).
\end{align}

For the $\{10,5\}$ lattice:
\begin{align}
    M_a^{\{10,5\}}=\left(
    \begin{array}{cccc}
    1 & 0 & 0 & 0  \\
    1 & -1 & 1 & -1 \\
    0 & 0 & 0 & -1 \\
    0 & 0 & -1 & 0
    \end{array}
    \right),\hspace{5mm}
    M_b^{\{10,5\}}=\left(
    \begin{array}{cccc}
    -1 & 0 & 0 & 0  \\
    -1 & 1 & -1 & 1 \\
    0 & 0 & 0 & 1 \\
    0 & 0 & 1 & 0
    \end{array}
    \right),\hspace{5mm}
    M_c^{\{10,5\}}=\left(
    \begin{array}{cccc}
    0 & 1 & 0 & 0  \\
    1 & 0 & 0 & 0 \\
    1 & -1 & 1 & -1 \\
    0 & 0 & 0 & -1
    \end{array}
    \right),
\end{align} 
\begin{align}
    M_P^{\{10,5\}}=\left(
\begin{array}{cccc}
 1 & -1 & 1 & -1 \\
 1 & 0 & 0 & 0 \\
 0 & 1 & 0 & 0 \\
 0 & 0 & 1 & 0 \\
\end{array}
\right),\hspace{5mm}
    M_Q^{\{10,5\}}=\left(
\begin{array}{cccc}
 0 & -1 & 0 & 0 \\
 0 & 0 & -1 & 0 \\
 0 & 0 & 0 & -1 \\
 1 & -1 & 1 & -1 \\
\end{array}
\right),\hspace{5mm}
    M_R^{\{10,5\}}=\left(
\begin{array}{cccc}
 -1 & 0 & 0 & 0 \\
 0 & -1 & 0 & 0 \\
 0 & 0 & -1 & 0 \\
 0 & 0 & 0 & -1 \\
\end{array}
\right).
\end{align}

For the $\{7,3\}$ lattice:
\begin{gather}
M_a^{\{7,3\}}=\left(\begin{array}{cccccc}
 0 & 0 & 0 & 0 & -1 & 0 \\
 0 & 0 & 0 & -1 & 0 & 0 \\
 0 & 0 & -1 & 0 & 0 & 0 \\
 0 & -1 & 0 & 0 & 0 & 0 \\
 -1 & 0 & 0 & 0 & 0 & 0 \\
 1 & 1 & 1 & 1 & 1 & 1
 \end{array}\right),
 \hspace{3mm}
 M_b^{\{7,3\}}=\left(\begin{array}{cccccc}
 -1 & 0 & 0 & 0 & 0 & 0 \\
 0 & 1 & 0 & 0 & 1 & 0 \\
 1 & 0 & 1 & 0 & 1 & 0 \\
 1 & 0 & 0 & 1 & 0 & 0 \\
 0 & 0 & 0 & 0 & -1 & 0 \\
 -1 & -1 & -1 & -1 & -1 & -1
\end{array}\right),
 \hspace{3mm}
M_c^{\{7,3\}}=\left(\begin{array}{cccccc}
 0 & 0 & 0 & 0 & 0 & -1 \\
 0 & 0 & 0 & 0 & -1 & 0 \\
 0 & 0 & 0 & -1 & 0 & 0 \\
 0 & 0 & -1 & 0 & 0 & 0 \\
 0 & -1 & 0 & 0 & 0 & 0 \\
 -1 & 0 & 0 & 0 & 0 & 0
\end{array}\right),
\end{gather}
\begin{gather}
M_P^{\{7,3\}}=\left(\begin{array}{cccccc}
 -1 & -1 & -1 & -1 & -1 & -1 \\
 1 & 0 & 0 & 0 & 0 & 0 \\
 0 & 1 & 0 & 0 & 0 & 0 \\
 0 & 0 & 1 & 0 & 0 & 0 \\
 0 & 0 & 0 & 1 & 0 & 0 \\
 0 & 0 & 0 & 0 & 1 & 0 \\
 \end{array}\right),
 \hspace{3mm}
 M_Q^{\{7,3\}}=\left(\begin{array}{cccccc}
 0 & 0 & 0 & 0 & 0 & 1 \\
 0 & -1 & 0 & 0 & -1 & 0 \\
 0 & -1 & 0 & -1 & 0 & -1 \\
 0 & 0 & -1 & 0 & 0 & -1 \\
 0 & 1 & 0 & 0 & 0 & 0 \\
 1 & 1 & 1 & 1 & 1 & 1 \\
\end{array}\right),
 \hspace{3mm}
M_R^{\{7,3\}}=\left(\begin{array}{cccccc}
 0 & 0 & 0 & 0 & 1 & 0 \\
 -1 & 0 & 0 & -1 & 0 & 0 \\
 -1 & 0 & -1 & 0 & -1 & 0 \\
 0 & -1 & 0 & 0 & -1 & 0 \\
 1 & 0 & 0 & 0 & 0 & 0 \\
 0 & 0 & 0 & 0 & 0 & -1 \\
\end{array}\right).
\end{gather}

For the $\{12,3\}$ lattice:
\begin{gather}
M_a^{\{12,3\}}=\left(
\begin{array}{cccccc}
 0 & 0 & 0 & -1 & 0 & 0 \\
 0 & 0 & -1 & 0 & 0 & 0 \\
 0 & -1 & 0 & 0 & 0 & 0 \\
 -1 & 0 & 0 & 0 & 0 & 0 \\
 -1 & 0 & 1 & 0 & -1 & 1 \\
 -1 & 1 & 1 & -1 & 0 & 1 \\
\end{array}
\right),
 \hspace{5mm}
 M_b^{\{12,3\}}=\left(
\begin{array}{cccccc}
 0 & 0 & 0 & 1 & 0 & 0 \\
 0 & 0 & 0 & 1 & -1 & 0 \\
 0 & 0 & 1 & 0 & -1 & 1 \\
 1 & 0 & 0 & 0 & 0 & 0 \\
 1 & -1 & 0 & 0 & 0 & 0 \\
 1 & -1 & 0 & 0 & 1 & -1 \\
\end{array}
\right),
 \hspace{5mm}
M_c^{\{12,3\}}=\left(
\begin{array}{cccccc}
 0 & 0 & 0 & 0 & -1 & 0 \\
 0 & 0 & 0 & -1 & 0 & 0 \\
 0 & 0 & -1 & 0 & 0 & 0 \\
 0 & -1 & 0 & 0 & 0 & 0 \\
 -1 & 0 & 0 & 0 & 0 & 0 \\
 -1 & 0 & 1 & 0 & -1 & 1 \\
\end{array}
\right),
\end{gather}
\begin{gather}
M_P^{\{12,3\}}=\left(
\begin{array}{cccccc}
 1 & 0 & -1 & 0 & 1 & -1 \\
 1 & 0 & 0 & 0 & 0 & 0 \\
 0 & 1 & 0 & 0 & 0 & 0 \\
 0 & 0 & 1 & 0 & 0 & 0 \\
 0 & 0 & 0 & 1 & 0 & 0 \\
 0 & 0 & 0 & 0 & 1 & 0 \\
\end{array}
\right),
 \hspace{5mm}
 M_Q^{\{12,3\}}=\left(
\begin{array}{cccccc}
 0 & -1 & 0 & 0 & 0 & 0 \\
 1 & -1 & 0 & 0 & 0 & 0 \\
 0 & 0 & 0 & 0 & -1 & 1 \\
 0 & 0 & 0 & 0 & -1 & 0 \\
 0 & 0 & 0 & 1 & -1 & 0 \\
 0 & 0 & -1 & 1 & 0 & -1 \\
\end{array}
\right),
 \hspace{5mm}
M_R^{\{12,3\}}=\left(
\begin{array}{cccccc}
 -1 & 0 & 0 & 0 & 0 & 0 \\
 0 & 0 & -1 & 0 & 1 & -1 \\
 0 & 0 & 0 & -1 & 1 & 0 \\
 0 & 0 & 0 & -1 & 0 & 0 \\
 0 & 0 & 1 & -1 & 0 & 0 \\
 0 & -1 & 1 & 0 & -1 & 0 \\
\end{array}
\right).
\end{gather}

For the $\{12,4\}$ lattice: 
\begin{gather}
M_a^{\{12,4\}}=\left(\begin{array}{cccccc}
 0 & 0 & 0 & 0 & 0 & -1 \\
 0 & 0 & 0 & 0 & -1 & 0 \\
 0 & 0 & 0 & -1 & 0 & 0 \\
 0 & 0 & -1 & 0 & 0 & 0 \\
 0 & -1 & 0 & 0 & 0 & 0 \\
 -1 & 0 & 0 & 0 & 0 & 0
 \end{array}\right),
 \hspace{3mm}
 M_b^{\{12,4\}}=\left(\begin{array}{cccccc}
 0 & 1 & -1 & 1 & -1 & 1 \\
 1 & 0 & -1 & 1 & -1 & 1 \\
 1 & -1 & 0 & 1 & -1 & 1 \\
 1 & -1 & 1 & 0 & -1 & 1 \\
 1 & -1 & 1 & -1 & 0 & 1 \\
 1 & -1 & 1 & -1 & 1 & 0
\end{array}\right),
 \hspace{3mm}
M_c^{\{12,4\}}=\left(\begin{array}{cccccc}
 1 & 0 & 0 & 0 & 0 & 0 \\
 0 & 0 & 0 & 0 & 0 & -1 \\
 0 & 0 & 0 & 0 & -1 & 0 \\
 0 & 0 & 0 & -1 & 0 & 0 \\
 0 & 0 & -1 & 0 & 0 & 0 \\
 0 & -1 & 0 & 0 & 0 & 0
\end{array}\right),
\end{gather}
\begin{gather}
M_P^{\{12,4\}}=\left(
\begin{array}{cccccc}
 0 & 0 & 0 & 0 & 0 & -1 \\
 1 & 0 & 0 & 0 & 0 & 0 \\
 0 & 1 & 0 & 0 & 0 & 0 \\
 0 & 0 & 1 & 0 & 0 & 0 \\
 0 & 0 & 0 & 1 & 0 & 0 \\
 0 & 0 & 0 & 0 & 1 & 0 \\
\end{array}
\right),
 \hspace{5mm}
 M_Q^{\{12,4\}}=\left(
\begin{array}{cccccc}
 0 & -1 & 1 & -1 & 1 & -1 \\
 1 & -1 & 1 & -1 & 1 & 0 \\
 1 & -1 & 1 & -1 & 0 & 1 \\
 1 & -1 & 1 & 0 & -1 & 1 \\
 1 & -1 & 0 & 1 & -1 & 1 \\
 1 & 0 & -1 & 1 & -1 & 1 \\
\end{array}
\right),
 \hspace{5mm}
M_R^{\{12,4\}}=\left(
\begin{array}{cccccc}
 -1 & 1 & -1 & 1 & -1 & 0 \\
 -1 & 1 & -1 & 1 & 0 & -1 \\
 -1 & 1 & -1 & 0 & 1 & -1 \\
 -1 & 1 & 0 & -1 & 1 & -1 \\
 -1 & 0 & 1 & -1 & 1 & -1 \\
 0 & -1 & 1 & -1 & 1 & -1 \\
\end{array}
\right).
\end{gather}
\end{widetext}

\section{Symmetry relations between Chern numbers}\label{app:Chern-constraints}

It is known that crystal symmetries constrain topological invariants. While previous work has elucidated point-group symmetry constraints on Chern numbers in Euclidean lattices~\cite{Fang:2012}, we here investigate symmetry relations between Chern numbers in the higher-dimensional BZ of hyperbolic lattices. 
Since the objects involved acquire multiple components (having two band indices and multiple coordinate indices), a suitable mathematical language is necessary to keep the derivations compact. The appropriate mathematical formalism here is that of matrix-valued differential forms.

The following discussion is divided into four parts. 
First, in Appendix~\ref{app:Chern-matrixforms} we provide a short overview of matrix-valued differential forms and their elementary manipulations~\cite{Fecko:2006,Baez:1994}. 
Then, in Appendix~\ref{app:Chern-defs} we introduce the Berry-Wilczek-Zee (BWZ) connection and curvature in the language of matrix-valued differential forms. We also verify there the well-known gauge covariance of the BWZ curvature and the gauge invariance of Chern characters and Chern numbers, with the goal of demonstrating the efficiency of the adopted mathematical language.
Finally, in Appendices~\ref{app:Chern-first} and~\ref{app:Chern-second} we deploy this mathematical formalism to identify symmetry constraints on first resp.~second Chern numbers. 
In particular, we derive there Eqs.~(\ref{eqn:C1-constraint}), (\ref{eqn:C2-constraint}), and (\ref{eqn:C2-constraint-6D}), which in combination with the point-group matrices listed in Appendix~\ref{app:derive-M-matrices} result in the Chern numbers and Chern matrices presented in Sec.~\ref{eqn:Chern-theory} of the main text.

\subsection{Matrix-valued differential forms}\label{app:Chern-matrixforms}

Our discussion adapts the language of \emph{matrix-valued differential forms}~\cite{Fecko:2006,Baez:1994}, i.e., 2D arrays (matrices) $X$ whose components $X^{ab} = \sum_{i\ldots \ell}  X^{ab}_{i\cdots \ell} \,dk^i \wedge \cdots \wedge dk^\ell$ are differential forms. 
Here, the coordinates ($k^i$) are momenta in the BZ torus, and the matrix rows and columns (indices $a,b,\ldots$) run over directions in the Hilbert space of Bloch Hamiltonians. 
The coefficients appearing in the expansion of differential forms (and of matrix-valued differential forms) into coordinate differentials are skew-symmetric under exchange of any two coordinate indices (labelled by Latin subscripts $i,j,k,\ell$), e.g., $X^{ab}_{ijk\ell\cdots} = - X^{ab}_{jik\ell\cdots}$.\footnote{Our convention is that band indices $a,b,\ldots$ always appear as superscripts, while coordinate indices $i,j,k,\ell,\ldots$ appear as subscripts (superscripts) if they are covariant (contravariant). Indices of differential forms are all covariant, while some contravariant coordinate indices arise from coordinate transformations---see, e.g.,~Eq.~(\ref{eqn:k-to-k-tilde}).}

There are two key operations on matrix-valued differential forms that we utilize and which we briefly review below: the wedge product ($\wedge$) and the exterior derivative ($d$).

The \emph{wedge product} (also called \emph{exterior product}) combines a differential form $\alpha$ of degree $\mathfrak{p}$ and a differential form $\beta$ of degree $\mathfrak{q}$ into a differential form $\alpha\wedge \beta$ of degree $\mathfrak{p}+\mathfrak{q}$. (The \emph{degree} counts the number of coordinate differentials appearing in the corresponding differential form.)
The wedge product obeys
\begin{equation}
    \alpha \wedge \beta = (-1)^{\mathfrak{p}\mathfrak{q}}\beta\wedge\alpha
\end{equation}
which is known as \emph{graded commutativity}.

Matrix-valued differential forms $X$ and $Y$ whose matrix dimensions are compatible with the standard matrix multiplication can also be multiplied with the wedge. In matrix components,
\begin{equation}
(X\wedge Y)^{ab} = \sum_{c} X^{ac} \wedge Y^{cb}.\label{eqn:wedge-matrices-def}
\end{equation}
We still call this the wedge (or exterior) product, even though it clearly combines \emph{both} the exterior product of differential forms and the scalar product of rows and columns of the matrices.
If any of the matrix-valued differential forms in a product is of degree zero, i.e., it is an ordinary matrix, then the corresponding product reduces to the usual matrix multiplication. Therefore, we sometimes drop the wedge ($\wedge$) symbol appearing before or after such a matrix.

The wedge product of matrix-valued differential forms does not obey graded commutativity, owing to the non-commutativity of matrix multiplication. 
However, the graded commutativity is restored for the \emph{trace} over the matrix components. Namely, the \emph{graded cyclic property} states that
\begin{eqnarray}
\tr [X \wedge Y] &=& \sum_a (X \wedge Y)^{aa} = \sum_{ac} X^{ac} \wedge Y^{ca} \nonumber \\
&=& \sum_{ac} (-1)^\mathfrak{pq} Y^{ca} \wedge X^{ac} = (-1)^\mathfrak{pq} \sum_c (Y\wedge X)^{cc} \nonumber \\
&=& (-1)^\mathfrak{pq} \tr [Y \wedge X]\label{eqn:graded-cyclic-trace}
\end{eqnarray}
where $X$ and $Y$ are matrix-valued differential forms of degree $\mathfrak{p}$ resp.~$\mathfrak{q}$. 
A similar graded property appears when taking the transpose of a wedge product of matrix-valued differential forms, namely
\begin{eqnarray}
\left\{(X\wedge Y)^\top\right\}^{ab} &=& (X\wedge Y)^{ba} = \sum_c X^{bc} \wedge Y^{ca} \nonumber \\
&=& \sum_c (-1)^{\mathfrak{pq}} Y^{ca} \wedge X^{bc} \nonumber \\
&=&  (-1)^{\mathfrak{pq}} \sum_c (Y^\top)^{ac} \wedge (X^\top)^{cb} \nonumber \\
&=& (-1)^{\mathfrak{pq}} \{Y^\top \wedge X^\top \}^{ab}.
\end{eqnarray}
It follows that 
\begin{equation}
(X \wedge Y)^\top = (-1)^{\mathfrak{pq}} Y^\top \wedge X^\top, \label{eqn:transpose-wedge}
\end{equation}
and similarly for Hermitian conjugation.\medskip 

For simplicity, we adopt in the following the \emph{Einstein summation convention} from now on, meaning that we always assume an implicit sum over repeated indices. 
We apply the convention both to the band indices and to the coordinate indices.
We further introduce the short-hand notation
\begin{equation}
X^{\wedge j} = \underbrace{X \wedge \ldots \wedge X}_{\textrm{$j$ copies of $X$}}    
\end{equation}
which we call the \emph{wedge power} of a matrix-valued differential form $X$. 
Note that the graded cyclic property implies that 
\begin{equation}
\tr[X^{\wedge (2j)}] =0 \qquad \textrm{if $X$ has odd degree}, \label{eqn:trace-odd-pmat}
\end{equation}
because the even-powered $X^{\wedge (2j)}$ can be expressed as a wedge product of two matrix-valued forms that are each of odd degree, namely $X^{\wedge (2j-1)}\wedge X$.
Another consequence of Eq.~(\ref{eqn:graded-cyclic-trace}) is that 
\begin{equation}
\begin{aligned}
\tr[Y^{-1} \wedge X \wedge Y] \equiv \tr [Y^{-1} X Y] = \tr[X] \\
\textrm{if $Y$ is of degree zero and invertible.}
\end{aligned}  \label{eqn:cyclic-trace}
\end{equation}
In the second expression above we have dropped the wedge symbols per the convention stated below Eq.~(\ref{eqn:wedge-matrices-def}).
\medskip

The second operator we need is the \emph{exterior derivative} $d$, which increases the degree of (matrix-valued) differential forms by $1$ and such that 
\begin{equation}
d^2 = 0 \label{eqn:nilpotent-d}
\end{equation}
(known as \emph{nilpotence of index $2$})~\cite{Fecko:2006}. 
On the wedge product of a differential $\mathfrak{p}$-form $\alpha$ with a $\mathfrak{q}$-form $\beta$, it obeys the \emph{graded Leibniz rule}, 
\begin{eqnarray}
d(\alpha \wedge \beta) &=& (d\alpha) \wedge \beta + (-1)^\mathfrak{p} \alpha \wedge (d \beta) \nonumber \\
&\equiv& d\alpha \wedge \beta + (-1)^\mathfrak{p} \alpha \wedge d \beta \label{eqn:exterior-d-def}  
\end{eqnarray}
By combining Eqs.~(\ref{eqn:exterior-d-def}) and~(\ref{eqn:wedge-matrices-def}) we find that the exterior derivative of the wedge product of matrix-valued differential forms likewise obeys the graded Leibniz rule,
\begin{equation}
d(X \wedge Y) = 
d X \wedge Y + (-1)^\mathfrak{p} X \wedge d Y, \label{eqn:exterior-d-matrix}  
\end{equation}
where $\mathfrak{p}$ is the degree of $X$. 
Since $d$ is a linear operator (i.e., ``derivative of sum'' is equal to ``sum of derivatives''), it follows that it commutes with the trace,
\begin{equation}
d \tr [X] = \tr [d X],   \label{eqn:d-tr-commute}
\end{equation}
which is also utilized in our discussion of symmetry constraints on Chern numbers.
\medskip

The final ingredient we need is the integration of differential forms on manifolds. 
A differential $\mathfrak{p}$-form $\alpha$ can be integrated over a $\mathfrak{p}$-dimensional manifold $\mathcal{M}^\textrm{p}$. The Stokes theorem states that~\cite{Fecko:2006,Baez:1994}
\begin{equation}
\int_{\mathcal{M}^{\mathfrak{p}+1}} d \alpha = \oint_{\partial \mathcal{M}^{\mathfrak{p}+1}} \alpha \label{eqn:Stokes}
\end{equation}
where $\partial$ is the boundary operator.
In particular, it follows from Eq.~(\ref{eqn:Stokes}) that the integral of an exact form $\alpha$ [i.e., that can be expressed as $\alpha = d \beta$ for some $(\mathfrak{p}-1)$-form $\beta$] on a closed manifold $\mathcal{M}$ (i.e., for which $\partial \mathcal{M}=\varnothing$) is necessarily zero.
We also encounter situations below where the differential form under the integral sign is evaluated at a transformed position, i.e., expressions like
\begin{equation}
\int_{\mathcal{M}^{\mathfrak{p}}} \alpha_{i\ldots \ell}(f\bs{k}) \, dk^i \wedge\ldots \wedge dk^\ell.
\end{equation}
The form under the integral sign is called the \emph{push-back} of $\alpha$ by $f$, and we denote it as $\tilde{\alpha}$.\footnote{Conventionally, the push-back of $\alpha$ by $f$ is labelled as $f^*\alpha$. However, our work reserves the ``${}^*$'' symbol for complex conjugation. Nevertheless, it should always be clear what map $f$ is used in the push-back considered, therefore dropping $f$ from the notation should not cause confusion. Therefore, we opt for the non-standard $\tilde{\alpha}$ notation.\label{foot:push-back}}  
A standard result in differential geometry~\cite{Fecko:2006} states that
\begin{equation}
\int_{\mathcal{M}^{\mathfrak{p}}} \tilde{\alpha} = \int_{f(\mathcal{M}^{\mathfrak{p}})} \alpha, \label{eqn:integerate:pushback}
\end{equation}
i.e., the transformation $f$ can be transferred from the integrand$^\textrm{\ref{foot:push-back}}$ $\tilde{\alpha}=f^*\alpha$ to the integration domain $f(\mathcal{M}^{\mathfrak{p}})$.

\subsection{Berry-Wilczek-Zee connection and Chern numbers}\label{app:Chern-defs}

We treat the occupied states of a (hyperbolic) Bloch Hamiltonian as a vector bundle $V \hookrightarrow E \stackrel{\pi}{\rightarrow} T^\textrm{d}$ over the Brillouin zone torus; more precisely, the $n$ occupied Bloch states at momentum $\bs{k}$ span the $n$-dimensional vector space $V$ that projects to momentum $\bs{k}$ in the $\textrm{d}$-dimensional torus ($T^\textrm{d}$). 
To relate the vector spaces at adjacent momenta, we define a rule of parallel transport by choosing a connection.
The standard choice for the characterization of energy bands in momentum space, motivated by the theory of polarization~\cite{KingSmith:1993,Zak:1989,Vanderbilt:2018}, is as follows. 
By arranging the occupied eigenstates at $\bs{k}$ as columns into a rectangular matrix $U$, the \emph{Berry-Wilczek-Zee (BWZ) connection}~\cite{Berry:1984,Wilczek:1984} is defined as 
\begin{equation}
A = U^\dagger dU,\label{eqn:BWZ-connex}
\end{equation}
and corresponds to a matrix-valued differential form of degree one. 
Note that $U$ is a \emph{rectangular} matrix of dimension $N \times n$: its $n$ columns correspond to the $n$ occupied Bloch eigenstates at $\bs{k}$, and its $N$ rows correspond to components of these eigenstates inside the $N$-dimensional Hilbert space of Bloch states at $\bs{k}$. For brevity, we do not explicitly display the $\bs{k}$-dependence of the mathematical objects on the two sides of Eq.~(\ref{eqn:BWZ-connex}).
The associated \emph{BWZ curvature} is a matrix-valued differential form of degree two, constructed as 
\begin{equation}
F = dA + A \wedge A. \label{eqn:BWZ-curvature}
\end{equation}
One easily verifies using Eqs.~(\ref{eqn:nilpotent-d}), (\ref{eqn:exterior-d-matrix}) and (\ref{eqn:BWZ-curvature}) that 
\begin{eqnarray}
d F &=& d A \wedge A - A \wedge d A \nonumber \\
&=& F \wedge A - A \wedge F, \label{eqn:Bianchi}
\end{eqnarray}
which is known as the Bianchi identity~\cite{Baez:1994}.

Let us comment on the symmetry of the matrix-valued differential forms $F$ and $A$ in band indices.
Since $U^\dagger U = \mathbb{1}$ for normalized eigenstates, it follows by taking the exterior derivative that $d U^\dagger U + U^\dagger dU =0$. 
We recognize from here that \begin{equation}
    A^\dagger = -A, \label{eqn:BWZ-A-skewHerm}
\end{equation} 
meaning that $A$ is skew-Hermitian in band indices. For $F$ we find that
\begin{eqnarray}
F^\dagger &=& (d A)^\dagger + (A \wedge A)^\dagger \stackrel{\textrm{(\ref{eqn:transpose-wedge})}}{=} d A^\dagger - A^\dagger \wedge A^\dagger \nonumber \\
&\stackrel{\textrm{(\ref{eqn:BWZ-A-skewHerm})}}{=}& -dA - A\wedge A = -F, \label{eqn:BWZ-F-skewHerm}
\end{eqnarray}
therefore $F$ is also skew-Hermitian in band indices. 
It follows from the result~(\ref{eqn:BWZ-F-skewHerm}) that the wedge power $F^{\wedge \textrm{n}}$ is skew-Hermitian and $\tr[F^{\wedge \textrm{n}}]$ is imaginary for $\textrm{n}$ odd, whereas the wedge power is Hermitian and the trace is real for $\textrm{n}$ even.

Building on the notions introduced above, the \emph{$\textrm{n}$-th Chern character} (with integer $\textrm{n}$ such that $2 \leq 2\textrm{n} \leq \textrm{d}$) is a differential form of degree $2\textrm{n}$ defined as
\begin{equation}
\textrm{ch}_{(\textrm{n})} = \frac{1}{\textrm{n}!}\frac{1}{\left(2\pi \mathrm{i}\right)^\textrm{n}} \tr\left[F^{\wedge \textrm{n}}\right].\label{eqn:characters}
\end{equation}
Note that the appearance of $\mathrm{i}^\mathrm{n}$ in the denominator implies that Chern characters are \emph{real}-valued differential forms. 
In addition, they are \emph{closed}, meaning that 
\begin{eqnarray}
d \textrm{ch}_{(\textrm{n})} &=& \frac{1}{\textrm{n}!}\frac{1}{\left(2\pi \mathrm{i}\right)^\textrm{n}} d \tr [F^{\wedge \textrm{n}}] \stackrel{\textrm{(\ref{eqn:d-tr-commute})}}{=} \frac{1}{\textrm{n}!}\frac{1}{\left(2\pi \mathrm{i}\right)^\textrm{n}} \tr [d(F^{\wedge \textrm{n}})] \nonumber \\
&\stackrel{\textrm{(\ref{eqn:exterior-d-matrix},\ref{eqn:graded-cyclic-trace})}}{=}& \frac{1}{\textrm{n}!}\frac{\textrm{n}}{\left(2\pi \mathrm{i}\right)^\textrm{n}} \tr [F^{\wedge (\textrm{n}-1)}\wedge dF] \nonumber \\
&\stackrel{\textrm{(\ref{eqn:Bianchi})}}{=}& \frac{1}{\textrm{n}!}\frac{\textrm{n}}{\left(2\pi \mathrm{i}\right)^\textrm{n}} \tr [F^{\wedge \textrm{n}}\wedge A - F^{\wedge (\textrm{n}-1)}\wedge A \wedge F] \nonumber \\
&\stackrel{\textrm{(\ref{eqn:graded-cyclic-trace})}}{=}& \frac{1}{\textrm{n}!}\frac{\textrm{n}}{\left(2\pi \mathrm{i}\right)^\textrm{n}} \tr [F^{\wedge \textrm{n}}\wedge A - F^{\wedge \textrm{n}}\wedge A ] = 0 \label{eqn:Chern-closed}
\end{eqnarray}
where in applications of Eq.~(\ref{eqn:graded-cyclic-trace}) in the last line we cyclically permuted the $2$-form $F$, implying a positive sign in the graded cyclic property.

The fact that Chern characters are closed forms suggests their integrals are topological invariants.
More precisely, the prefactor in the definition~(\ref{eqn:characters}) is chosen such that the integral
of the $\textrm{n}^\textrm{th}$ Chern character over a closed $(2\textrm{n})$-dimensional submanifold $\mathcal{M}^{2\mathrm{n}}$ of $T^d$ gives an integer-valued topological invariant~\cite{Nakahara:1990},
\begin{equation}
C_{\mathcal{M}^{2\mathrm{n}}} = \oint_{\mathcal{M}^{2\mathrm{n}}} \textrm{ch}_{(\textrm{n})} \in \mathbb{Z},  \label{eqn:Chern-numbers-def} 
\end{equation}
known as the \emph{$\textrm{n}^\textrm{th}$ Chern number} (of bundle $E$ restricted to submanifold $\mathcal{M}^{2\mathrm{n}}$). 
Note that if $2\textrm{n}<\textrm{d}$, there are multiple topologically inequivalent choices of submanifold $\mathcal{M}\subset T^\textrm{d}$. 
In particular, we find it convenient to arrange $\textrm{n}^\textrm{th}$ Chern numbers \begin{equation}
C_{(\mathrm{n}),i\ldots \ell} \equiv C_{T^{2\mathrm{n}}_{i\ldots\ell}} = \oint_{T^\textrm{d}_{i\ldots \ell}} \textrm{ch}_{(\textrm{n})} \label{eqn:Chern-tori}
\end{equation}
computed over $(2\textrm{n})$-dimensional subtori of $T^\textrm{d}$ spanned by $2\mathrm{n}$ momenta $k^i,\ldots,k^\ell$ into an array $C_{(\mathrm{n})}$. 
This array is skew-symmetric under exchange of any two momentum arguments, because it is obtained by integrating a differential form which is skew-symmetric under the same exchange (see the first paragraph of Appendix~\ref{app:Chern-matrixforms}). 
In particular, first Chern numbers computed over 2D subtori of $T^\textrm{d}$ can be arranged into an skew-symmetric matrix $C_{(1)}$. 
Similarly, second Chern numbers numbers computed over 4D subtori of $T^\textrm{d}$ can be arranged into an skew-symmetric tensor of rank $4$. 
In our applications with $\textrm{d}=4$ and $\textrm{d}=6$, we find it convenient to utilize the Hodge dual, $\mathscr{C}_{(2)}=\star C_{(2)}$, with components 
\begin{equation}
\mathscr{C}_{(2)}^{k\cdots \ell} = \frac{1}{4!}C_{(2),i\cdots j}\epsilon^{i\cdots jk\cdots \ell} \label{eqn:C2-Hodge}
\end{equation}
where $\epsilon^{i\cdots \ell}$ (with $\textrm{d}$ indices in the superscript) is the completely skew-symmetric Levi-Civita symbol. 
The dual $\mathscr{C}_{(2)}$ is a single number for $\textrm{d}=4$ and a skew-symmetric matrix for $\textrm{d}=6$.
\medskip

Before analyzing symmetry constraints on the Chern numbers in Appendices~\ref{app:Chern-first} and~\ref{app:Chern-second}, let us illustrate the convenience of adopting the formalism of matrix-valued differential forms by considering the effect of gauge transformation. 
A \emph{gauge transformation} of the occupied states corresponds to a right multiplication of the matrix $U$ of occupied Bloch states by a ($\bs{k}$-dependent) \emph{gauge matrix} (sometimes also called \emph{sewing matrix}) $B \in\mathsf{U}(n)$,
\begin{equation}
U\mapsto \;U' = U B,    
\end{equation}
where we use the prime ($'$) to denote the gauge-transformed quantities. The transformed BWZ connection is
\begin{eqnarray}
A' = U'^\dagger dU' &=& B^\dagger U^\dagger \left[dU B+U dB\right] \nonumber \\
&=& B^\dagger A  B + B^\dagger dB
\end{eqnarray}
where we utilized that $U^\dagger U = \mathbb{1}$. 

\begin{widetext}
For the transformed BWZ curvature we find
\begin{eqnarray}
F' 
&=& d A' + A' \wedge A' \nonumber \\
&=& d\left[B^\dagger A B + B^\dagger  dB\right] + \left[B^\dagger A B + B^\dagger  dB\right]\wedge \left[B^\dagger A B + B^\dagger  dB\right] \nonumber \\
&=&  dB^\dagger \wedge A B \;+\; B^\dagger (dA) B \;-\; B^\dagger A \wedge  dB \;+\; d B^\dagger \wedge  dB \nonumber \\
&\phantom{=}& \;+\;(B^\dagger A B)\wedge  (B^\dagger A B) \;+\; (B^\dagger A B)\wedge (B^\dagger  dB) \;+\; (B^\dagger  dB) \wedge (B^\dagger A B) \;+\; (B^\dagger  dB) \wedge (B^\dagger  dB) =: \text{\large \textcircled{\small 1}}
\end{eqnarray}
where the minus sign in the third term follows from the graded Leibniz rule. From the unitarity of matrix $B$ we have 
\begin{equation}
    B^\dagger B = \mathbb{1}    \label{eqn:rule-1}
\end{equation}
from which after applying $d$ we further obtain 
\begin{equation}
\quad B^\dagger  dB= -d B^\dagger  B. \label{eqn:rule-2}
\end{equation}
We apply Eq.~(\ref{eqn:rule-1}) to the fifth and sixth terms in $\text{\large \textcircled{\small 1}}$ and Eq.~(\ref{eqn:rule-2}) to the last two terms therein. Then
\begin{eqnarray}
\text{\large \textcircled{\small 1}} &=&  dB^\dagger \wedge A B \;+\; B^\dagger (dA) B \;-\; B^\dagger A \wedge  dB \;+\; d B^\dagger \wedge  dB \nonumber \\
&\phantom{=}& \;+\;B^\dagger (A \wedge A) B \;+\; B^\dagger A \wedge  dB \;-\;  dB^\dagger BB^\dagger \wedge A B \;-\;  dB^\dagger  BB^\dagger \wedge  dB =: \text{\large \textcircled{\small 2}}.
\end{eqnarray}
We observe that in $\text{\large \textcircled{\small 2}}$ six of the eight terms pairwise cancel, leaving behind
\begin{equation}
\text{\large \textcircled{\small 2}} = B^\dagger (dA) B + B^\dagger (A \wedge A) B = B^\dagger F B,   \label{eqn:F-covar} 
\end{equation}
i.e., the BWZ curvature transforms covariantly.
One further easily verifies that the gauge matrices trivially cancel under the trace [see also Eq.~(\ref{eqn:cyclic-trace})], 
\begin{equation}
\textrm{ch}'_{(\textrm{n})} = \frac{1}{\textrm{n}!} \frac{1}{(2\pi \imi)^\textrm{n}} \tr\left[F'^{\wedge \textrm{n}}\right] = \frac{1}{\textrm{n}!} \frac{1}{(2\pi \imi)^\textrm{n}} \tr\left[(B^\dagger F B)^{\wedge \textrm{n}}\right] = \frac{1}{\textrm{n}!} \frac{1}{(2\pi \imi)^\textrm{n}} \tr\left[F^{\wedge \textrm{n}}\right] = \textrm{ch}_{(\textrm{n})},\label{eqn:ch-invar}
\end{equation}
meaning that Chern characters (as well as their integrals, i.e., Chern numbers) are invariant under gauge transformations.
\end{widetext}

\subsection{Symmetry constraints on first Chern numbers}\label{app:Chern-first}

We finally deploy the mathematical techniques reviewed in Appendix~\ref{app:Chern-matrixforms} to study the differential and topological characteristics described in Appendix~\ref{app:Chern-first}.
As discussed in Appendix~\ref{app:derive-M-matrices}, a general symmetry $g$ of the hyperbolic lattice transforms the momentum vector as
\begin{align}
    \bs{k} \mapsto \;  
    \tilde{\bs{k}} = M_g \bs{k},\quad\; \textrm{or\; $\tilde{k}^i = (M_g)^i_{\phantom{i}j} k^j$\; in components,}\label{eqn:k-to-k-tilde}
\end{align} 
where we call $M_g\in \mathsf{GL}(2\mathfrak{g},\mathbb{Z})$ the \emph{point-group matrix} of symmetry $g$. 
We presently assume that $g$ is unitary (i.e., it does not involve time reversal), and we will comment on the antiunitary case at the end of this section.
Then there is a unitary representation $V_g(\bs{k})\in \mathsf{U}(N)$ of symmetry $g$ that linearly transforms the Hilbert space spanned by the hyperbolic Bloch states at $\bs{k}$ to those at $M_g\bs{k}$. 
In particular, $V_g(\bs{k})$ also transforms \emph{eigen}states of the hyperbolic Bloch Hamiltonian at $\bs{k}$ to those at $M_g\bs{k}$.
Since a general choice of gauge may not respect this unitary relation, one is led to introduce the sewing matrix
\begin{equation}
B^\dagger(\bs{k}) = U^\dagger(M_g \bs{k}) V_g(\bs{k}) U(\bs{k})\label{eqn:sewing-def}
\end{equation}
which is a $\mathsf{U}(n)$ matrix that encodes the gauge transformation from $V_g(\bs{k})U(\bs{k})$ to $U(M_g\bs{k})$. 
By inverting Eq.~(\ref{eqn:sewing-def}) we find that eigenstates at $M_g\bs{k}$ are generally related to eigenstates at $\bs{k}$ as~\cite{Fang:2012}
\begin{equation}
U(M_g\bs{k}) = V_g(\bs{k}) U(\bs{k}) B(\bs{k}).\label{eqn:states-k-vs-Mk}
\end{equation}
Our goal is to understand how Eq.~(\ref{eqn:states-k-vs-Mk}) relates the differential objects defined in Appendix~\ref{app:Chern-defs} at $\bs{k}$ to those at $M_g\bs{k}$, and especially how it constrains the Chern numbers on subtori of the higher-dimensional BZ.

The considerations of Eq.~(\ref{eqn:states-k-vs-Mk}) can be divided into two steps. 
First, analyze the effect of the right multiplication with sewing matrix $B(\bs{k})$, and second, of the left-multiplication with the unitary representation $V_g(\bs{k})$ of the symmetry $g$ in the Hilbert space. 
Note that the first step has been already analyzed in great generality in Appendix~\ref{app:Chern-defs}, where we revealed the invariance of Chern characters and Chern numbers under gauge transformations [see Eq.~(\ref{eqn:F-covar})].
We can therefore consider this step to be completed [e.g., by dropping $B(\bs{k})$ from Eq.~(\ref{eqn:states-k-vs-Mk}) altogether], and focus solely on the left multiplication with $V_g(\bs{k})$.

The complexity of the second step (especially if contrasted to a related discussion in Appendix~B of Ref.~\onlinecite{Fang:2012}) comes mainly from the two following aspects of hyperbolic lattices.
First, the point-group matrices $M_g \in\mathsf{GL}(2\mathfrak{g},\mathbb Z)$ may not be orthogonal. Second, we find that the unitary representations $V_g$ of some hyperbolic lattice symmetries (in particular for $g\in\{b,Q,R\}$ in most of the lattices discussed, see Fig.~\ref{fig:pd}) are non-symmorphic in the following sense: given an arbitrary unit cell, we find it is not mapped by the symmetry onto a unit cell related by an element of the translation group. Instead, various parts of the unit cell are mapped to pieces of several distinct unit cells which are obtained by various elements of the translation group. 
As a consequence, the representation $V_g$ is necessarily $\bs{k}$-dependent.
For those reasons, computing the BWZ connection and curvature generates additional terms involving $d V_g(\bs{k})$, which turn out to be more subtle to manipulate than the $d B(\bs{k})$-terms encountered during the discussion of gauge transformations in Appendix~\ref{app:Chern-defs}.

Having clarified the key idiosyncracies of hyperbolic momentum space, we proceed with the actual derivations.
As our analysis considers one symmetry $g$ at a time, we drop the subscript $g$ in the equations. We recall that the BWZ connection is given by Eq.~(\ref{eqn:BWZ-connex}). 
Then, computing the connection at $M\bs{k}$ and combining the expression with Eq.~(\ref{eqn:states-k-vs-Mk}) (where we drop the sewing matrix for the reasons discussed above), we find
\begin{eqnarray}
A(M{\bs{k}}) &=&U^\dagger(M{\bs{k}})dU(M{\bs{k}}) \nonumber \\
&=& U^\dagger(\bs{k})V^\dagger (\bs{k}) d[V(\bs{k}) U(\bs{k})].\label{eqn:BWZ-connex-Mk}
\end{eqnarray}
Note that in Eq.~(\ref{eqn:BWZ-connex-Mk}) we express the BWZ connection at $M\bs{k}$ solely in terms of variables at $\bs{k}$ (including the differentials $d \equiv d/d\bs{k}$ of coordinates $\bs{k}$). 
Using the language of Appendix~\ref{app:Chern-matrixforms}, the left-hand side corresponds to a pull-back by $M$.
Since this is true for most subsequent equations, we will drop the momentum arguments of all quantities. 
Quantities at $M\bs{k} \equiv \tilde{\bs{k}}$, usually appearing on the left-hand side of the equations, will be henceforth decorated with a tilde ($\,\tilde{}\,$).

Manipulating Eq.~(\ref{eqn:BWZ-connex-Mk}) further, we obtain
\begin{eqnarray}
\tilde{A} 
&=& U^\dagger V^\dagger dV U + U^\dagger V^\dagger V dU \nonumber \\
&=& Z + \underbrace{U dU}_A
\end{eqnarray}
where we recognized the BWZ connection $A$ at $\bs{k}$, and we introduced the matrix-valued $1$-form 
\begin{align}
    Z = U^\dagger V^\dagger dV U. \label{eqn:Z-form}
\end{align}
Plugging Eq.~(\ref{eqn:BWZ-connex-Mk}) into Eq.~(\ref{eqn:BWZ-curvature}), we find that the BWZ curvature at $\tilde{\bs{k}}$ is expressed via quantities and differentials at~$\bs{k}$~as
\begin{eqnarray}
\tilde{F}
&=& d(Z+A)+ (Z+A)\wedge(Z+A) \nonumber \\
&=& \underbrace{dA + A\wedge A}_F + dZ + A\wedge Z + Z\wedge A  + Z\wedge Z, \label{eqn:BWZ-tilde-Mk}
\end{eqnarray}
where we recognized the BWZ curvature $F$ at $\bs{k}$. 
Focusing presently on first Chern numbers, we next extract the trace,
\begin{eqnarray}
\tr[\tilde{F}
] &=& \tr[F] + \underbrace{\tr[d Z]}_{\textrm{$d\tr [Z]$ per Eq.~(\ref{eqn:d-tr-commute})}} + \underbrace{\tr[Z^{\wedge 2}]}_{\textrm{vanishes per Eq.~(\ref{eqn:trace-odd-pmat})}} \nonumber \\
&\phantom{=}& + \underbrace{\tr[A\wedge Z] + \tr[Z\wedge A].}_{\textrm{cancel per Eq.~(\ref{eqn:graded-cyclic-trace})}}  \label{eqn:BWZ-curv-Mk}
\end{eqnarray}
The term $d \tr[Z]$ in Eq.~(\ref{eqn:BWZ-curv-Mk}) does not generally vanish; therefore, the Chern characters at $\bs{k}$ and $\tilde{\bs{k}}$ are related as
\begin{equation}
\tilde{\textrm{ch}}_{(1)} 
= \textrm{ch}_{(1)} + \frac{1}{2\pi \imi} d \tr[Z] \label{eqn:ch1-relation}
\end{equation}
i.e., they differ by an amount set by $d \tr[Z(\bs{k})]/(2\pi\imi)$. 

We next integrate both sides of Eq.~(\ref{eqn:ch1-relation}) over the 2D torus $T^2_{ij}$ spanned by momenta $(k^i,k^j)$.
On the right-hand side, we recognize that the integration of the first term is precisely the Chern number $C_{(1),ij}$ in Eq.~(\ref{eqn:Chern-tori}), whereas the integral of the second term vanishes by Stokes' theorem (\ref{eqn:Stokes}).
On the left-hand side, we apply the rule (\ref{eqn:integerate:pushback}) for integrating push-backs of differential forms, leading to
\begin{equation}
\oint_{T^2_{ij}}\tilde{\textrm{ch}}_{(1)} = \oint_{M(T^2_{ij})}\textrm{ch}_{(1)}. \label{eqn:confusing-integral-explicit}
\end{equation}
The next key observation is \emph{geometrical}, namely that $M(T^2_{ij})$ can be continuously deformed into a union of tori $T^2_{k\ell}$, each torus appearing with multiplicity $M^{k}_{\phantom{k}i} M^{\ell}_{\phantom{\ell}j}$. 
To understand this, note that by Eq.~(\ref{eqn:k-to-k-tilde}) the transformed torus $M(T^2_{ij})$ is spanned by momenta $(Mk^i,Mk^j)$, and observe that as $k^j$ winds once around the BZ torus, the $\ell$-th component of $Mk^j$ winds around the BZ torus $\partial \tilde{k}^\ell/\partial k^j = M^{\ell}_{\phantom{\ell}j}$ times (while the factor $M^{k}_{\phantom{k}i}$ is obtained similarly by considering the progression of $\tilde{k}^k$ upon increasing $k^i$).
Since the integral which we compute is topological [see~Eq.~(\ref{eqn:Chern-closed})] and the Hamiltonian spectrum is presumed gapped throughout the entire BZ, its value is invariant under such a continuous deformation of the integration domain; therefore,
\begin{eqnarray}
\textrm{(\ref{eqn:confusing-integral-explicit})} = M^{k}_{\phantom{k}i} M^{\ell}_{\phantom{\ell}j} \oint_{T^2_{k\ell}} \textrm{ch}_{(1)} =M^{k}_{\phantom{k}i} C_{(1),k\ell} M^{\ell}_{\phantom{\ell}j}.&
\end{eqnarray}
The last expression is naturally interpreted as a matrix multiplication. 
Combining this result with the integration of the right-hand side of Eq.~(\ref{eqn:ch1-relation}), we conclude that the matrix of first Chern numbers must obey the constraint
\begin{equation}
C_{(1)} = M_g^\top C_{(1)} M_g \label{eqn:C1-relations}
\end{equation}
if there is a unitary symmetry $g$ that transforms momenta with $M_g\in\textsf{GL}(2\mathfrak{g},\mathbb{Z})$.

Let us finally comment on the case of an antiunitary symmetry $g$, such as those which involve time reversal. 
In this case there is an antiunitary representation $V_g(\bs{k}) \mathcal{K}$ that linearly transforms the Hilbert space spanned by hyperbolic Bloch states at $\bs{k}$ to those at $M_g(\bs{k})$, and Eq.~(\ref{eqn:states-k-vs-Mk}) is altered to
\begin{equation}
U(M_g\bs{k}) = V_g(\bs{k}) U^*(\bs{k}) B(\bs{k}).\label{eqn:states-k-vs-Mk-anti}
\end{equation}
One easily verifies that derivations in Eqs.~(\ref{eqn:BWZ-connex-Mk}) to~(\ref{eqn:ch1-relation}) are essentially unchanged, except that objects $A,F,\textrm{ch}_{(1)}$ evaluated at $\bs{k}$ become complex-conjugated, and Eq.~(\ref{eqn:Z-form}) is replaced by $Z = U^\top V^\dagger  dV  U^*$.
Crucially, we showed in Eq.~(\ref{eqn:BWZ-F-skewHerm}) that $F$ is skew-Hermitian with imaginary trace; therefore, $\textrm{ch}_{(1)}^* = -\textrm{ch}_{(1)}$, leading to the appearance of a relative minus sign in Eq.~(\ref{eqn:C1-relations}). 
In total, we find that symmetry $g$ constrains the matrix of first Chern numbers as  
\begin{equation}
C_{(1)} = \varsigma_g M_g^\top C_{(1)} M_g^{\phantom{\top}}\label{eqn:C1-constraint-supp}
\end{equation}
where $\varsigma_g$ is $+1$ ($-1$) for unitary (antiunitary) symmetry $g$. This results corresponds to Eq.~(\ref{eqn:C1-constraint}) in the main text.

\subsection{Symmetry constraints on second Chern numbers}\label{app:Chern-second}

We proceed to investigate symmetry-imposed constraints on second Chern numbers. 
These are obtained by integrating $\textrm{ch}_{(2)} = -\tfrac{1}{8\pi^2}\tr[F\wedge F]$ as prescribed in Eq.~(\ref{eqn:Chern-numbers-def}). 
We again first consider unitary symmetries, and comment on the antiunitary case near the end of this section.
Starting from Eq.~(\ref{eqn:BWZ-tilde-Mk}), we find that 
\begin{widetext}
\begin{align}
\tilde{F}\wedge\tilde{F} & =\begin{array}[t]{lllll}
\ \ F\wedge F & +F\wedge  d Z    & +F\wedge A\wedge  Z   & +F\wedge  Z  \wedge A & +F\wedge  Z  \wedge  Z  \\
+  d Z   \wedge F & +  d Z   \wedge  d Z    & +  d Z   \wedge A\wedge  Z   & +  d Z   \wedge  Z  \wedge A & +  d Z   \wedge  Z  \wedge  Z  \\
+A\wedge  Z  \wedge F & +A\wedge  Z  \wedge  d Z    & +A\wedge  Z  \wedge A\wedge  Z   & +A\wedge  Z  \wedge  Z  \wedge A & +A\wedge  Z  \wedge  Z  \wedge  Z  \\
+ Z  \wedge A\wedge F & + Z  \wedge A\wedge  d Z    & + Z  \wedge A\wedge A\wedge  Z   & + Z  \wedge A\wedge  Z  \wedge A & + Z  \wedge A\wedge  Z  \wedge  Z  \\
+ Z  \wedge  Z  \wedge F & + Z  \wedge  Z  \wedge  d Z    & + Z  \wedge  Z  \wedge A\wedge  Z   & + Z  \wedge  Z  \wedge  Z  \wedge A & + Z  \wedge  Z  \wedge  Z  \wedge  Z,  
\end{array}\label{eqn:horror}
\end{align}
in terms of $A, F$ and $Z$ previously defined.
Although at first sight intimidating, many terms in Eq.~(\ref{eqn:horror}) trivially vanish after taking the trace. 
First of all, observe that all terms on the right-hand side can be expressed as wedge products of two matrix-valued $2$-forms. 
By commuting these matrix-valued $2$-forms under the trace, we can reduce the list of terms to just the upper triangular part of  Eq.~(\ref{eqn:horror}), namely
\begin{align}
\tr[\tilde{F}\wedge\tilde{F}] & =\begin{array}[t]{lllll}
\ \ \tr[F\wedge F] & \underline{+2\tr[F\wedge  d Z ]} & \underline{+2\tr[F\wedge A\wedge  Z]}   & \underline{+2\tr[F\wedge  Z  \wedge A]}\dotuline{} & \uwave{+2\tr[F\wedge  Z  \wedge  Z]}  \\
& \dotuline{+\tr[  d Z   \wedge  d Z ]}   & \uwave{+2\tr[  d Z   \wedge A\wedge  Z ]}  & \uwave{+2\tr[  d Z   \wedge  Z  \wedge A]} & \dotuline{+2\tr[  d Z   \wedge  Z  \wedge  Z]} \\
& & \dashuline{+\tr[A\wedge  Z  \wedge A\wedge  Z]}   & \uwave{+2\tr[A\wedge  Z  \wedge  Z  \wedge A]} & \dashuline{+2\tr[A\wedge  Z  \wedge  Z  \wedge  Z]} \\
& & & \dashuline{+\tr[Z  \wedge A\wedge  Z  \wedge A]} & \dashuline{+2\tr[Z  \wedge A\wedge  Z  \wedge  Z]}  \\
& & & & \dashuline{+\tr[Z  \wedge  Z  \wedge  Z  \wedge  Z]},  
\end{array}\label{eqn:milder-horror}
\end{align}
where the meaning of the various types of underlining is clarified in the following.
\end{widetext}

Applying the graded cyclic property (\ref{eqn:graded-cyclic-trace}) on some of the terms, we identify the cancellations
\begin{eqnarray}
2\tr[A\wedge Z\wedge Z\wedge Z]+2\tr[Z\wedge A\wedge Z\wedge Z] =0,\\
\tr[A\wedge Z\wedge A\wedge Z]+\tr[Z\wedge A\wedge Z\wedge A] =0,
\end{eqnarray}
while 
\begin{equation}
\tr[Z\wedge Z\wedge Z\wedge Z] = \tr [Z^{\wedge 4}] = 0 
\end{equation}
follows from Eq.~(\ref{eqn:trace-odd-pmat}). The three equations above imply the vanishing of the terms indicated with a dashed underline in Eq.~(\ref{eqn:milder-horror}).

We next find that certain terms in Eq.~(\ref{eqn:milder-horror}) are total derivatives (i.e., exact forms), implying that they yield zero upon integration on closed manifolds. We can therefore drop them from the right-hand side of Eq.~(\ref{eqn:milder-horror}). 
In particular, 
\begin{eqnarray}
\tr[ d Z  \wedge  d Z ] = d \tr[Z \wedge  d Z ], \label{eqn:exact-1} \\
2 \tr[ d Z  \wedge Z \wedge Z] = \tfrac{2}{3} d \tr[Z^{\wedge 3}], \label{eqn:exact-2}
\end{eqnarray}
which correspond to terms indicated with dotted underline in Eq.~(\ref{eqn:milder-horror}). Further terms produce exact forms only if suitably combined, especially
\begin{eqnarray}
&\phantom{=}& 2\tr[F\wedge A \wedge Z]+ 2\tr[F\wedge Z \wedge A] + 2\tr[F\wedge  d Z ] \nonumber \\
&\stackrel{\textrm{(\ref{eqn:graded-cyclic-trace})}}{=}& 2\tr[F \wedge A \wedge Z - A \wedge F \wedge Z + F \wedge  d Z ] \nonumber \\
&\stackrel{\textrm{(\ref{eqn:Bianchi})}}{=}& 2\tr[dF\wedge Z + F\wedge  d Z ] \stackrel{\textrm{(\ref{eqn:exterior-d-matrix})}}{=} 2 \tr[d(F\wedge Z)]\nonumber \\
&\stackrel{\textrm{(\ref{eqn:d-tr-commute})}}{=}& 2 d \tr[F\wedge Z]. \label{eqn:exact-3}
\end{eqnarray}
Therefore, these three terms together yield zero upon integration on a closed manifold, allowing us to drop them from Eq.~(\ref{eqn:milder-horror}) where they are indicated with a solid underline.
We finally deal with the four terms indicated with squiggly underline in Eq.~(\ref{eqn:milder-horror}). 
By performing cyclic permutations (\ref{eqn:graded-cyclic-trace}), it is possible to combine them to form
\begin{eqnarray}
&\phantom{=}& \!\!\!\!\!\!\!\! 2\tr\Big[(F-A\wedge A)\wedge Z\wedge Z - A \wedge [ d Z \wedge Z - Z\wedge  d Z ]\Big] \nonumber \\    
&\stackrel{\textrm{(\ref{eqn:BWZ-curvature},\ref{eqn:exterior-d-matrix})}}{=}&  2\tr[dA\wedge Z \wedge Z - A\wedge d(Z\wedge Z)] \nonumber \\
&\stackrel{\textrm{(\ref{eqn:exterior-d-matrix},\ref{eqn:d-tr-commute})}}{=}&2 d \tr[A\wedge Z \wedge Z]. \label{eqn:exact-4}
\end{eqnarray}
The exactness implies that the four terms indicated with squiggly underline together yield zero upon integration over closed manifolds, and can likewise be dropped from Eq.~(\ref{eqn:milder-horror}). By collecting the results above, we succeeded in showing that 
\begin{equation}
\tilde{\textrm{ch}}_{(2)} = \textrm{ch}_{(2)} - \frac{1}{8\pi^2} d (\cdots), \label{eqn:ch-2-rel}
\end{equation}
where the dots comprise the right-hand sides of Eqs.~(\ref{eqn:exact-1}), (\ref{eqn:exact-2}), (\ref{eqn:exact-3}), and (\ref{eqn:exact-4}).

We next integrate both sides of Eq.~(\ref{eqn:ch-2-rel}) over the 4D torus $T^4_{{i_1}{i_2}{i_3}{i_4}}$ that is spanned by momenta $(k^{i_1},k^{i_2},k^{i_3},k^{i_4})$. 
Integration of the first-term on the right-hand side yields $C_{(2),{i_1}{i_2}{i_3}{i_4}}$ by definition (\ref{eqn:Chern-tori}), while integration of the second term yields zero by Stokes' theorem~(\ref{eqn:Stokes}). 
On the left-hand side we use the push-back rule in Eq.~(\ref{eqn:integerate:pushback}), which implies
\begin{equation}
\oint_{T^4_{{i_1}{i_2}{i_3}{i_4}}} \tilde{\textrm{ch}}_{(2)} = \oint_{M(T^4_{{i_1}{i_2}{i_3}{i_4}})} {\textrm{ch}}_{(2)}.\label{eqn:C2-rotate} 
\end{equation}
We next utilize the same geometric arguments as below Eq.~(\ref{eqn:C1-relations}), which allow us to express
\begin{equation}
\textrm{(\ref{eqn:C2-rotate})} = M^{j_1}_{\phantom{j_1}i_1} M^{j_2}_{\phantom{j_2}i_2} M^{j_3}_{\phantom{j_3}i_3} M^{j_4}_{\phantom{j_4}i_4} \oint_{T^4_{{j_1}{j_2}{j_3}{j_4}}} \textrm{ch}_{(2)}.
\end{equation}
Therefore, the integration of both sides of Eq.~(\ref{eqn:ch-2-rel}) results in
\begin{equation}
C_{(2),{i_1}{i_2}{i_3}{i_4}}= M^{j_1}_{\phantom{j_1}i_1} M^{j_2}_{\phantom{j_2}i_2} M^{j_3}_{\phantom{j_3}i_3} M^{j_4}_{\phantom{j_4}i_4} C_{(2),{j_1}{j_2}{j_3}{j_4}}\label{eqn:C2-relation}
\end{equation}
which we analyze below.

Let us briefly pause to consider the action of an \emph{antiunitary} symmetry $g$. 
The additional complex conjugation, displayed explicitly in Eq.~(\ref{eqn:states-k-vs-Mk-anti}), leads again to complex conjugation of objects $A,F,F \wedge F,\textrm{ch}_{(2)}$ that appear on the right-hand sides of expressions for $\tilde{A},\tilde{F},\tilde{F}\wedge\tilde{ F},\tilde{\textrm{ch}}_{(2)}$, and also implies certain minor modification of $Z$.
Crucially, since $F \wedge F$ is Hermitian [see the text below Eq.~(\ref{eqn:BWZ-F-skewHerm})], it follows that its trace is real, and therefore $\textrm{ch}_{(2)}^* = \textrm{ch}_{(2)}$. 
Thus, we conclude that the relative sign in Eq.~(\ref{eqn:C2-relation}) is positive both for unitary and for antiunitary symmetries. The following analysis therefore applies equally to both categories of symmetries.

To study the implications of Eq.~(\ref{eqn:C2-relation}), it is convenient to compute the Hodge dual $\mathscr{C}_{(2)}=\star C_{(2)}$, defined component-wise in Eq.~(\ref{eqn:C2-Hodge}). 
As the prescription for the Hodge dual explicitly depends on the dimension of the manifold considered, we need to discuss separately the case of 4D~resp.~6D~BZ.

For models with 4D BZ, $C_{(2),{i_1}{i_2}{i_3}{i_4}}$ has only one independent component, which is turned into a pseudoscalar 
\begin{equation}
    \mathscr{C}_{(2)} = \frac{1}{4!}  C_{(2),{i_1}{i_2}{i_3}{i_4}} \epsilon^{{i_1}{i_2}{i_3}{i_4}}\label{eqn:C2-Hodge-4D}
\end{equation} 
under the Hodge map, where $\epsilon$ is the Levi-Civita symbol. 
The inverse relation reads 
\begin{equation}
C_{(2),{j_1}{j_2}{j_3}{j_4}} =  \mathscr{C}_{(2)} \epsilon_{{j_1}{j_2}{j_3}{j_4}} .    \label{eqn:C2-Hodge-4D-inverse}
\end{equation}
To identify the implications of Eq.~(\ref{eqn:C2-relation}) on $\mathscr{C}_{(2)}$, we multiply that equation by $\tfrac{1}{4!}\epsilon^{{i_1}{i_2}{i_3}{i_4}}$, and apply the definition~(\ref{eqn:C2-Hodge-4D}) on the left-hand side and the inverse relation~(\ref{eqn:C2-Hodge-4D-inverse}) on the right-hand side; this results in
\begin{equation}
\mathscr{C}_{(2)} 
= \frac{1}{4!}\epsilon^{{i_1}{i_2}{i_3}{i_4}} \epsilon_{{j_1}{j_2}{j_3}{j_4}}  M^{j_1}_{\phantom{j_1}i_1} M^{j_2}_{\phantom{j_2}i_2} M^{j_3}_{\phantom{j_3}i_3} M^{j_4}_{\phantom{j_4}i_4} \mathscr{C}_{(2)}. \label{eqn:I-feel-in-shape}
\end{equation}
In Eq.~(\ref{eqn:I-feel-in-shape}) we recognize the standard expression for the determinant of a $4\times 4$ matrix in terms of its components and two Levi-Civita symbols~\cite{Fecko:2006}. It follows that the second Chern number in the 4D BZ is constrained by
\begin{equation}
\mathscr{C}_{(2)} = (\det M_g) \mathscr{C}_{(2)}\label{eqn:C2-constraint-app}
\end{equation}
in the presence of (unitary or antiunitary) symmetry $g$ that acts in momentum-space with point-group matrix $M_g$. This result corresponds to Eq.~(\ref{eqn:C2-constraint}) in the main text.

Finally, for models with 6D BZ, $C_{(2),{i_1}{i_2}{i_3}{i_4}}$ has 15 independent components, which are via
\begin{equation}
\mathscr{C}_{(2)}^{{k_5}{k_6}}=\frac{1}{4!} C_{(2),{i_1}{i_2}{i_3}{i_4}}  \epsilon^{{i_1}{i_2}{i_3}{i_4}{k_5}{k_6}} \label{eqn:C2-Hodge-6D}
\end{equation}
arranged into a skew-symmetrix $6\times 6$ matrix. 
The inverse relation reads
\begin{equation}
C_{(2),{j_1}{j_2}{j_3}{j_4}} = \frac{1}{2!} \mathscr{C}_{(2)}^{{\ell_5}{\ell_6}} \epsilon_{{j_1}{j_2}{j_3}{j_4}{\ell_5}{\ell_6}}. \label{eqn:C2-Hodge-6D-inverse}    
\end{equation}
To find the implications of Eq.~(\ref{eqn:C2-relation}) on elements of matrix $\mathscr{C}_{(2)}$, we first multiply that equation by $\tfrac{1}{4!}\epsilon^{{i_1}{i_2}{i_3}{i_4}{j_5}{j_6}}$, and we then apply the definition~(\ref{eqn:C2-Hodge-6D}) on the left-hand side and the inverse relation~(\ref{eqn:C2-Hodge-6D-inverse}) on the right-hand side; this produces
\begin{widetext}
\begin{equation}
\mathscr{C}_{(2)}^{{k_5}{k_5}} 
\!=\! \frac{1}{4!}\frac{1}{2!} \epsilon^{{i_1}{i_2}{i_3}{i_4}{k_5}{k_6}} \epsilon_{{j_1}{j_2}{j_3}{j_4}{\ell_5}{\ell_6}} 
M^{j_1}_{\phantom{j_1}i_1} M^{j_2}_{\phantom{j_2}i_2} M^{j_3}_{\phantom{j_3}i_3} M^{j_4}_{\phantom{j_4}i_4}
\mathscr{C}_{(2)}^{{\ell_5}{\ell_6}}.\label{eqn:C2-not-yet-result}
\end{equation}
To simplify the above equation, we use the identity~\cite{mathSE:4638444}
\begin{equation}
\frac{1}{(\textrm{d}-m)!\times m!}
\epsilon^{i_1 \cdots i_{\textrm{d}-m} k_{\textrm{d}-m+1}\cdots k_\textrm{d}}
\epsilon_{j_1 \cdots j_{\textrm{d}-m} \ell_{\textrm{d}-m+1}\cdots \ell_\textrm{d}}
M^{j_1}_{\phantom{j_1}i_1} \cdots M^{j_{\textrm{d}-m}}_{\phantom{j_{\textrm{d}-m}}i_{\textrm{d}-m}} \alpha^{\ell_{\textrm{d}-m+1}\cdots \ell_\textrm{d}} = (\det M)(M^{-1})^{k_{\textrm{d}-m+1}}_{\phantom{k_{\textrm{d}-m+1}}\ell_{\textrm{d}-m+1}}\cdots (M^{-1})^{k_{\textrm{d}}}_{\phantom{k_{\textrm{d}}}\ell_{\textrm{d}}} \alpha^{\ell_{\textrm{d}-m+1}\cdots \ell_\textrm{d}}
\end{equation}
\end{widetext}
that holds for any invertible matrix $M$ and $m$-form $\alpha$ (i.e., a fully skew-symmetric array $\alpha$ of degree $m$).
Knowing that $\mathscr{C}_{(2)}$ is skew-symmetric in its superscript indices, one can use the above identity to rewrite Eq.~(\ref{eqn:C2-not-yet-result}) into
\begin{equation}
\mathscr{C}_{(2)}^{{k_5}{k_6}} = (\det M)(M^{-1})^{k_5}_{\phantom{k_5}\ell_5} (M^{-1})^{k_6}_{\phantom{k_6}\ell_6} \mathscr{C}_{(2)}^{{\ell_5}{\ell_6}}.\label{eqn:C2-almost-there}
\end{equation}
Since for $M\in\mathsf{GL}(2\mathfrak{g},\mathbb{Z})$ we have $(\det M)^2=1$, we can move the determinant to the other side of the equation. 
We finally multiply Eq.~(\ref{eqn:C2-almost-there}) with $M^{i_5}_{\phantom{i_5}k_5} M^{i_6}_{\phantom{i_6}k_6}$, and bring it into matrix multiplication form. 
We find that
\begin{equation}
\mathscr{C}_{(2)} = (\det M_g) M_g \mathscr{C}_{(2)} M_g^\top \label{eqn:C2-constraint-6D-supp}
\end{equation}
in the presence of a (unitary or antiunitary) symmetry $g$ that acts in momentum-space with point-group matrix $M_g$. This result corresponds to Eq.~(\ref{eqn:C2-constraint-6D}) in the main text.

\section{Symmetry-constrained Chern matrices}\label{app:complete-Chern-matrices}

We here evaluate the general constraints on the Chern numbers inside the hyperbolic BZ.
This is achieved by combining the general equations derived in Appendix~\ref{app:Chern-constraints} with the specific form of the point-group matrices $M_g$ listed in Appendix~\ref{app:derive-M-matrices}. 
The discussion for hyperbolic Haldane models is subdivided into Appendix~\ref{app:complete-C1-matrices} focused on matrices of first Chern numbers, followed by Appendix~\ref{app:complete-C2-matrices} that discusses second Chern numbers.
A concise summary of these results is offered in Table~\ref{table:Cherns} in the main text.
Finally, in Appendix~\ref{app:6-4-other} we present the matrices of first Chern numbers of the generalized Haldane model with modified flux patterns on $\{6,4\}$ lattice, considered in the discussion in Sec.~\ref{sec:flux-patterns} of the main text.

\subsection{Matrices of first Chern numbers}\label{app:complete-C1-matrices}

We first study how Eq.~(\ref{eqn:C1-relations}) constrains the matrix of first Chern numbers for the Haldane models on $\{p,q\}$ lattices with a 4D BZ.
In the absence of sublattice mass, we evaluate using \textsc{Wolfram Mathematica} that equation for $g\in\{a\mcT,b\mcT,c\mcT\}$, obtaining the following results.
For the $\{8,3\}$ and $\{8,4\}$
Haldane models:
\begin{equation}
C^{\{8,3\}}_{(1),m=0}=C^{\{8,4\}}_{(1),m=0}=
\left(\begin{array}{cccc}
0       &   C_{12}      &   -C_{12}     &   C_{12}  \\
-C_{12} &   0           &   C_{12}      &   -C_{12} \\
C_{12}  &   -C_{12}     &   0           &   C_{12}  \\
-C_{12} &   C_{12}      &   -C_{12}     &   0       \\
\end{array}\right),\label{eqn:C83,M=0}
\end{equation}
for the $\{6,4\}$ Haldane model:
\begin{equation}
C^{\{6,4\}}_{(1),m=0}=\left(\begin{array}{cccc}
0       &   C_{12}      &   0     &   C_{12}  \\
-C_{12} &   0           &  0    &   0 \\
0 &   0     &   0           &   C_{12}  \\
-C_{12} &   0      &   -C_{12}     &   0       \\
\end{array}\right),\label{eqn:C64,M=0}
\end{equation}
and for the $\{10,5\}$ Haldane model: 
\begin{equation}
C^{\{10,5\}}_{(1),m=0} = C^{\{10,5\}}_{(1),m \neq 0} = \left(\begin{array}{cccc}
0       &   C_{12}      &   C_{13}     &   C_{12}  \\
-C_{12} &   0           &  -C_{13}    &   C_{13} \\
-C_{13} &   C_{13}     &   0           &   C_{12}  \\
-C_{12} &   -C_{13}      &   -C_{12}     &   0       \\
\end{array}\right).   \label{eqn:C105,both-M}
\end{equation}

We next consider the same lattices in the presence of sublattice mass $m\neq 0$, in which case Eq.~(\ref{eqn:C1-relations}) has to be evaluated with $g\in\{P^2,b\mcT,c\mcT\}$. 
The inclusion of sublattice mass corresponds to a reduction in symmetry, and is found to increase the number of independent first Chern numbers in the matrix by one. For the $\{8,3\}$ Haldane model, we obtain:
\begin{equation}
C^{\{8,3\}}_{(1),m\neq 0}\!=\!
\left(\begin{array}{cccc}
0       &   C_{12}      &   C_{13}     &   \!\!-C_{12}{-}2C_{13}\!\!\!  \\
-C_{12} &   0           &   \!\!\!-C_{12}{-}2C_{13}\!\!\!      &   C_{13} \\
-C_{13}  &   \!\!\!C_{12}{+}2C_{13}\!\!\!     &   0           &   C_{12}  \\
\!\!C_{12}{+}2C_{13}\!\! &   -C_{13}      &   -C_{12}     &   0       \\
\end{array}\right),\label{eqn:C83,M!=0}
\end{equation}
from which one obtains Eq.~(\ref{eqn:C83,M=0}) by setting $C_{13}=-C_{12}$. 
For the $\{6,4\}$ Haldane model:
\begin{equation}
C^{\{6,4\}}_{(1),m\neq 0}=\left(\begin{array}{cccc}
0       &   C_{12}      &   0     &   C_{14}  \\
-C_{12} &   0           &  -C_{12}{+}C_{14}    &   0 \\
0 &   C_{12}{-}C_{14}    &   0           &   C_{12}  \\
-C_{14} &   0      &   -C_{12}     &   0       \\
\end{array}\right),\label{eqn:C64,M!=0}
\end{equation}
from which one obtains Eq.~(\ref{eqn:C64,M=0}) by setting $C_{14}=C_{12}$. 
For the $\{8,4\}$ Haldane model: 
\begin{equation}
C^{\{8,4\}}_{(1),m\neq 0}=
\left(\begin{array}{cccc}
0       &   C_{12}      &   C_{13}     &   C_{12}  \\
-C_{12} &   0           &   C_{12}      &   -2C_{12}{-}C_{13} \\
-C_{13}  &   -C_{12}     &   0           &   C_{12}  \\
-C_{12} &   2C_{12}{+}C_{13}      &   -C_{12}     &   0       \\
\end{array}\right),\label{eqn:C84,M!=0}
\end{equation}
from which one obtains Eq.~(\ref{eqn:C83,M=0}) by setting $C_{13}=C_{12}$.
For the $\{10,5\}$ Haldane model, the inclusion of the mass term does not alter the general form of the Chern matrix, as explicitly indicated in~Eq.~(\ref{eqn:C105,both-M}).

We next summarize the symmetry constraints on the matrix of first Chern numbers for the lattices with a 6D BZ.
For the $\{7,3\}$ Haldane model:
\begin{equation}
C^{\{7,3\}}_{(1)} = \left(\begin{array}{cccccc}
0       &   0       &   C_{13}  &   -C_{13} &   C_{13}  &   0       \\
0       &   0       &   C_{13}  &   0       &   0       &   C_{13}  \\
-C_{13} &   -C_{13} &   0       &   -C_{13} &   0       &   -C_{13} \\
C_{13}  &   0       &   C_{13}  &   0       &   C_{13}  &   C_{13}  \\
-C_{13} &   0       &   0       &   -C_{13} &   0       &   0       \\
0       &   -C_{13} &   C_{13}  &   -C_{13} &   0       &   0
\end{array}\right).\label{eqn:C73}
\end{equation}
On this lattice one cannot include the sublattice mass because $p$ is odd. 
For the $\{12,3\}$ Haldane model we find:\begin{widetext}
\begin{equation}
C^{\{12,3\}}_{(1),m=0} = \left(
\begin{array}{cccccc}
 0              & C_{12}        & 0                 & C_{14}            & 0                 & C_{12} \!    \\
 \! -C_{12}     & 0             & C_{12}            & \!\!-C_{12}{-}C_{14} \!\! & C_{14}            & 0         \\
 0              & -C_{12}       & 0                 & C_{    12}           & \!\!-C_{12}{-}C_{14} \!\! & C_{14} \!   \\
 \! -C_{14}     & \!\! C_{12}{+}C_{14} \!\!& -C_{12}         & 0                 & C_{12}            & 0         \\
 0              & -C_{14}       & \!\!C_{12}{+}C_{14} \!\!  & -C_{12}           & 0                 & C_{12} \!   \\
 \! -C_{12}     & 0             & -C_{14}            & 0                 & -C_{12}            & 0         \\
\end{array}
\right),\label{eqn:C1mat_m=0}
\end{equation}
in the absence of sublattice mass, and
\begin{equation}
C^{\{12,3\}}_{(1),m\neq 0} = \left(
\begin{array}{cccccc}
 0 & C_{12} & C_{13} & C_{14} & C_{13} & C_{12} \\
 \!-C_{12} & 0 & C_{12}{-}C_{13} & \!\!-C_{12}{-}C_{13}{-}C_{14}\!\! & C_{13}{+}C_{14} & -C_{13} \\
 \!-C_{13} & C_{13}{-}C_{12} & 0 & C_{12} & \!\!-C_{12}{-}C_{13}{-}C_{14}\!\! & 2 C_{13}{+}C_{14}\! \\
 \!-C_{14} & \!\!C_{12}{+}C_{13}{+}C_{14}\!\! & -C_{12} & 0 & C_{12}{+}C_{13} & -C_{13} \\
 \!-C_{13} & -C_{13}{-}C_{14} & \!\!C_{12}{+}C_{13}{+}C_{14}\!\! & -C_{12}{-}C_{13} & 0 & C_{12} \\
 \!-C_{12} & C_{13} & -2 C_{13}{-}C_{14} & C_{13} & -C_{12} & 0
   \\
\end{array}
\right),\label{eqn:C1mat_m!=0}
\end{equation}
in the presence of $m$. One reduces Eq.~(\ref{eqn:C1mat_m!=0}) to Eq.~(\ref{eqn:C1mat_m=0}) by setting $C_{13}=0$. 
For the $\{12,4\}$ Haldane model we find:
\begin{equation}
C^{\{12,4\}}_{(1),m=0} = \left(
\begin{array}{cccccc}
 0 & C_{12} & C_{13} & -C_{12}-2 C_{13} & C_{13} & C_{12} \\
 -C_{12} & 0 & C_{12} & C_{13} & -C_{12}-2 C_{13} & C_{13} \\
 -C_{13} & -C_{12} & 0 & C_{12} & C_{13} & -C_{12}-2 C_{13} \\
 C_{12}+2 C_{13} & -C_{13} & -C_{12} & 0 & C_{12} & C_{13} \\
 -C_{13} & C_{12}+2 C_{13} & -C_{13} & -C_{12} & 0 & C_{12} \\
 -C_{12} & -C_{13} & C_{12}+2 C_{13} & -C_{13} & -C_{12} & 0 \\
\end{array}
\right),\label{eqn:C1mat_m=0_124}
\end{equation}
in the absence of sublattice mass, and
\begin{equation}
C^{\{12,4\}}_{(1),m\neq0} = \left(
\begin{array}{cccccc}
 0 & C_{12} & C_{13} & C_{14} & C_{13} & C_{12} \\
 -C_{12} & 0 & C_{12} & -C_{12}-C_{13}-C_{14} & C_{14} & -C_{12}-C_{13}-C_{14} \\
 -C_{13} & -C_{12} & 0 & C_{12} & C_{13} & C_{14} \\
 -C_{14} & C_{12}+C_{13}+C_{14} & -C_{12} & 0 & C_{12} & -C_{12}-C_{13}-C_{14} \\
 -C_{13} & -C_{14} & -C_{13} & -C_{12} & 0 & C_{12} \\
 -C_{12} & C_{12}+C_{13}+C_{14} & -C_{14} & C_{12}+C_{13}+C_{14} & -C_{12} & 0 \\
\end{array}
\right),\label{eqn:C1mat_m!=0_124}
\end{equation}
in the presence of $m$. One reduces Eq.~(\ref{eqn:C1mat_m!=0_124}) to Eq.~(\ref{eqn:C1mat_m=0_124}) by setting $C_{14}=-C_{12}-2C_{13}$. 

\subsection{Matrices of second Chern numbers}\label{app:complete-C2-matrices}

We next analyze the second Chern numbers. 
First, for models with a 4D BZ there is a single independent second Chern number, which is constrained by symmetries as given by Eq.~(\ref{eqn:C2-constraint-app}). 
We have discussed in Sec.~\ref{eqn:Chern-theory} of the main text that this equation poses no actual restriction for the particular cases of $\{8,3\}$, $\{8,4\}$, $\{6,4\}$, and $\{10,5\}$, i.e., the second Chern number for Haldane models on each of those lattices is not constrained by symmetry.

Finally, we consider the matrix of second Chern numbers for Haldane models on lattices with a 6D BZ.
By combining Eq.~(\ref{eqn:C2-constraint-6D}) with the point-group matrices $M_g$ listed in Appendix~\ref{app:derive-M-matrices}, we find the following symmetry-constrained forms of the matrices of second Chern numbers.
For the $\{7,3\}$ lattice, we obtain:
\begin{equation}
\mathscr{C}^{\{7,3\}}_{(2)} = \left(\begin{array}{cccccc}
0       &   \mathscr{C}^{12}   &   \mathscr{C}^{12}  &   0 &   0  &   -\mathscr{C}^{12}      \\
-\mathscr{C}^{12}   &   0       &   \mathscr{C}^{12}  &   \mathscr{C}^{12}       &   0       &   0  \\
-\mathscr{C}^{12} &   -\mathscr{C}^{12} &   0       &   \mathscr{C}^{12} &   \mathscr{C}^{12}       &   0 \\
0  &   -\mathscr{C}^{12}       &   -\mathscr{C}^{12}  &   0       &   \mathscr{C}^{12}  &   \mathscr{C}^{12}  \\
0 &  0       &   -\mathscr{C}^{12}       &   -\mathscr{C}^{12} &   0       &   \mathscr{C}^{12}       \\
\mathscr{C}^{12}       &   0 &   0  &   -\mathscr{C}^{12} &   -\mathscr{C}^{12}       &   0
\end{array}\right).\label{eqn:secondC73}
\end{equation}
For $\{12,3\}$ in the absence of the sublattice mass, we obtain:
\begin{equation}
\mathscr{C}_{(2),m=0}^{\{12,3\}}=\left(
\begin{array}{cccccc}
 0 & \mathscr{C}^{12} & \mathscr{C}^{12} & \mathscr{C}^{14} & \mathscr{C}^{14}-\mathscr{C}^{12} & \mathscr{C}^{14}-\mathscr{C}^{12} \\
 -\mathscr{C}^{12} & 0 & \mathscr{C}^{12} & \mathscr{C}^{12} & \mathscr{C}^{14} & \mathscr{C}^{14}-\mathscr{C}^{12} \\
 -\mathscr{C}^{12} & -\mathscr{C}^{12} & 0 & \mathscr{C}^{12} & \mathscr{C}^{12} & \mathscr{C}^{14} \\
 -\mathscr{C}^{14} & -\mathscr{C}^{12} & -\mathscr{C}^{12} & 0 & \mathscr{C}^{12} & \mathscr{C}^{12} \\
 \mathscr{C}^{12}-\mathscr{C}^{14} & -\mathscr{C}^{14} & -\mathscr{C}^{12} & -\mathscr{C}^{12} & 0 & \mathscr{C}^{12} \\
 \mathscr{C}^{12}-\mathscr{C}^{14} & \mathscr{C}^{12}-\mathscr{C}^{14} & -\mathscr{C}^{14} & -\mathscr{C}^{12} & -\mathscr{C}^{12} & 0 \\
\end{array}
\right),\label{eqn:C2mat_m=0}
\end{equation}
whereas in the presence of $m$ we find
\begin{equation}
\mathscr{C}_{(2),m\neq 0}^{\{12,3\}}=
\left(
\begin{array}{cccccc}
 0 & \mathscr{C}^{12} & \mathscr{C}^{13} & \mathscr{C}^{14} & -2 \mathscr{C}^{12}+\mathscr{C}^{13}+\mathscr{C}^{14} & \mathscr{C}^{14}-\mathscr{C}^{12} \\
 -\mathscr{C}^{12} & 0 & \mathscr{C}^{12} & 2 \mathscr{C}^{12}-\mathscr{C}^{13} & \mathscr{C}^{14} & \mathscr{C}^{14}-\mathscr{C}^{13} \\
 -\mathscr{C}^{13} & -\mathscr{C}^{12} & 0 & \mathscr{C}^{12} & \mathscr{C}^{13} & \mathscr{C}^{14} \\
 -\mathscr{C}^{14} & \mathscr{C}^{13}-2 \mathscr{C}^{12} & -\mathscr{C}^{12} & 0 & \mathscr{C}^{12} & 2 \mathscr{C}^{12}-\mathscr{C}^{13} \\
 2 \mathscr{C}^{12}-\mathscr{C}^{13}-\mathscr{C}^{14} & -\mathscr{C}^{14} & -\mathscr{C}^{13} & -\mathscr{C}^{12} & 0 & \mathscr{C}^{12} \\
 \mathscr{C}^{12}-\mathscr{C}^{14} & \mathscr{C}^{13}-\mathscr{C}^{14} & -\mathscr{C}^{14} & \mathscr{C}^{13}-2 \mathscr{C}^{12} & -\mathscr{C}^{12} & 0 \\
\end{array}
\right).\label{eqn:C2mat_m!=0}
\end{equation}
One observes that the effect of setting $m=0$ on the $\{12,3\}$ lattice is that $\mathscr{C}^{13}=\mathscr{C}^{12}$. 
Note that the structure of the matrices $\mathscr{C}_{(2),m=0}^{\{12,3\}}$ and $\mathscr{C}_{(2),m\neq 0}^{\{12,3\}}$ of second Chern numbers is \emph{different} from the structure of matrices $C^{\{12,3\}}_{(1),m = 0}$ and $C^{\{12,3\}}_{(1),m\neq 0}$ of first Chern numbers [see~Eqs.~(\ref{eqn:C1mat_m=0}) and~(\ref{eqn:C1mat_m!=0})].

Finally, for $\{12,4\}$ in the absence of the sublattice mass we obtain:
\begin{equation}
\mathscr{C}_{(2),m=0}^{\{12,4\}}=\left(
\begin{array}{cccccc}
 0 & \mathscr{C}_{12} & \mathscr{C}_{13} & 2 \mathscr{C}_{13}-\mathscr{C}_{12} & \mathscr{C}_{13} & \mathscr{C}_{12} \\
 -\mathscr{C}_{12} & 0 & \mathscr{C}_{12} & \mathscr{C}_{13} & 2 \mathscr{C}_{13}-\mathscr{C}_{12} & \mathscr{C}_{13} \\
 -\mathscr{C}_{13} & -\mathscr{C}_{12} & 0 & \mathscr{C}_{12} & \mathscr{C}_{13} & 2 \mathscr{C}_{13}-\mathscr{C}_{12} \\
 \mathscr{C}_{12}-2 \mathscr{C}_{13} & -\mathscr{C}_{13} & -\mathscr{C}_{12} & 0 & \mathscr{C}_{12} & \mathscr{C}_{13} \\
 -\mathscr{C}_{13} & \mathscr{C}_{12}-2 \mathscr{C}_{13} & -\mathscr{C}_{13} & -\mathscr{C}_{12} & 0 & \mathscr{C}_{12} \\
 -\mathscr{C}_{12} & -\mathscr{C}_{13} & \mathscr{C}_{12}-2 \mathscr{C}_{13} & -\mathscr{C}_{13} & -\mathscr{C}_{12} & 0 \\
\end{array}
\right),\label{eqn:C2mat_m=0_124}
\end{equation}
whereas in the presence of $m$ we find
\begin{equation}
\mathscr{C}_{(2),m\neq 0}^{\{12,4\}}=
\left(
\begin{array}{cccccc}
 0 & \mathscr{C}_{12} & \mathscr{C}_{13} & \mathscr{C}_{14} & \mathscr{C}_{13} & \mathscr{C}_{12} \\
 -\mathscr{C}_{12} & 0 & \mathscr{C}_{12} & \mathscr{C}_{12}-\mathscr{C}_{13}+\mathscr{C}_{14} & \mathscr{C}_{14} & \mathscr{C}_{12}-\mathscr{C}_{13}+\mathscr{C}_{14} \\
 -\mathscr{C}_{13} & -\mathscr{C}_{12} & 0 & \mathscr{C}_{12} & \mathscr{C}_{13} & \mathscr{C}_{14} \\
 -\mathscr{C}_{14} & -\mathscr{C}_{12}+\mathscr{C}_{13}-\mathscr{C}_{14} & -\mathscr{C}_{12} & 0 & \mathscr{C}_{12} & \mathscr{C}_{12}-\mathscr{C}_{13}+\mathscr{C}_{14} \\
 -\mathscr{C}_{13} & -\mathscr{C}_{14} & -\mathscr{C}_{13} & -\mathscr{C}_{12} & 0 & \mathscr{C}_{12} \\
 -\mathscr{C}_{12} & -\mathscr{C}_{12}+\mathscr{C}_{13}-\mathscr{C}_{14} & -\mathscr{C}_{14} & -\mathscr{C}_{12}+\mathscr{C}_{13}-\mathscr{C}_{14} & -\mathscr{C}_{12} & 0 \\
\end{array}
\right).\label{eqn:C2mat_m!=0_124}
\end{equation}
One observes that the effect of setting $m=0$ in the $\{12,4\}$ lattice amonths to the condition $\mathscr{C}^{14}=2\mathscr{C}^{13}-2\mathscr{C}^{12}$. 
As above, the structure of the matrices $\mathscr{C}_{(2),m=0}^{\{12,4\}}$ and $\mathscr{C}_{(2),m\neq 0}^{\{12,4\}}$ of second Chern numbers is again \emph{different} from the structure of matrices $C^{\{12,4\}}_{(1),m = 0}$ and $C^{\{12,4\}}_{(1),m\neq 0}$ of first Chern numbers [see~Eqs.~(\ref{eqn:C1mat_m=0_124}) and~(\ref{eqn:C1mat_m!=0_124})].
\end{widetext}

\subsection{Generalized Haldane models on \texorpdfstring{$\{6,4\}$}{(6,4)} lattice}\label{app:6-4-other}

In Sec.~\ref{sec:flux-patterns} of the main text we consider generalized Haldane models on the $\{6,4\}$ lattice, which are characterized by three $\ztwo=\{\pm 1\}$ numbers $(s_a,s_b,s_c)$. 
These numbers indicate the composition ($s_j=+1$) resp.~the absence of a composition ($s_j = -1$) of reflections $a,b,c$ with the operator of time reversal $\mcT$, and they relate the magnetic flux through adjacent Schwarz triangles as illustrated in Fig.~\ref{fig:flux_config} of the main text. 

Let us use the symbol $C_{(1),m=0}^{\{6,4\},(s_a,s_b,s_c)}$ to indicate the symmetry-constrained matrix of first Chern number for the generalized Haldane model on the $\{6,4\}$ lattice in the absence of sublattice mass and for a flux pattern characterized by signs $(s_a,s_b,s_c)$. 
By properly adjusting the signs $\varsigma_j = -s_j = +1$ resp.~$\varsigma_{j\mcT} = -s_j = -1$ in Eq.~(\ref{eqn:C1-constraint-supp}), we find the following nontrivial results~\cite{Chen:2023:SDC}. 
With generators $a\mcT,b,c\mcT$ we obtain 
\begin{equation}
C^{\{6,4\},(+,-,+)}_{(1),m=0}=\left(\begin{array}{cccc}
0       &   0      &   C_{13}     &   0  \\
0 &   0           &  0    &   C_{13} \\
-C_{13} &   0     &   0           &   0  \\
0 &   -C_{13}      &   0     &   0       \\
\end{array}\right),\label{eqn:C64+-+}
\end{equation}
with generators $a\mcT,b,c$ we find the result
\begin{equation}
C^{\{6,4\},(+,-,-)}_{(1),m=0}=\left(\begin{array}{cccc}
0       &   0      &   C_{13}     &   0  \\
0 &   0           &  0    &   -C_{13} \\
-C_{13} &   0     &   0           &   0  \\
0 &   C_{13}      &   0     &   0       \\
\end{array}\right),\label{eqn:C64+--}
\end{equation}
whereas with generators $a,b\mcT,c\mcT$ we derive
\begin{equation}
C^{\{6,4\},(-,+,+)}_{(1),m=0}=\left(\begin{array}{cccc}
0 &   C_{12}     &   0     &   -C_{12}  \\
-C_{12} &   0     &   -2C_{12}     &   0 \\
0 &   2C_{12}     &   0     &   C_{12}  \\
C_{12} &   0     &   -C_{12}     &   0       \\
\end{array}\right).\label{eqn:C64-++}
\end{equation}
The results in Eqs.~(\ref{eqn:C64+-+}--\ref{eqn:C64-++}) are nontrivial in the sense that some of the momentum-space Chern numbers can be nonzero despite the presence of reflection symmetry in the corresponding models, which is known to enforce vanishing real-space Chern number.
In contrast, the obtained results are trivial for the the other five combinations of $(s_a,s_b,s_c)$, namely: when the triplet of signs is either of $(+,+,-)$, $(-,+,-)$, $(-,-,+)$, $(-,-,-)$, then the momentum-space Chern number are all zero as intuitively expected in the presence of reflection symmetry; on the other hand, for $(+,+,+)$ there are no reflection symmetries and we get back to the result in Eq.~(\ref{eqn:C64,M=0}).

\FloatBarrier
\clearpage

\bibliography{main}

\end{document}